\documentclass[11pt]{article}

\usepackage[authoryear]{natbib}
\usepackage{amsmath}
\usepackage{graphicx}
\usepackage{caption}
\usepackage{subcaption}
\usepackage{comment}
\usepackage{amsthm, amsmath}
\usepackage{amsmath}
\usepackage{amsthm}
\usepackage{amsmath, amssymb}

\usepackage[dvips]{epsfig}
\usepackage{delarray}
\usepackage{hyperref}
\usepackage{url}
\usepackage{anysize}
\usepackage{multirow}
\usepackage{enumerate}

\usepackage{graphicx}
\usepackage{verbatim}
\usepackage{multirow}

\marginsize{1.2in}{1.2in}{.8in}{.8in}

\newcommand{\sgn}{\mbox{sgn}}

\newcommand{\real}{I\kern-0.37emR}
\newcommand{\bx}{\mathbf{x}}
\newcommand{\by}{\boldsymbol{y}}
\newcommand{\ba}{\mathbf{a}}
\newcommand{\bd}{\mathbf{d}}

\newcommand{\bX}{\mathbf{X}}
\newcommand{\bG}{\mathbf{G}}

\newcommand{\bB}{\mathbf{B}}
\newcommand{\bD}{\mathbf{D}}
\newcommand{\bQ}{\mathbf{Q}}
\newcommand{\bP}{\mathbf{P}}

\newcommand{\bI}{\mathbf{I}}
\newcommand{\bH}{\mathbf{H}}
\newcommand{\bM}{\mathbf{M}}

\newcommand{\bb}{\mathbf{b}}
\newcommand{\bz}{\mathbf{z}}

\newcommand{\bv}{\mathbf{v}}

\newcommand{\be}{\mathbf{e}}
\newcommand{\bzero}{\mathbf{0}}

\newcommand{\bbeta}{\boldsymbol{\beta}}

\newcommand{\bmu}{\boldsymbol{\mu}}

\newcommand{\btheta}{\boldsymbol{\theta}}

\newcommand{\bepsilon}{\boldsymbol{\varepsilon}}

\newcommand{\hbbeta}{\widehat{\bbeta}}
\newcommand{\hbmu}{\widehat{\bmu}}

\newcommand{\beq}{\begin{equation}}
\newcommand{\eeq}{\end{equation}}
\newcommand{\bed}{\begin{definition}}
\newcommand{\eed}{\end{definition}}

\newcommand{\cA}{\mathcal{A}}
\newcommand{\cB}{\mathcal{B}}

\newcommand{\cM}{\mathcal{M}}

\newcommand{\sameA}{\overset{\cA}{\sim}}

\newcommand{\blue}[1]{{\color{blue}#1}}

\newtheorem{theorem}{Theorem}[section]

\newtheorem{corollary}{Corollary}[section]
\newtheorem{proposition}{Proposition}[section]
\newtheorem{condition}{Condition}[section]

\newtheorem{definition}{Definition}[section]

\allowdisplaybreaks
\begin{document}
\title{Homogeneity in Regression}
\date{}

\author{Tracy Ke\thanks{The research was partially supported by the National Institute of General Medical Sciences of the National Institutes of Health through Grant Numbers R01-GM072611 and R01-GM100474 and National Science Foundation grants DMS-1206464.}\;,
Jianqing Fan$^*$ and Yichao Wu$^\dag$
\medskip\\{\normalsize $^*$Department of Operations Research and Financial Engineering,  Princeton University}
\medskip\\{\normalsize $^\ddag$ Department of Statistics, North Carolina State University}}

\date{}

\maketitle

\sloppy


\begin{abstract}
This paper explores the homogeneity of coefficients in high-dimensional regression, which extends the sparsity concept and is more general and suitable for many applications.  Homogeneity arises when one expects regression coefficients corresponding to neighboring geographical regions or a similar cluster of covariates to be approximately the same.
Sparsity corresponds to a special case of homogeneity with a known atom zero.
In this article, we propose a new method called clustering algorithm in regression via data-driven segmentation (CARDS) to explore  homogeneity.  New mathematics are provided on the gain that can be achieved by exploring  homogeneity.  Statistical properties of two versions of CARDS are analyzed.
In particular, the asymptotic normality of our proposed CARDS estimator is established, which reveals better estimation accuracy for homogeneous parameters than that without homogeneity exploration.  When our methods are combined with sparsity exploration, further efficiency can be achieved beyond the exploration of  sparsity alone.  This provides additional insights into the power of exploring low-dimensional strucuture in high-dimensional regression:  homogeneity and sparsity.  The newly developed method is further illustrated by simulation studies and applications to real data.
\end{abstract}


\newpage

\section{Introduction}

Driven by applications in genetics, image processing, etc., high dimensionality has become one of the major themes in statistics. To overcome the difficulty of fitting high dimensional models, one usually assumes that the true parameters lie in a low dimensional subspace. For example, many papers focus on {\it sparsity}, i.e., only a small fraction of coefficients are nonzero. In this article, we consider a more general type of low dimensional structure: {\it homogeneity}, i.e., the coefficients share only a few common clusters of values. A motivating example is the gene network analysis, where it is assumed that genes cluster into groups which play similar functions in molecular processes. It can be modeled as a linear regression problem with groups of homogeneous coefficients.  Similarly, in diagnostic lab tests, one often counts the number of positive results in a battery of medical tests, which implicitly assumes that their regression coefficients (impact) in the joint models are approximately the same.  In spatial-temporal studies, it is not unreasonable to assume the dynamics of neighboring geographical regions are similar, namely, their regression coefficients are clustered.  In the same vein, financial returns of similar sectors of industry share similar loadings on risk factors.

Homogeneity is a more general assumption than sparsity, where the latter can be viewed as a special case of the former with a large group of $0$-value coefficients.  In addition, the atom 0 is known to data analysts.  One advantage of assuming homogeneity rather than sparsity is that it enables us to select more than $n$ variables ($n$ is the sample size). Moreover, identifying the homogeneous groups naturally provides a structure in the covariates, which can be helpful in scientific discoveries.

Regression under the homogeneity setting has been studied in a few literature.  First of all, the fused lasso \citep{Tib05, FHHT2007} can be regarded as an effort of exploring homogeneity, with the assistance of neighborhoods defined according to either time or location.
The difference of our studies is that we do not assume such a neighborhood to be known \textit{a priori}.  The clustering of homogeneous coefficients is completely data-driven.
For example, in the fused Lasso, where given a complete ordering of the covariates, \cite{Tib05} add $L_1$ penalties to the pair of adjacent coordinates; in the case without a complete ordering, they suggest penalizing the pair of `neighboring' nodes in the sense of a general distance measure.
\cite{OSCAR} propose the method OSCAR where a special octagonal shrinkage penalty is applied to each pair of coordinates to promote equal-value solutions.
\cite{ShenHuang10} develop an algorithm called Gouping Pursuit, where they add truncated $L_1$ penalties to the pairwise differences for all pairs of coordinates. However, these methods depend either on a known ordering of the covariates, which is usually not available, or exhaustive pairwise penalties, which may increase the computation complexity when the dimension $p$ is large.

In this article, we propose a new method called Clustering Algorithm in Regression via Data-driven Segmentation (CARDS) to explore  homogeneity. The main idea of CARDS is to take advantage of available estimates without homogeneity structure and shrink those coefficients, that are estimated ``close", further towards homogeneity. In the basic version of CARDS, it first builds an ordering of covariates from a preliminary estimate, then runs a penalized least squares with fused penalties in the new ordering. The number of penalty terms is only $(p-1)$, compared to $p(p-1)/2$ in the exhaustive pairwise penalties. In an advanced version of CARDS, it builds an ``ordered segmentation" on the covariates, which can be viewed as a generalized ordering, and  imposes so-called ``hybrid pairwise penalties", which can be viewed as a generalization of fused penalties. This version of CARDS is more tolerant on possible misorderings in the preliminary estimate.
Compared with other methods for homogeneity, CARDS can successfully deal with the case of unordered covariates. At the same time, it avoids using exhaustive pairwise penalties and can be computationally more efficient than the Grouping Pursuit and OSCAR.

We also provide theoretical analysis on CARDS. It reveals that the sum of squared errors of estimated coefficients is $O_p (K/n)$, where $K$ is the number of true homogeneous groups. Therefore, the smaller the number of true groups is, the better precision it can achieve.  In particular, when $K = p$, there is no homogeneity to explore and the result reduces to the case without grouping.  Moreover, in order to exactly recover the true groups with high probability, the minimum signal strength (the gaps between different groups) is of the order $\max_{k}\{\sqrt{|A_k|\log(p)/n}\}$ where $|A_k|$'s are sizes of true groups.
In addition, the asymptotic normality of our proposed CARDS estimator is established, which reveals better estimation accuracy than that without homogeneity exploration.  Furthermore, our results can be further combined with the sparsity results to provide additional insights on the power of the low-dimensional structure in high-dimensional regression:  homogeneity and sparsity.  Our analysis on the basic version of CARDS also establishes a framework for analyzing the fused type of penalties, which is to our knowledge new to the literature.

Throughout this paper, we consider the following linear regression setting
\beq  \label{model_reg}
\by = \bX \bbeta^{0} + \bepsilon,
\eeq
where $\bX=(\bx_1,\cdots,\bx_p)$ is an $n\times p$ design matrix, $\by = (y_1,\cdots,y_n)^T$ is an $n\times 1$ vector of response,
$\bbeta^0=(\beta_1^0,\cdots,\beta_p^0)^T$ denotes the true parameters of interest, and $\bepsilon=(\varepsilon_1, \cdots, \varepsilon_n)^T$ with $\varepsilon_i$'s being independent and identically distributed noises with
$E(\varepsilon_i)= 0$ and $E(\varepsilon_i^2)= \sigma^2$. We assume further that there is a partition of $\{1, 2, \cdots, p\}$ denoted as $\cA=(A_0, A_1, \cdots, A_K)$ such that
\beq  \label{model_homo}
\beta_i^0=\beta^0_{A,k} \qquad \text{ for all } i\in A_k,
\eeq
where $\beta^0_{A,k}$ is the common value shared by all indices in $A_k$. By default, $\beta^0_{A,0}=0$, so $A_0$ is the group of $0$-value coefficients. This allows us to explore  homogeneity and sparsity simultaneously.  Write $\bbeta^0_A=(\beta^0_{A,1},\cdots, \beta^0_{A,K})^T$.
Without loss of generality, we assume $\beta^0_{A,1} < \beta^0_{A,2} < \cdots < \beta^0_{A,K}$.

Our theory and methods are stated for the standard least-squares problem although they can be adapted to other more sophisticated models.  For example, when forecasting housing appreciation
in the United States \citep{FLQ2011}, one builds the spatial-temporal model
\beq  \label{model_spacetime}
  Y_{it} = \bX_{it}^T \bbeta_i + \varepsilon_{it},
\eeq
in which $i$ indicates a spatial location and $t$ indicates time.  It is expected that $\bbeta_i's$ are approximately the same for neighboring zip codes $i$ and this type of  homogeneity can be explored in a similar fashion.  Similarly, when $Y_{it}$ represents the returns of a stock and $\bX_{it}=\bX_t$ stands for risk factors, one can assume certain degree of homogeneity within a sector of industry; namely, the factor loading vector $\bbeta_i$ is approximately the same.

Throughout this paper, $\mathbb{R}$ denotes the set of real numbers, and for a positive integer $p$, $\mathbb{R}^p$ denotes the $p$-dimensional real Euclidean space. For any positive sequences $\{a_n\}$ and $\{b_n\}$, we write $a_n\gg b_n$ if $a_n/b_n$ tends to infinity as $n$ increases to infinity.
Given $1\leq q\leq \infty$, for any vector $\bx$, $\Vert \bx \Vert_q = (\sum_{j} |x_j|^q)^{1/q}$ denotes the $L_q$-norm of $\bx$. In particular, $\Vert \bx \Vert_{\infty}=\max\{|x_j|\}$.
For any matrix $\bM$, $\Vert \bM\Vert_q = \max_{\bx: \Vert \bx\Vert_q=1}\Vert \bM\bx\Vert_q$ denotes the matrix
$L_q$-norm of $\bM$. In particular, $\Vert \bM\Vert_{\infty}$ is the maximum absolute row sum of $\bM$.
We omit the subscript $q$ when $q=2$.
$\Vert \bM\Vert_{\max} = \max\{|M_{ij}|\}$ denotes the maxtrix max norm.
When $\bM$ is symmetric, $\lambda_{\max}(\bM)$ and $\lambda_{\min}(\bM)$ denote the maximum and minimum eigenvalues of $\bM$, respectively.

The rest of the paper is organized as follows. Section 2 describes CARDS, including the basic and advanced versions. Section 3 states theoretical properties of the basic version of CARDS, and Section 4 analyzes the advanced version. Sections 5 and 6 present the results of simulation studies and real data analysis. Section 7 contains concluding remarks. Proofs can be found in Section 8.

\section{CARDS: a data-driven pairwise shrinkage procedure} \label{sec:method}

\subsection{Basic version of CARDS}

Without considering the homogeneity assumption \eqref{model_homo}, there are many methods available for fitting model \eqref{model_reg}. Let $\widetilde{\bbeta}$ be such a preliminary estimator. A very simple idea to generate homogeneity is as follows: first, rearrange the coefficients in $\widetilde{\bbeta}$ in the ascending order; second, group together those adjacent indices whose coefficients in $\widetilde{\bbeta}$ are close; finally, force indices in each estimated group to share a common coefficient and refit model \eqref{model_reg}. A main problem of this naive procedure is how to group the indices. Alternatively, we can run a penalized least squares to simultaneously extract the grouping structure and estimate coefficients. To shrink coefficients of adjacent indices (after reordering) towards homogeneity, we can add fused penalties, i.e., $\{|\beta_{i+1}-\beta_i|, i=1,\cdots, p-1\}$ are penalized. This leads to the following two-stage procedure:

\begin{itemize}
\item {\bf Preordering}: Construct the rank statistics $\{\tau(j): 1\leq j\leq p\}$ such that $\tilde{\beta}_{\tau(j)}$ is the $j$-th smallest value in $\{\tilde{\beta}_i, 1\leq i\leq p\}$, i.e.,
    \beq  \label{tau}
    \tilde{\beta}_{\tau(1)}\leq \tilde{\beta}_{\tau(2)}\leq \cdots \leq \tilde{\beta}_{\tau(p)}.
    \eeq
\item {\bf Estimation}: Given a folded concave penalty function $p_\lambda(\cdot)$ \citep{FL2001} with a regularization parameter $\lambda$, let
\beq \label{nCARDS}
\widehat{\bbeta} = \arg\min_{\bbeta} \Big\lbrace \frac{1}{2n}\Vert \by - \bX\bbeta\Vert^2 +  \sum_{j = 1}^{p-1} p_\lambda(|\beta_{\tau(j+1)} - \beta_{\tau(j)}|)  \Big\rbrace.
\eeq
\end{itemize}

We call this two-stage procedure the basic version of CARDS (bCARDS). In the first stage, it establishes a data-driven rank mapping $\tau(\cdot)$ from the preliminary estimator $\widetilde{\bbeta}$. In the second stage, only ``adjacent" coefficient pairs under the order $\tau$ are penalized, resulting in only $(p-1)$ penalty terms in total. In addition, note that \eqref{nCARDS} does not require  that $\beta_{\tau(j)}\leq \beta_{\tau(j+1)}$. This allows coordinates in $\widehat{\bbeta}$ to have a different order of increasing values from that in $\widetilde{\bbeta}$.

With an appropriately large tuning parameter $\lambda$, $\widehat{\bbeta}$ is a piecewise constant vector in the order of $\tau(\cdot)$ and consequently its elements have homogeneous groups.
In Section \ref{sec:theory_basic}, we shall show that,  if $\tau$ is from a rank consistent estimate of $\bbeta^0$, namely
\beq  \label{cond_tau}
\beta^0_{\tau(1)}\leq \beta^0_{\tau(2)}\leq \cdots \leq \beta^0_{\tau(p)},
\eeq
then under some regularity conditions, $\widehat{\bbeta}$ can consistently estimate the true coefficient groups of $\bbeta^0$ with high probability.

When $p_\lambda(\cdot)$ is a folded-concave penalty function (e.g. SCAD, MCP), \eqref{nCARDS} is a non-convex optimization problem. It is generally difficult  to compute the global minimum. The local linear approximation (LLA) algorithm can be applied to produce a certain local minimum for any fixed initial solution; see
\cite{LLA, Lingzhou12} and references therein for details.

\subsection{Advanced version of CARDS}  \label{subsec:CARDS}

To guarantee the success of CARDS, \eqref{cond_tau} is an essential condition. To be more specific, \eqref{cond_tau} requires that within each true group $A_k$, the order of the coordinates can be arbitrarily shuffled, but for $(i, j)$ belonging to different true groups, if $\beta^0_i < \beta^0_j$, $\tau(i)< \tau(j)$ must hold. This imposes fairly strong conditions on the preliminary estimator $\widetilde{\bbeta}$. For example, \eqref{cond_tau} can be easily violated if $\Vert \widetilde{\bbeta} - \bbeta^0\Vert_{\infty}$ is larger than the minimum gap between groups. 
To relax such a restrictive requirement, we now introduce an advanced version of CARDS, where the main idea is to use less information from $\widetilde{\bbeta}$ and to add more penalty terms in \eqref{nCARDS}.


We first introduce the \emph{ordered segmentation}, which can be viewed as a generalized ordering.
It is similar to letter grades assigned to a class.
\bed \label{def:seg}
For a positive integer $L$, the mapping $\Upsilon: \{1,\cdots, p\}\to \{1,\cdots. L\}$ is called an ordered segmentation if the sets $B_l\equiv \{1\leq j\leq p: \Upsilon(j)=l\}$, $1\leq l\leq L$, form a partition of $\{1,\cdots, p\}$.
\eed
\noindent Each set $B_l$ is called a {\it segment}. When $L=p$, $\Upsilon$ is a one-to-one mapping and it defines a complete ordering. When $L<p$, only the segments $\{B_1,\cdots, B_L\}$ are ordered, but the order of coordinates within each segment is not defined.

In the basic version of CARDS, the preliminary estimator $\widetilde{\bbeta}$ produces a complete rank mapping $\tau$. Now in the advanced version of CARDS, instead of extracting a complete ordering, we only extract an ordered segmentation $\Upsilon$ from $\widetilde{\bbeta}$.  The analogue is similar to grading an exam:  overall score rank (percentile rank) versus letter grade.  Let $\delta>0$ be a predetermined parameter.  First, obtain the rank mapping $\tau$ as in \eqref{tau} and find all indices $i_2 < i_3 < \cdots < i_L$ such that the gaps
\[
\tilde{\beta}_{\tau(j)}-\tilde{\beta}_{\tau(j-1)} > \delta, \qquad j=i_2,\cdots, i_L.
\]
Then, construct the segments
\beq  \label{segments}
B_l = \{ \tau(i_l), \tau(i_{l}+1),  \cdots, \tau(i_{l+1}-1) \}, \qquad l=1,\cdots, L,
\eeq
where $i_1=1$ and $i_{L+1}=p+1$.  This process is indeed similar to the letter grade that we assign.
The intuition behind this construction is that when $\tilde{\beta}_{\tau(k+1)}\leq \tilde{\beta}_{\tau(k)}+\delta$, i.e., the estimated coefficients of two ``adjacent coordinates" differ by only a small amount, we do not trust the ordering between them and group them into a same segment. Compared to the complete ordering $\tau$, the ordered segments $\{B_1,\cdots, B_L\}$ utilize less information from $\widetilde{\bbeta}$.

Given an ordered segmentation $\Upsilon$, how can we design the penalties so that we can take advantage of the ordering of segments $B_1,\cdots, B_L$ and at the same time allow flexibility of order shuffling within each segment? Towards this goal, we introduce the \emph{hybrid pairwise penalty}.
\bed
Given a penalty function $p_\lambda(\cdot)$ and tuning parameters $\lambda_1$ and $\lambda_2$, the hybrid pairwise penalty corresponding to an ordered segmentation $\Upsilon$ is
\beq \label{hybrid}
P_{\Upsilon,\lambda_1,\lambda_2}(\bbeta) = \sum_{l=1}^{L-1}\sum_{i\in B_l, j\in B_{l+1}} p_{\lambda_1} (|\beta_i -\beta_j|)
+  \sum_{l=1}^{L} \sum_{i,j\in B_l} p_{\lambda_2} (|\beta_i -\beta_j|).
\eeq
\eed
\noindent In \eqref{hybrid}, we call the first part \emph{between-segment} penalty and the second part \emph{within-segment} penalty. The within-segment penalty penalizes all pairs of indices in each segment, hence, it does not rely on any ordering within the segment. The between-segment penalty penalizes pairs of indices from two adjacent segments, and it can be viewed as a ``generalized" fused penalty on segments.

When $L=p$,  each $B_l$ is a singleton and \eqref{hybrid} reduces to the fused penalty in \eqref{nCARDS}. On the other hand, when $L=1$, there is only one segment $B_1=\{1,\cdots, p\}$, and \eqref{hybrid} reduces to the exhaustive pairwise penalty
\beq  \label{TV}
P^{TV}_{\lambda}(\bbeta) =  \sum_{1\leq i,j\leq p} p_\lambda(|\beta_i -\beta_j|).
\eeq
It is also called the total variation penalty, and the case with $p_\lambda(\cdot)$ being a truncated $L_1$ penalty is studied in \cite{ShenHuang10}.
Thus, the penalty \eqref{hybrid} is a generalization of both the fused penalty and the total variation penalty, which explains the name ``hybrid".

Now, we discuss how the condition \eqref{cond_tau} can be relaxed.
Parallel to the definition that $\tau$ preserves the order of $\bbeta^0$, we make the following definition.
\bed \label{def:cond_seg}
An ordered segmentation $\Upsilon$ preserves the order of $\bbeta^0$ if $\max_{j\in B_l}\beta^0_j\leq \min_{j\in B_{l+1}} \beta^0_j$, for $l=1,\cdots, L-1$.
\eed

By the construction \eqref{segments}, even if $\tau$ does not preserve the order of $\bbeta^0$, it is still possible that the resulting $\Upsilon$ does.
Consider a toy example where $p=4$, and $\beta^0_{\tau(1)}=\beta^0_{\tau(2)} = \beta^0_{\tau(4)} < \beta^0_{\tau(3)}$ so that $\{\tau(1),\tau(2),\tau(4)\}$ and $\{\tau(3)\}$ are two true homogeneous groups in $\bbeta^0$. By definition of $\tau$, $\tau$ ranks  $\beta_4^0$ wrongly ahead of $\beta_3^0$ based on the preliminary estimate $\widetilde{\bbeta}$. It is obvious that $\tau$ does not preserve the order of $\bbeta^0$. However, as long as $\tilde{\beta}_{\tau(4)}\leq \tilde{\beta}_{\tau(3)}+\delta$, $\tau(3)$ and $\tau(4)$ are grouped into the same segment in \eqref{segments}, say, $B_1=\{\tau(1), \tau(2)\}$ and $B_2=\{\tau(3), \tau(4)\}$. Then  $\Upsilon$ still preserves the order of $\bbeta^0$ according to the above definition.

Now we formally introduce the advanced version of Clustering Algorithm in Regression via Data-driven Segmentation (aCARDS). It consists of three steps, where the first two steps are very similar to the way that we assign letter grades based on an exam (preliminary estimate).
\begin{itemize}
\item {\bf Preliminary Ranking}:  Given a preliminary estimate $\widetilde{\bbeta}$, generate the rank statistics $\{\tau(j): 1\leq j\leq p\}$ such that $\tilde{\beta}_{\tau(1)}\leq \tilde{\beta}_{\tau(2)}\leq \cdots \leq \tilde{\beta}_{\tau(p)}$.
\item {\bf Segmentation}: For a tuning parameter $\delta>0$, construct an ordered segmentation $\Upsilon$ as described in \eqref{segments}.
\item {\bf Estimation}: For tuning parameters $\lambda_{1}$ and $\lambda_{2}$, compute the solution $\hbbeta$ that minimizes
\beq   \label{CARDS}
Q_n(\bbeta)= \frac{1}{2n}\Vert \by-\bX\bbeta \Vert^2
+ P_{\Upsilon,\lambda_1,\lambda_2}(\bbeta).
\eeq
\end{itemize}

In Section \ref{sec:theory}, we shall show that if $\Upsilon$ preserves the order of $\bbeta^0$, under certain conditions, $\widehat{\bbeta}$ recovers the true homogeneous groups of $\bbeta^0$ with high probability. Therefore, to guarantee the success of this advanced version of CARDS, we need the existence of  a $\delta>0$ for the initial estimate $\widetilde{\bbeta}$ such that the associated $\Upsilon$ preserves the order of $\bbeta^0$. We see from the toy example that even when \eqref{cond_tau} fails, this condition can still hold. So the advanced version of CARDS requires weaker conditions on $\widetilde{\bbeta}$. The main reason is that the hybrid penalty contains penalty terms corresponding to more pairs of indices. Hence, it is more robust to possible mis-ordering in $\tau$. In fact, the basic version of CARDS is a special case with $\delta=0$.

\subsection{CARDS under sparsity}  \label{subsec:sCARDS}

In applications, we may need to explore homogeneity and sparsity simultaneously. Often the preliminary estimator $\widetilde{\bbeta}$ takes into account the sparsity, namely it is obtained with a penalized least-squares method \citep{FL2001, Tib05} or sure independence screening \citep{FanLv2008}. Suppose $\widetilde{\bbeta}$ has the sure screening property, i.e., $S_0\subset \widetilde{S}$ with high probability, where $\widetilde{S}$ and $S_0$ denote the support of $\widetilde{\bbeta}$ and $\bbeta^0$, respectively. We modify CARDS as follows:
In the first two steps, using the non-zero elements of $\widetilde{\bbeta}$, we can similarly construct data-driven hybrid penalties only on coefficients of variables in $\widetilde{S}$. In the third step, we fix $\hbbeta_{\widetilde{S}^c}=\bzero$ and obtain $\hbbeta_{\widetilde{S}}$ by minimizing the following penalized least squares
\beq  \label{sCARDS}
Q_n^{sparse}(\bbeta) = \frac{1}{2n}\Vert \by-\bX_{\widetilde{S}}\bbeta_{\widetilde{S}} \Vert^2
+ P_{\Upsilon,\lambda_1,\lambda_2}(\bbeta_{\widetilde{S}}) + \sum_{j\in\widetilde{S}} p_\lambda(|\beta_j|),
\eeq
where $\bX_{\widetilde{S}}$ is the submatrix of $\bX$ restricted to columns in $\widetilde{S}$. In \eqref{sCARDS}, the second term is the hybrid penalty to encourage homogeneity among coefficients of variables already selected in $\widetilde{\bbeta}$, and the third term is the element-wise penalty to help further filter out falsely selected variables.
We call this modified version the shrinkage-CARDS (sCARDS).

\section{Analysis of the basic CARDS} \label{sec:theory_basic}

In this section, we analyze theoretical properties of the basic CARDS. For simplicity, we assume that there is no group of $0$, i.e., the usual sparsity is not explicitly explored.
We first provide heuristics to two essential questions: (1) How does it help reduce the convergence rate of $\Vert \hbbeta-\bbeta^0\Vert$ by taking advantage of homogeneity? (2) What is the order of minimum signal strength required for recovering the true groups with high probability?
We then formally state our main results.  After that, we will give conditions under which the ordinary least squares can provide a good preliminary estimator, as well as the effect of mis-ranking on CARDS.

\subsection{Heuristics}

Consider an ideal case of orthogonal design $\bX^T\bX=n\bI_p$ (necessarily $p\leq n$). The ordinary least-square estimator $\hbbeta^{ols} = (\bX^T\bX)^{-1}\bX^T\by$ has the decomposition
\[
\widehat{\beta}_j^{ols} = \beta^0_j + \epsilon_j, \qquad \epsilon_j \overset{i.i.d.}{\sim} N(0, n^{-1}),\qquad j=1,\cdots, p.
\]
It is clear by the square-root law that $\Vert \hbbeta^{ols} - \bbeta^0\Vert = O_P (\sqrt{p/n})$. Now, if there are $K$ homogeneous groups in $\bbeta^0$ and we know the true groups, the original model \eqref{model_reg} can be rewritten as
\[
\by = \bX_A\bbeta^0_A + \bepsilon,
\]
where $\bbeta^0_A=(\beta^0_{A,1},\cdots, \beta^0_{A,K})^T$ contains distinct values in $\bbeta^0$, and $\bX_A=(\bx_{A,1},\cdots, \bx_{A,K})$ with $\bx_{A,k}=\sum_{j\in A_k}\bx_j$. The corresponding ordinary least-squares estimator $\hbbeta^{ols}_A=(\bX_A^T\bX_A)^{-1}\bX_A^T\by$ has the decomposition
\beq  \label{heuristics}
\widehat{\beta}_{A,k}^{ols} = \beta^0_{A,k} + \bar{\epsilon}_k, \qquad \bar{\epsilon}_k \sim N(0, \frac{1}{n|A_k|}), \text{ and $\bar{\epsilon}_k$'s are independent.}
\eeq
Here $\bar{\epsilon}_k= \tfrac{1}{|A_k|} \sum_{j\in A_k} \epsilon_j$ is the noise averaged over group $k$. The oracle estimator $\hbbeta^{oracle}$ is defined such that $\widehat{\beta}^{oracle}_j= \widehat{\beta}^{ols}_{A,k}$ for all $j\in A_k$. Then, by the square-root law,
\begin{eqnarray*}
\Vert \hbbeta^{oracle} - \bbeta^0\Vert^2 &=& \sum_{k=1}^K |A_k| |\widehat{\beta}^{ols}_{A,k} - \beta^0_{A,k}|^2\\
 &=& O_p\left(  \sum_{k=1}^K |A_k|\frac{1}{n|A_k|} \right) = O_p\big(K/n),
\end{eqnarray*}
which implies immediately that $\Vert \hbbeta^{oracle}-\bbeta^0\Vert= O_p(\sqrt{K/n})$.

The surprises of the results are two fold: First, the rate $\sqrt{K/n}$ is for $\Vert \hbbeta^{oracle}-\bbeta^0\Vert$ instead of $\Vert \widehat{\bbeta}^{ols}_A - \bbeta^0_A \Vert$. The former can be viewed as duplicate counts of the terms in the latter, hence it can be much larger than the latter. However, since there are $K$ parameters in $\hbbeta_A^{ols}$,
common heuristics in regression analysis give $\Vert \widehat{\bbeta}^{ols}_A - \bbeta^0_A\Vert=O_p (\sqrt{K/n})$, and so the convergence rate of $\Vert \hbbeta^{oracle}-\bbeta^0\Vert$ should be much larger than $\sqrt{K/n}$. The above results seem to be counter-intuitive. The point is that in \eqref{heuristics} the noises are averaged, and so the rate of $\Vert \widehat{\bbeta}^{ols}_A - \bbeta^0_A \Vert$ is much smaller than $\sqrt{K/n}$.
In fact, by taking advantage of homogeneity, we not only estimate much fewer parameters, but also reduce the noise level.

The second surprise is that the rate has nothing to do with the sizes of true homogeneous groups. No matter whether we have $K$ groups of equal size, or one dominating group and $(K-1)$ very small groups, the rate is always the same in the oracle situation. This is also a consequence of noise averaging.

Next, we discuss when the CARDS estimator equals the oracle estimator $\hbbeta^{oracle}$ that is based on the knowledge of the true grouping structure. For simplicity, we still consider the case of orthogonal design  $\bX^T\bX=n\bI_p$, and assume the preliminary ordering $\tau$ preserves the order of $\bbeta^0$ so that the basic version of CARDS works. Write $\tau(j)=j$ without loss of generality. CARDS finds a local solution of
\begin{eqnarray*}
Q_n(\bbeta)&=& \frac{1}{2n}\Vert \by- \bX\bbeta \Vert^2 + \sum_{j=1}^{p-1}p_\lambda(|\beta_{j+1}-\beta_j|)\\
&=& \frac{1}{2n}\Vert \by - \bX\bz\Vert^2 + \frac{1}{2}\Vert \bz - \bbeta\Vert^2 + \sum_{j=1}^{p-1}p_\lambda(|\beta_{j+1}-\beta_j|),
\end{eqnarray*}
where $\bz= n^{-1}\bX^T\by$ is the vector of marginal correlations (when $\by$ is also normalized).
As a result, if the estimator produced by CARDS is the oracle estimator $\hbbeta^{oracle}$, necessarily $\hbbeta^{oracle}$ has to satisfy the KKT condition
\beq  \label{KKT}
\left\lbrace
\begin{array}{lr}
- (z_1- \widehat{\beta}^{oracle}_1) -  \bar{p}_\lambda(d\widehat{\beta}^{oracle}_2) = 0, &  \\
- (z_j- \widehat{\beta}^{oracle}_j) +  \bar{p}_\lambda(d\widehat{\beta}^{oracle}_{j}) -  \bar{p}_\lambda(d\widehat{\beta}^{oracle}_{j+1}) = 0 &  2\leq j\leq p-1, \\
- (z_p- \widehat{\beta}^{oracle}_p) +  \bar{p}_\lambda(d\widehat{\beta}^{oracle}_{p}) = 0,
\end{array}
\right.
\eeq
where $d\hbbeta^{oracle}_j = \hbbeta^{oracle}_{j}-\hbbeta^{oracle}_{j-1}$ for $2\leq j\leq p$; and $\bar{p}_\lambda(t)=p_\lambda'(|t|)\sgn(t)$ with $\sgn(t)=1$ for $t>0$, $-1$ for $t<0$, and any value on $[-1,1]$ for $t=0$.
Write the true groups as $A_k = \{j_k, j_k+1, \cdots, j_{k+1}-1\}$, $1\leq k\leq K$, for some $1=j_1<j_2<\cdots <j_K<j_{K+1}=p+1$. It is not hard to show that the sufficient and necessary conditions for \eqref{KKT} to hold are
\beq  \label{KKT_equiv}
\left\lbrace
\begin{array}{l}
\bar{p}_\lambda (d\widehat{\beta}^{oracle}_{j_k}) =0, \qquad 2\leq k \leq  K, \\
\big|\sum_{i=j_k}^j (\widehat{\beta}_i^{oracle} - z_i)\big|\leq p'_{\lambda}(0+), \quad 1\leq k\leq K, \ j_k< j\leq j_{k+1}-1.\
\end{array}
\right.
\eeq
Here $d\hbbeta^{oracle}_{j_{k}}$ is the estimated coefficient gap between groups $A_{k-1}$ and $A_k$ in the oracle estimator, and it is equal to $d\beta^0_{j_k}+\bar{\epsilon}_{k}-\bar{\epsilon}_{k-1}$, where $d\beta^0_{j_k}$ is the true coefficient gap between groups $A_{k-1}$ and $A_k$. Also, $\widehat{\beta}_j^{oracle} - z_j = \bar{\epsilon}_k - \epsilon_j$ for $j\in A_k$, which is purely determined by the noises.
Therefore, to guarantee \eqref{KKT_equiv}, the penalty function $p_\lambda(\cdot)$ must have flat tails, i.e., $p'(|t|)=0$ when $|t|> a\lambda$ ($a>0$ is a constant); furthermore, the true coefficient gaps $\{d\beta^0_{j_k}: k=2,\cdots, K\}$, the tuning parameter $\lambda$ and the noises $\{\epsilon_j\}$ need to satisfy, say,
\beq \label{KKT_suffcond}
\left\lbrace
\begin{array}{l}
\min_{2\leq k\leq K} |d\beta^0_{j_k}| \geq 2(a+1)\lambda, \\
\max_{1\leq k\leq K} |\bar{\epsilon}_k| \leq \lambda, \\
\max_{1\leq k\leq K}\max_{j \in A_k} \big|\sum_{i=j_k}^j (\epsilon_j-\bar{\epsilon}_k)\big| \leq p'_{\lambda}(0+).
\end{array}
\right.
\eeq
Note that $p'_{\lambda}(0+)=\lambda$ for most sparsity penalty functions; and $\bar{\epsilon}_k$ is much smaller than $\max_{j\in A_k}|\epsilon_j|$ with high probability. So \eqref{KKT_suffcond} requires that the minimum true coefficient gap between groups satisfies
\beq \label{KKT_suffcond2}
\min_{2\leq k\leq K} |d\beta^0_{j_k}| > C \max_{1\leq k\leq K}\max_{j \in A_k} \big|\sum_{i=j_k}^j \epsilon_j\big|.
\eeq
Using results in \cite{DarlErdos}, the right hand side of \eqref{KKT_suffcond2} is upper bounded by $C \max_k\big\{ \sqrt{\tfrac{|A_k|\log(K \vee \log(|A_k|))}{n}}\big\}$ with high probability, for a sufficiently large constant $C$.
Therefore, for CARDS to produce the oracle estimator, the minimum coefficient gap between true groups should be at least in that order. Up to a logarithmic factor, we write this order as $\max_k \{\sqrt{|A_k|\log(p)/n}\}$.

\subsection{Notations and regularity conditions}  \label{subsec:simple_L2}

Let $\cM_A$ be the subspace of $\mathbb{R}^p$ defined by 
\[
\cM_{A}=\{\bbeta\in\mathbb{R}^{p}: \beta_i=\beta_j, \text{ for any } i,j\in A_k, 1\leq k\leq K\}.
\]
For each $\bbeta\in\cM_A$, we can always write $\bX\bbeta=\bX_A\bbeta_A$, where $\bX_A$ is an $n\times K$ matrix with $\bX_A(i,k)=\sum_{j\in A_k}X(i,j)$ with $X(i,j)$ denoting the $(i,j)$-element of $\bX$, and $\bbeta_A$ is a $K\times 1$ vector with its $k$th component $\beta_{A,k}$ being the common coefficient in group $A_k$. Define the matrix $\bD=\text{diag}(|A_1|^{1/2}, \cdots, |A_K|^{1/2})$.
We introduce the following conditions on the design matrix $\bX$:
\begin{condition}
$\Vert\bx_j\Vert=\sqrt{n}$, for $1\leq j\leq p$. The eigenvalues of the matrix $\tfrac{1}{n}\bD^{-1}\bX_A^T\bX_A\bD^{-1}$ are bounded below by $c_1>0$ and bounded above by $c_2>0$.
\end{condition}
\noindent In the case of orthogonal design, i.e., $\tfrac{1}{n}\bX^T\bX=\bI_p$, the matrix $\tfrac{1}{n}\bD^{-1}\bX_A^T\bX_A\bD^{-1}$ simplifies to $\bI_K$, and $c_1=c_2=1$.


Let $\rho(t)=\lambda^{-1}p_{\lambda}(t)$ and $\bar{\rho}(t)=\rho'(|t|)\sgn(t)$.
We assume that the penalty function $p_\lambda(\cdot)$ satisfies the following condition.
\begin{condition}
$p_\lambda(\cdot)$ is a symmetric function and it is non-descreasing and concave on $[0,\infty)$. $\rho'(t)$ exists and is continuous except for a finite number of $t$ with $\rho'(0+)=1$. There exists a constant $a>0$ such that $\rho(t)$ is a constant for all $|t|\geq a\lambda$.
\end{condition}

We also assume that the noise vector $\bepsilon=(\epsilon_1,\cdots, \epsilon_p)^T$ has sub-Gaussian tails.
\begin{condition}
For any vector $\ba\in\mathbb{R}^n$ and $x>0$, $P(|\ba^T\bepsilon|>\Vert \ba\Vert x)\leq 2e^{-c_3x^2}$, where $c_3$ is a positive constant.
\end{condition}

Given the design matrix $\bX$, let $\bX_k$ be its submatrix formed by including columns in $A_k$, for $1\leq k\leq K$. For any vector $\bv\in\mathbb{R}^q$, let $\mbox{DC}(\bv)=\max_{1\leq i\leq q}|v_i- q^{-1}\sum_{j=1}^qv_j|$ be the ``deviation from centrality". Define
\beq  \label{sigma&nu}
\sigma_k = \lambda_{\max}(\tfrac{1}{n}\bX_k^T\bX_k) \qquad  \mbox{and} \quad \nu_k = \max_{\bmu\in\cM_A: \Vert \bmu\Vert=1} \mbox{DC}(\tfrac{1}{n}\bX_k^T\bX\bmu),
\eeq
where $\lambda_{\max}(\cdot)$ denotes the largest eigenvalue operator.
In the case of orthogonal design, $\sigma_k=1$ and $\nu_k=0$.
Let $b_n=\frac{1}{2}\min_{1\leq k<l \leq K}|\beta^0_{A,k}-\beta^0_{A,l}|$ denote the minimal gap between two groups in $\bbeta^0$, and $\lambda=\lambda_n$ the tuning parameter in the penalty function.

\subsection{Main results}

When the true groups $A_1,\cdots, A_K$ are known, the oracle estimator is
\[
\hbbeta^{oracle}= \arg\min_{\bbeta\in\cM_A}\Big\lbrace \frac{1}{2n}\Vert \by-\bX\bbeta \Vert^2 \Big\rbrace.
\]

\begin{theorem} \label{thm:fuse}
Suppose Conditions 3.1-3.3 hold, $K=o(n)$, and the preliminary estimate $\widetilde{\bbeta}$ generates an order $\tau$ that preserves the order of $\bbeta^0$ with probability at least $1-\epsilon_0$. If $b_n>a\lambda_n$ and
\beq  \label{lam_order}
\lambda_n \gg \max_k \left\lbrace \sqrt{\sigma_k|A_k|\log(p)/n} + (1+\nu_k|A_k|)\sqrt{K\log(n)/n} \right\rbrace,
\eeq
then with probability at least $1-\epsilon_0-n^{-1}K - 2p^{-1}$, $\hbbeta^{oracle}$ is a strictly local minimum of \eqref{nCARDS}. Moreover, $\Vert \hbbeta^{oracle} - \bbeta^0\Vert= O_p(\sqrt{K/n})$.
\end{theorem}

\noindent Theorem \ref{thm:fuse} shows that there exists a local minimum of \eqref{nCARDS} which is equal to the oracle estimator with overwhelming probability. This strong oracle property is a stronger result than the oracle property in \citep{FL2001}.

The bCARDS formulation \eqref{nCARDS} is a non-convex problem and it may have multiple local minima. In practice, we apply the Local Linear Approximation algorithm (LLA) \citep{LLA} to solve it: start from an initial solution $\hbbeta^{(0)}=\hbbeta^{initial}$; at step $m$, update solution by
\[
\widehat{\bbeta}^{(m)} = \arg\min_{\bbeta} \Big\lbrace \frac{1}{2n}\Vert \by - \bX\bbeta\Vert^2 +  \sum_{j = 1}^{p-1} p_\lambda'\big(|\hat{\beta}^{(m-1)}_{\tau(j+1)} - \hat{\beta}^{(m-1)}_{\tau(j)}|\big)\cdot|\beta_{\tau(j+1)} - \beta_{\tau(j)}|  \Big\rbrace.
\]
Given $\hbbeta^{initial}$, this algorithm produces a unique sequence of estimators which converge to a certain local minimum. Theorem \ref{thm:LLA} shows that under certain conditions, the sequence of estimators produced by the LLA algorithm converge to the oracle estimator.

\begin{theorem} \label{thm:LLA}
Under conditions of Theorem \ref{thm:fuse},  suppose $\rho'(\lambda_n)\geq a_0$ for some constant $a_0>0$, and there exists an initial solution $\hbbeta^{initial}$ of \eqref{nCARDS} satisfying $\Vert \hbbeta^{initial} - \bbeta^0\Vert_\infty\leq \lambda_n/2$. Then with probability at least $1-\epsilon_0-n^{-1}K - 2p^{-1}$, the LLA algorithm yields $\hbbeta^{oracle}$ after one iteration, and it converges to $\hbbeta^{oracle}$ after two iterations.
\end{theorem}

The $L_1$ penalty $\rho(t)=|t|$ is widely used in high-dimensional penalization methods partially due to its convexity.
For example, it can be used here to get the initial solution $\hbbeta^{initial}$ for the LLA algorithm. However, this penalty function is excluded in Condition 3.2, and consequently Theorem \ref{thm:fuse} does not apply. Now, we discuss the $L_1$ penalty in more details.

We first relax the requirement that $\tau$ preserves the order of $\bbeta^0$. Instead, we consider the case that $\tau$ is ``consistent" with coefficient groups in $\bbeta^0$, that is, for any two variables in the same true group, variables ranked between them are also in this group (if $\tau$ preserves the order of $\bbeta^0$, $\tau$ belongs to this class). Note that we do not require $\beta^0_{\tau(i)}\leq \beta^0_{\tau(j)}$ for all $i<j$. In this case, recovering the true groups is equivalent to locating jumps (which can have positive or negative magnitudes) in $\bbeta^0$.

Below we introduce an ``irrepresentability" condition.
For $k=1,\cdots, K-1$, write $d\beta^0_{A,k} = \beta^0_{A,k+1}-\beta^0_{A,k}$. Define the $K$-dimensional vector $\bd_0$ by $d^0_1= \sgn(d\beta^0_{A,1})$, $d^0_K = - \sgn(d\beta^0_{A,K-1})$ and
\[
d_k^0 = \sgn(d\beta^0_{A,k}) - \sgn(d\beta^0_{A,k-1}), \qquad 2\leq k \leq K-1.
\]
Here $\bd^0$ is the adjacent difference of the sign vector of jumps in $\bbeta^0$. For example, suppose $K=4$ and the common coefficients in 4 groups satisfy $\beta^0_{A,2}-\beta^0_{A,1}>0$, $\beta^0_{A,3}- \beta^0_{A,2}<0$ and $\beta^0_{A,4}-\beta^0_{A,3}>0$. Then $\bd^0= (1, -2, 2, -1)$.
Also, define the $p$-dimensional vector
\[
\bb^0=\bX^T\bX_A(\bX_A^T\bX_A)^{-1}\bd^0.
\]
In the case of orthogonal design $\bX^T\bX=n\bI_p$, $\bb^0\in\cM_A$ and it has the form $b^0_j=1/|A_k|$ for $j\in A_k$.
For each $j\in A_k$, let
\[
A^1_{kj}=\{\tau(i)\in A_k: i\leq j\}, \qquad A^2_{kj}=\{\tau(i)\in A_k: i>j\}.
\]
Namely, $A^{1,j}_k$ contain indices in group $k$ that have ranks $\leq j$ in the mapping $\tau$, and $A^{2,j}_k$ contain those have ranks $>j$. Write $\theta_{kj} = |A^{1,j}_k|/|A_k|$ as the proportion of indices in group $k$ which is mapped in front of (and including) $\tau(j)$.
Denote $\overline{b}_{kj} = \tfrac{1}{|A^{1,j}_k|}\sum_{\tau(i)\in A^{1,j}_{k}}b^0_{\tau(i)}$ the average of elements in $\bb^0$ over the indices in $A_k^{1,j}$, and $\underline{b}_{kj} = \tfrac{1}{|A^{2,j}_k|}\sum_{\tau(i)\in A^{1,j}_{k}}b^0_{\tau(i)}$ the average of elements in $\bb^0$ over the indices in $A^{2,j}_k$.
The following inequality is called the ``irrepresentability" condition on $\bX$ and $\bbeta^0$: for any $1\leq k\leq K$ and $j\in A_k$, $j\neq j_{k+1}-1$,
\begin{eqnarray}  \label{irrep}
&& 1 - \omega_n \geq    \\
&& \left\lbrace
\begin{array}{lr}
\big| \theta_{1j} \sgn(d\beta^0_{A,1})
+ |A_1|^2 \theta_{1j}(1-\theta_{1j}) \big( \overline{b}_{1j} - \underline{b}_{1j}\big)\big|, \\
\big|(1-\theta_{kj}) \sgn(d\beta^0_{A,k-1}) + \theta_{kj} \sgn(d\beta^0_{A,k})
+ |A_k|^2 \theta_{kj}(1-\theta_{kj}) \big( \overline{b}_{kj} - \underline{b}_{kj}\big)\big|, & 2\leq k\leq K-1, \\
\big|(1-\theta_{Kj}) \sgn(d\beta^0_{A,K-1})
+ |A_K|^2 \theta_{Kj}(1-\theta_{Kj}) \big( \overline{b}_{Kj} - \underline{b}_{Kj}\big)\big|.
\end{array}
\right.  \nonumber
\end{eqnarray}
Here $\{\omega_n\}$ is a positive sequence, which can go to $0$.
In the case of orthogonal design, $\bb^0\in\cM_A$ and $\overline{b}_{kj} - \underline{b}_{kj}=0$ holds for all $k$ and $j\in A_k$. The ``irrepresentability" condition reduces to
\[
1 - \omega_n \geq
\left\lbrace
\begin{array}{lr}
\big| \theta_{1j} \sgn(d\beta^0_{A,1})\big|, \\
\big| (1-\theta_{kj}) \sgn(d\beta^0_{A,k-1}) + \theta_{kj} \sgn(d\beta^0_{A,k})\big|, & 2\leq k\leq K-1, \\
\big|(1-\theta_{Kj}) \sgn(d\beta^0_{A,K-1})\big|.
\end{array}
\right.
\]
This is possible only when
\beq \label{jumpsign}
\sgn(d\beta^0_{A,k-1})\neq \sgn(d\beta^0_{A,k}), \qquad 2\leq k\leq K-1.
\eeq
Noting that $1/|A_k|\leq \theta_{kj}\leq 1-1/|A_k|$, the associated $\omega_n$ can be chosen as $\min_{k}\{1/|A_k|\}$ when \eqref{jumpsign} holds.

\begin{theorem}  \label{thm:fuseL1}
Suppose Conditions 3.1 and 3.3 hold, the ``irrepresentability" condition \eqref{irrep} is satisfied, $K=o(n)$, and the preliminary estimate $\widetilde{\bbeta}$ generates an order $\tau$ that is consistent with $\bbeta^0$ with probability at least $1-\epsilon_0$. If $b_n$ and $\lambda_n$ satisfy
\beq  \label{b_lam_order}
b_n \gg \sqrt{K\log(n)/n} + \lambda_n \Big(\sum_{k=1}^K \tfrac{1}{|A_k|^2}\Big)^{1/2}, \quad
\lambda_n \gg \omega_n^{-1} \max_{k}\left\lbrace \sqrt{\sigma_k|A_k|\log(p)/n} \right\rbrace,
\eeq
then with probability at least $1-\epsilon_0- n^{-1}K - 2p^{-1}$, \eqref{nCARDS} has a unique global minimum $\hbbeta$ such that $\hbbeta\in \cM_A$ and it satisfies the sign restrictions $\sgn(\widehat{\beta}_{A, k+1}-\widehat{\beta}_{A,k})= \sgn(\beta^0_{A, k+1}-\beta^0_{A,k})$, $k=1,\cdots, K-1$. Moreover, $\Vert \hbbeta-\bbeta^0 \Vert = O_p (\sqrt{K/n} + \gamma_n)$, where $\gamma_n=  \lambda_n\big(\sum_{k=1}^K \tfrac{1}{|A_k|}\big)^{1/2}$.
\end{theorem}

Compared to Theorem \ref{thm:fuse}, there is an extra bias term in the $L_2$ estimation error. We consider an ideal case where the sizes of all groups have the same order $s/K$, the sequence $\omega_n\geq \omega$ for some positive constant $\omega$, and $\max_k\sigma_k\leq C$. From \eqref{b_lam_order}, the magnitude of the bias term is $\sqrt{K\log(p)/n}$, which is much larger than $\sqrt{K/n}$. So in the $L_1$ penalty case, it is generally hard to guarantee both exact recovery of the true grouping structure and the $\sqrt{K/n}$-convergence rate of $\Vert \hbbeta-\bbeta^0 \Vert$. Moreover, the ``irrepresentability" condition is very restrictive, even in the orthogonal design case.
From \eqref{jumpsign}, in order to exactly locate all jumps, necessarily all consecutive jumps (in the ordering $\tau$) have opposite signs. However, this is sometimes hard to guarantee. Especially when $\tau$ preserves the order of $\bbeta^0$, all the jumps have positive signs.


\subsection{Preliminary estimator, effects of mis-ranking}

We now give sufficient conditions under which the least-squares estimator induces an order-preserving rank.   When sparsity is explored, after the model selection consistency \citep{FanLv2011, Lingzhou12}, the problem becomes a dense problem.  Hence, the fundamental insights can be gained when the coefficients are not sparse and it will be the case that we focus upon next.

The ordinary least squares estimator
\[
\hbbeta^{ols} = \arg\min_{\bbeta\in\mathbb{R}^p} \Big\lbrace \frac{1}{2n}\Vert \by - \bX\bbeta \Vert^2 \Big\rbrace,
\]
can be used as the preliminary estimator. The following theorem shows that it induces a rank preserving mapping that satisfies Theorem~\ref{thm:fuse}.

\begin{theorem}  \label{thm:ols}
Under Condition 3.3, suppose $p<n$ and $\Vert (\bX^T\bX)^{-1}\Vert_{\max}\leq c_4n^{-1}$ for some constant $c_4>0$. If $b_n>\sqrt{(2c_4/c_3)\log(p)/n}$, then with probability at least $1-2p^{-1}$, the order generated from $\hbbeta^{ols}$ preserves the order of $\bbeta^0$.
\end{theorem}

When the order $\tau$ extracted from $\widetilde{\bbeta}$ does not preserve the order of $\bbeta^0$, the penalty in \eqref{nCARDS} is no longer a ``correct" penalty for promoting the true grouping structure.
There is no hope that local minima of \eqref{nCARDS} exactly recover the true groups. However, if there are not too many misordering in $\tau$, it is still possible to control  $\Vert \hbbeta-\bbeta^0\Vert$.

Given an order $\tau$, define $K^*(\tau)=\sum_{j=1}^{p-1}1\{\beta^0_{\tau(j)}\neq \beta^0_{\tau(j+1)}\}$, which is the number of jumps in $\bbeta^0$ in the ordering $\tau$. These jumps define subgroups $A'_1, A'_2,\cdots, A'_{K^*}$, each being a subset of one true group. Although different subgroups may share the same true coefficients,  consecutive subgroups, $A'_{k}$ and $A'_{k+1}$, have a gap in coefficient values. As a result, the above results apply to this \emph{subgrouping structure}. The following theorem is a direct application of the proof of Theorem \ref{thm:fuse} and its details are omitted.

\begin{theorem} \label{thm:misorder}
Suppose Conditions 3.1-3.3 hold, $K^*(\tau)=o(n)$, $b_n>a\lambda_n$ and $\lambda_n$ satisfies \eqref{lam_order}. Then with probability tending to $1$, there is a strictly local minimum $\hbbeta$ of \eqref{nCARDS} such that $\Vert \hbbeta - \bbeta^0 \Vert= O_p (\sqrt{K^*(\tau)/n})$.
\end{theorem}

\section{Analysis of the advanced CARDS}  \label{sec:theory}

In this section, we analyze the advanced version of CARDS described, as well as its variate the shrinkage-CARDS.

\subsection{Main results} \label{subsec:prop_aCARDS}

To guarantee the success of the advanced CARDS, a key condition is that the ordered segmentation preserves the order of $\bbeta^0$. This implies restrictions on how much the ordering (in terms of increasing values) of coordinates in $\widetilde{\bbeta}$ deviates from that of $\bbeta^0$. This is reflected on how the segments $\{B_1,\cdots, B_L\}$ intersect with the true groups $\{A_1,\cdots, A_K\}$. Write $V_{kl}= A_k\cap B_l$. We have the following proposition:

\begin{proposition}  \label{prop:A&B}
When $\Upsilon$ preserves the order of $\bbeta^0$, for each $k$, there exist $d_k$ and $u_k$ such that $A_k = \cup_{d_k \leq l\leq u_k} V_{kl}$, and $V_{kl}= B_l$ for $d_k<l<u_k$. For each $l$, there exist $a_l$ and $b_l$ such that $B_l = \cup_{a_l \leq k \leq b_l} V_{kl}$, and $V_{kl}=A_k$ for $a_l< k< b_l$.
\end{proposition}

Proposition \ref{prop:A&B} indicates that there are two cases for each $A_k$: either $A_k$ is contained in a single $B_l$ or it is contained in some consecutive $B_l$'s where except the first and last one, all the other $B_l$'s are fully occupied by $A_k$.  Similarly, there are two cases for each $B_l$: either it is contained in a single $A_k$ or it is contained in some consecutive $A_k$'s where except the first and last one, all the other $A_k$'s are fully occupied by $B_l$.

\begin{theorem}  \label{thm:rate}
Suppose Conditions 3.1-3.3 hold, $K=o(n)$, and the preliminary estimate $\widetilde{\bbeta}$ and the tuning parameter $\delta_n$ together generate an ordered segmentation $\Upsilon$ that preserves the order of $\bbeta^0$ with probability at least $1-\epsilon_0$. If $b_n>a\max\{\lambda_{1n},\lambda_{2n}\}$,
\beq  \label{lam1_order}
\lambda_{1n} \gg \max_{k,h}\left\lbrace |V_{kh}|^{-2}\left[ \sqrt{\sigma_k|A_k|\log(p)/n} + (1+\nu_k |A_k|)\sqrt{K\log(n)/n}\right] \right\rbrace,
\eeq
and
\beq \label{lam2_order}
\lambda_{2n} \gg  \max_{k}\left\lbrace  \sqrt{\log(p)/(n|A_k|)} + \nu_k \sqrt{K\log(n)/(n|A_k|)} \right\rbrace,
\eeq
then with probability at least $1-\epsilon_0-O(n^{-1})$, $\hbbeta^{oracle}$ is a strictly local minimum of \eqref{CARDS}. Moreover, $\Vert \hbbeta^{oracle}-\bbeta^0\Vert= O_p (\sqrt{K/n})$.
\end{theorem}

\noindent Compared to Theorem \ref{thm:fuse}, the advanced version of CARDS not only imposes less restrictive conditions on $\widetilde{\bbeta}$, but also requires a smaller minimum gap between true coefficients.

Next, we establish the asymptotic normality of the CARDS estimator. By Theorem \ref{thm:rate}, with probability tending to $1$, the advanced CARDS performs as if the oracle. In the oracle situation, for example, if $p=5$ and $\beta_1=\beta_4$, $\beta_3=\beta_5$, the accuracy of estimating $\bbeta$ is the same as if we know the model:
$$
    Y = \beta_1 (X_1+X_4) + \beta_2 X_2 + \beta_3 (X_3 + X_5) + \varepsilon.
$$

\begin{theorem}  \label{thm:normality}
Let $\hbbeta$ be any local minimum of \eqref{CARDS} such that $\Vert \hbbeta-\bbeta^0\Vert\leq C\sqrt{K\log(n)/n}$ for a large constant $C>0$ with probability at least $1-o(1)$. Under conditions of Theorem \ref{thm:rate}, if $\Vert \bX_A(\bX_A^T\bX_A)^{-1/2}\Vert_\infty = O(1)$, then for a fixed positive integer $q$, and any sequence $\{\bB_n\}$ such that $\bB_n\in\mathbb{R}^{q\times K}$, $\Vert \bB_n^T\Vert_{2,\infty}=o(1)$ and $\bB_n\bB_n^T\to\bH$, where $\bH$ is a fixed $q\times q$ positive definite matrix, we have
\begin{equation*}
\bB_n (\bX^T_A\bX_A)^{1/2}(\hbbeta_A - \bbeta^0_A )\overset{d}{\to} N(\boldsymbol{0},\bH),
\end{equation*}
where $\hbbeta_A$ is the $K$-dimensional vector of distinct values in $\hbbeta$.
\end{theorem}

In the case of orthogonal design $\bX^T\bX=n\bI$, the matrix $\bX_A(\bX_A^T\bX_A)^{-1/2}$ has orthonormal columns, so it is reasonable to assume $\Vert \bX_A(\bX_A^T\bX_A)^{-1/2}\Vert_\infty = O(1)$. In addition, when all the entries of $\bB_n$  have the same order, $\Vert \bB_n^T\Vert_{2,\infty}= O(1/\sqrt{K})=o(1)$, as long as $K\to\infty$.

To compare the asymptotic variance of $\hbbeta$ and $\hbbeta^{ols}$, we introduce the following corollary.
\begin{corollary} \label{coro:compare}
Suppose conditions of Theorem \ref{thm:normality} hold and let $\hbbeta^{ols}$ and $\hbbeta$ be the ordinary least squares estimator and CARDS estimator respectively.
Let $\bM_n$ be the $p\times K$ matrix with $M_n(j,k)=(1/|A_k|^{1/2}) 1\{j\in A_k\}$. For any sequence of $p$-dimensional vectors $\ba_n$,
\[
v_{1n}^{-1/2}\ba_n^T(\hbbeta^{ols}-\bbeta^0) \overset{d}{\to} N(0, 1) \quad \text{ and }\quad
v_{2n}^{-1/2}\ba_n^T(\hbbeta-\bbeta^0) \overset{d}{\to} N(0, 1).
\]
where $v_{1n}=\ba_n^T(\bX^T\bX)^{-1}\ba_n$ and $v_{2n}=\ba_n^T\bM_n^T(\bM_n^T\bX^T\bX\bM_n)^{-1}\bM_n\ba_n$. In addition,  $v_{1n}\geq v_{2n}$.
\end{corollary}

\subsection{CARDS under sparsity} \label{subsec:prop_sCARDS}

In Section \ref{subsec:sCARDS}, we introduced the shrinkage-CARDS (sCARDS) to explore both homogeneity and sparsity. In sCARDS, given a preliminary estimator $\widetilde{\bbeta}$ and a parameter $\delta$, we extract segments $B_1,\cdots,B_L$ such that $\cup_{l=1}^L B_l=\widetilde{S}$, where $\widetilde{S}$ is the support of $\widetilde{\bbeta}$. Denote $B_0=\{j: \widetilde{\beta}_j = 0\}$. In this case, we say $\Upsilon =\{B_0, B_1,\cdots, B_L\}$ preserves the order of $\bbeta^0$ if $\max_{j\in B_0}|\beta^0_j|=0$, and $\max_{j\in B_l}\beta^0_j\leq \min_{j\in B_{l+1}} \beta^0_j$, for $l=1,\cdots, L-1$. This implies that $\widetilde{\bbeta}$ has the sure screening property; and on those preliminarily selected variables, the data-driven segments preserve the order of true coefficients. In particular, from Proposition \ref{prop:A&B}, those falsely selected variables, i.e., $\{j: \beta_j^0=0, \widetilde{\beta}_j\neq 0\}$, should be contained in either a single segment or some consecutive segments.

Suppose there is a group of zero coefficients in $\bbeta^0$, namely, $\cA =(A_0, A_1, \cdots, A_K)$. Let $\cM^*_A$ be the subspace of $\mathbb{R}^p$ defined by 
\[
\cM^*_{A}=\{\bbeta\in\mathbb{R}^{p}: \beta_i=0, \text{ for any } i\in A_0;\ \beta_i=\beta_j,\text{ for any } i,j\in A_k, 1\leq k\leq K\}.
\]
Denote the support of $\bbeta^0$ as $S$ and $s=|S|$. The following theorem is proved in Section \ref{sec:proof}.

\begin{theorem}  \label{thm:cards+sparse}
Suppose Conditions 3.1-3.3 hold, $s= o(n)$, $\log(p)=o(n)$, and the preliminary estimate $\widetilde{\bbeta}$ and the tuning parameter $\delta_n$ together generate an ordered segmentation $\Upsilon$ that preserves the order of $\bbeta^0$ with probability at least $1-\epsilon_0$. If $b_n > a\max\{\lambda_{1n}, \lambda_{2n}\}$, $\min\{|\beta_j^0|: \beta_j^0\neq 0\}> 2a\lambda_n$, $\lambda_{1n}$ and $\lambda_{2n}$ satisfy \eqref{lam1_order}-\eqref{lam2_order} and $\lambda_n\gg\sqrt{\log(p)/n}$,
then with probability at least $1-\epsilon_0- n^{-1}K - 2p^{-1}$, $\hbbeta^{oracle}$ is a strictly local minimum of \eqref{sCARDS}. Moreover, $\Vert \hbbeta^{oracle} - \bbeta^0\Vert= O_p(\sqrt{K/n})$.
\end{theorem}

The preliminary estimator $\widetilde{\bbeta}$ can be chosen, for example, as the SCAD estimator
\beq  \label{SCAD}
\hbbeta^{scad}\in \arg\min\Big\{ \frac{1}{2n}\Vert \by  -\bX\bbeta\Vert^2 + \sum_{j=1}^p p_{\lambda'}(|\beta_j|) \Big\},
\eeq
where $p_{\lambda'}(\cdot)$ is the SCAD penalty function \cite{FL2001}. The following theorem is a direct result of Theorem 2 in \cite{FanLv2011}, and the proof is omitted.

\begin{theorem}
Under Condition 3.1 and 3.3, if $s=o(n)$, $\lambda'_n\gg n^{-1/2} [\log (n)]^2$ and $\min\{|\beta^0_j|: \beta^0_j\neq 0\} \gg n^{-1/2} \max\left\lbrace \sqrt{\log p}, \Vert\frac{1}{n}\bX^T_{S^c}\bX_{S}\Vert_\infty\sqrt{\log n} \right\rbrace$,  then with probability at least $1- o(1)$, there exists a strictly local minimum $\hbbeta^{scad}$ and $\delta_n=O(\log(n)/n)$ which together generate a segmentation preserving the order of $\bbeta^0$.
\end{theorem}

\section{Simulation studies}

We conduct numerical experiments to implement two versions of CARDS and their variate sCARDS. The goal is to investigate the performance of CARDS under different situations: Experiment 1 and 2 are based on the linear regression setting $Y_i=\bX_i^T\bbeta^0 + \epsilon_i$, where in Experiment 1 only the homogeneity is explored, and in Experiment 2 the homogeneity and sparsity are explored simultaneously. Experiment 3 is based on the spatial-temporal model $Y_{it}=\bX_{t}^T\bbeta^0_i + \epsilon_{it}$.

In all experiments, $\{\bX_i: 1\leq i\leq n\}$ or $\{\bX_t: 1\leq t\leq T\}$ are generated independently and identically from the multivariate standard Gaussian distributions, and $\{\epsilon_i: 1\leq i\leq n\}$ or $\{\epsilon_{it}: 1\leq i\leq p, 1\leq t\leq T\}$ are IID samples of $N(0,1)$.
All results are based on $100$ repetitions.

{\bf Example 1}: Consider the linear regression setting with $p=60$ and $n=100$. Predictors are divided into four groups with each group having a size of 15. The four different values of the true regression coefficients are $-2r$, $-r$, $r$ and $2r$, respectively. Here different values of $r>0$ lead to various signal-to-noise ratios.

We compare the performance of six different methods: Oracle, ordinary least squares (OLS), bCARDS, aCARDS, total variations (TV), fused Lasso (fLasso). Oracle is the least squares estimator knowing the true groups. aCARDS and bCARDS are described in Section \ref{sec:method}; here we let the penalty function $p_{\lambda}(\cdot)$ be the SCAD penalty with $a=3.7$, and take the OLS estimator as the preliminary estimator. TV uses the exhaustive pairwise penalty \eqref{TV} with $p_{\lambda}(\cdot)$ being the same as that in aCARDS and bCARDS. The fused Lasso is based on an order generated from ranking the OLS coefficients. Tuning parameters of all these methods are selected via Bayesian information criteria (BIC).

Performance is evaluated in terms of the average prediction error over an independent test set of size $10,000$. In addition, to measure how close the estimated grouping structure approaches the true one, we introduce the normalized mutual information (NMI), which is a common measure for similarity between clusterings \cite{FredJain}. Suppose $\mathbb{C}=\{C_1,C_2, \cdots\}$ and $\mathbb{D}=\{D_1, D_2, \cdots,\}$ are two sets of disjoint clusters of $\{1,\cdots, p\}$, define
\[
\text{NMI}(\mathbb{C}, \mathbb{D}) = \frac{I(\mathbb{C}; \mathbb{D})}{[H(\mathbb{C})+ H(\mathbb{D})]/2},
\]
where $I(\mathbb{C}; \mathbb{D})=\sum_{k,j}(|C_k\cap D_j|/p)\log(p|C_k\cap D_j|/|C_k||D_j|)$ is the mutual information between $\mathbb{C}$ and $\mathbb{D}$, and $H(\mathbb{C})= \sum_k (|C_k|/p)\log(|C_k|/p)$ is the entropy of $\mathbb{C}$. $\text{NMI}(\mathbb{C},\mathbb{D})$ takes values on $[0,1]$, and large NMI implies that the two grouping structures are close.

\begin{figure}[b]
\begin{subfigure}{1 \textwidth}
\centering
\includegraphics[width= 0.99\textwidth]{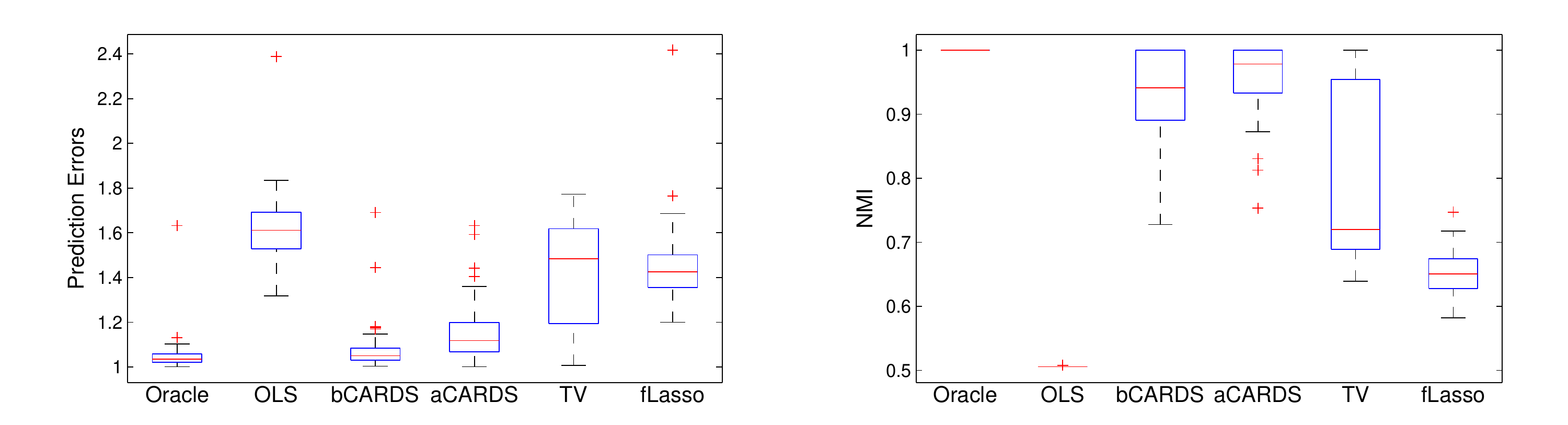}
\caption{r=1}
\end{subfigure} \\
\begin{subfigure}{1 \textwidth}
\centering
\includegraphics[width= 0.99\textwidth]{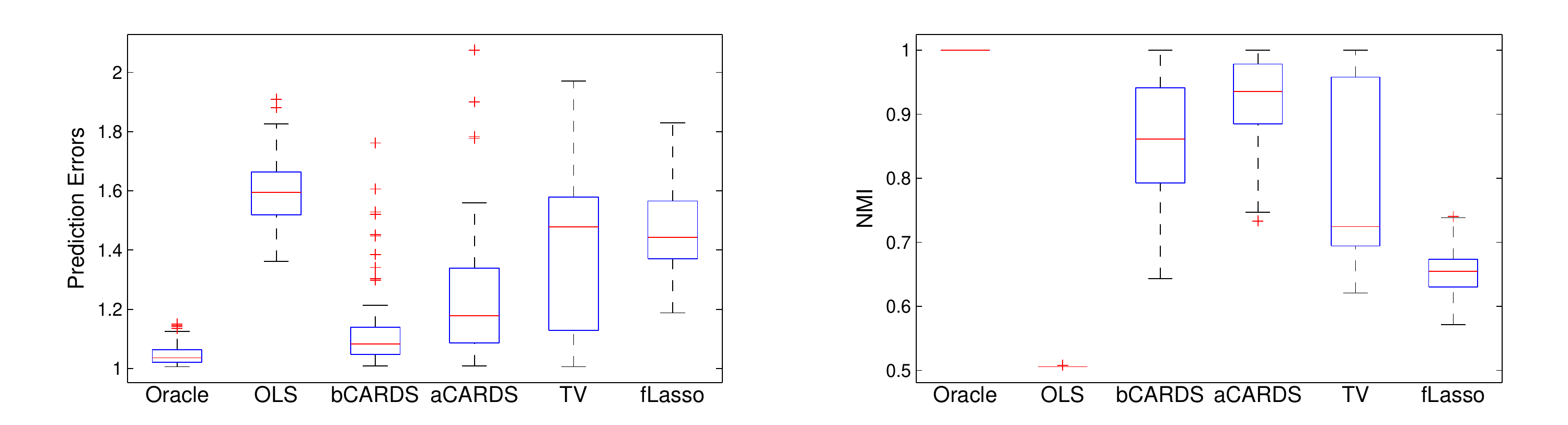}
\caption{r=0.8}
\end{subfigure}
\begin{subfigure}{1 \textwidth}
\centering
\includegraphics[width= 0.99\textwidth]{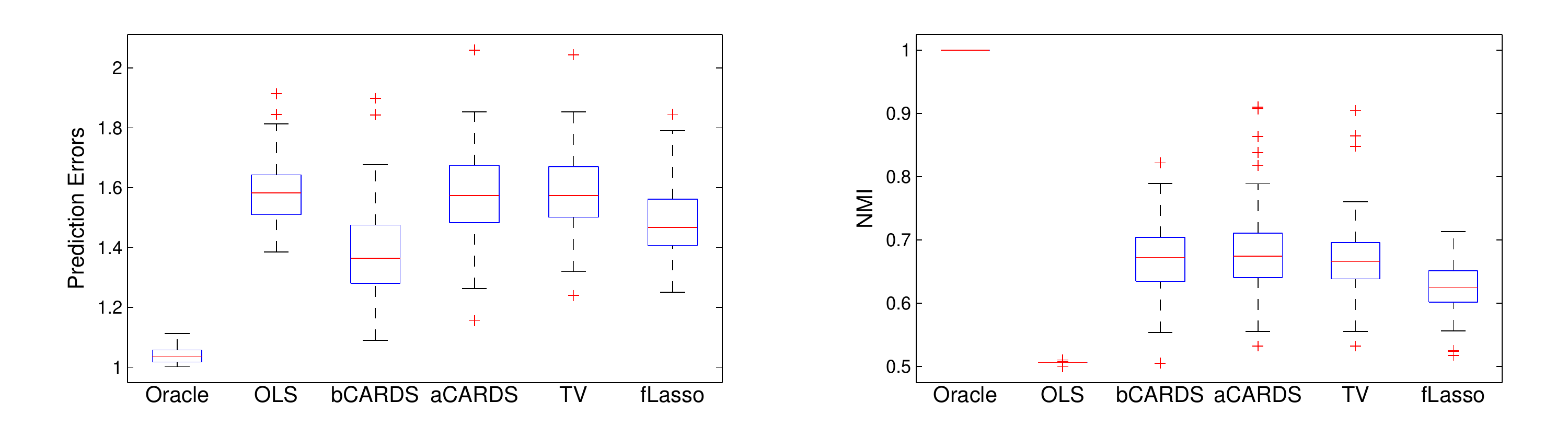}
\caption{r=0.5}
\end{subfigure}
\caption{Boxplots of the average prediction error and normalized mutual information over $100$ repetitions in Experiment 1.} \label{fig:simu1box}
\end{figure}

Table \ref{tb:simu1Err} shows medians of the average prediction error for six different methods under various values of $r$. Table \ref{tb:simu1NMI} shows medians of NMI. The boxplots are displayed in Figure \ref{fig:simu1box}. We see that except for the case of weak signals ($r=0.5$), two versions of CARDS outperform other methods in terms of smaller prediction error and larger NMI. bCARDS is especially good in achieving low prediction errors, even in the case $r=0.5$. aCARDS has a better performance in NMI, which shows that it is good in recovering the true grouping structure.

\begin{table} [tb]
\begin{center}
\caption{Medians of the average prediction error over 100 repetitions for Experiment 1.}\label{tb:simu1Err}
\begin{tabular}{c|cccccc}
\hline
 & Oracle & OLS & bCARDS & aCARDS & TV & fLasso \\
\hline
r=1  &  1.0355  &  1.6112  &  1.0504  &  1.1182  &  1.4847  &  1.4253\\
r=0.9  &  1.0273  &  1.5885  &  1.0479  &  1.1048  &  1.4608  &  1.4186\\
r=0.8  &  1.0359  &  1.5947  &  1.0826  &  1.1786  &  1.4777  &  1.4427\\
r=0.7  &  1.0311  &  1.6038  &  1.1250  &  1.2830  &  1.5591  &  1.4625\\
r=0.6  &  1.0370  &  1.6054  &  1.3172  &  1.4586  &  1.5795  &  1.4824\\
r=0.5  &  1.0347  &  1.5826  &  1.3645  &  1.5734  &  1.5734  & 1.4668\\
\hline
\end{tabular}
\end{center}
\end{table}

\begin{table} [tb]
\begin{center}
\caption{Medians of NMI over 100 repetitions for Experiment 1.}\label{tb:simu1NMI}
\begin{tabular}{c|cccccc}
\hline
 & Oracle & OLS & bCARDS & aCARDS & TV & fLasso \\
\hline
r=1 &  1.0000  &  0.5059  &  0.9414  &  0.9784  &  0.7203  &  0.6503\\
r=0.9 &  1.0000  &  0.5059  &  0.9414  &  0.9784  &  0.7167  &  0.6521\\
r=0.8 &  1.0000  &  0.5059  &  0.8609  &  0.9355  &  0.7245  &  0.6549\\
r=0.7 & 1.0000  &  0.5059  &  0.7912  &  0.8989  &  0.6991  &  0.6458\\
r=0.6 & 1.0000  &  0.5059  &  0.7008  &  0.8763  &  0.6808  &  0.6373 \\
r=0.5 & 1.0000  &  0.5059  &  0.6722  &  0.6741  &  0.6654  &  0.6251\\
\hline
\end{tabular}
\end{center}
\end{table}

{\bf Experiment 2}: Consider the linear regression setting with $p=100$ and $n=150$. Among the 100 predictors, 60 are important ones and their coefficients are the same as those in Experiment 1. Besides, there are 40 unimportant predictors whose coefficients are all equal to $0$.

We implemented sCARDS in this setting and compared its performance to different oracle estimators, Oralce, Oracle0 and OracleG, as well as ordinary least squares (OLS) and the SCAD estimator.
The three oracles are defined with different prior information: The Oracle knows both the important predictors and the true groups among them; the Oracle0 only knows which are important predictors; and the OracleG only knows the true groups (it treats all unimportant predictors as one group with unknown coefficients). 
sCARDS is as described in Section \ref{sec:method}; when implementing it, we take the SCAD estimator as the preliminary estimator.

\begin{figure}[tb]
\begin{subfigure}{0.48 \textwidth}
\centering
\includegraphics[width= 0.99\textwidth]{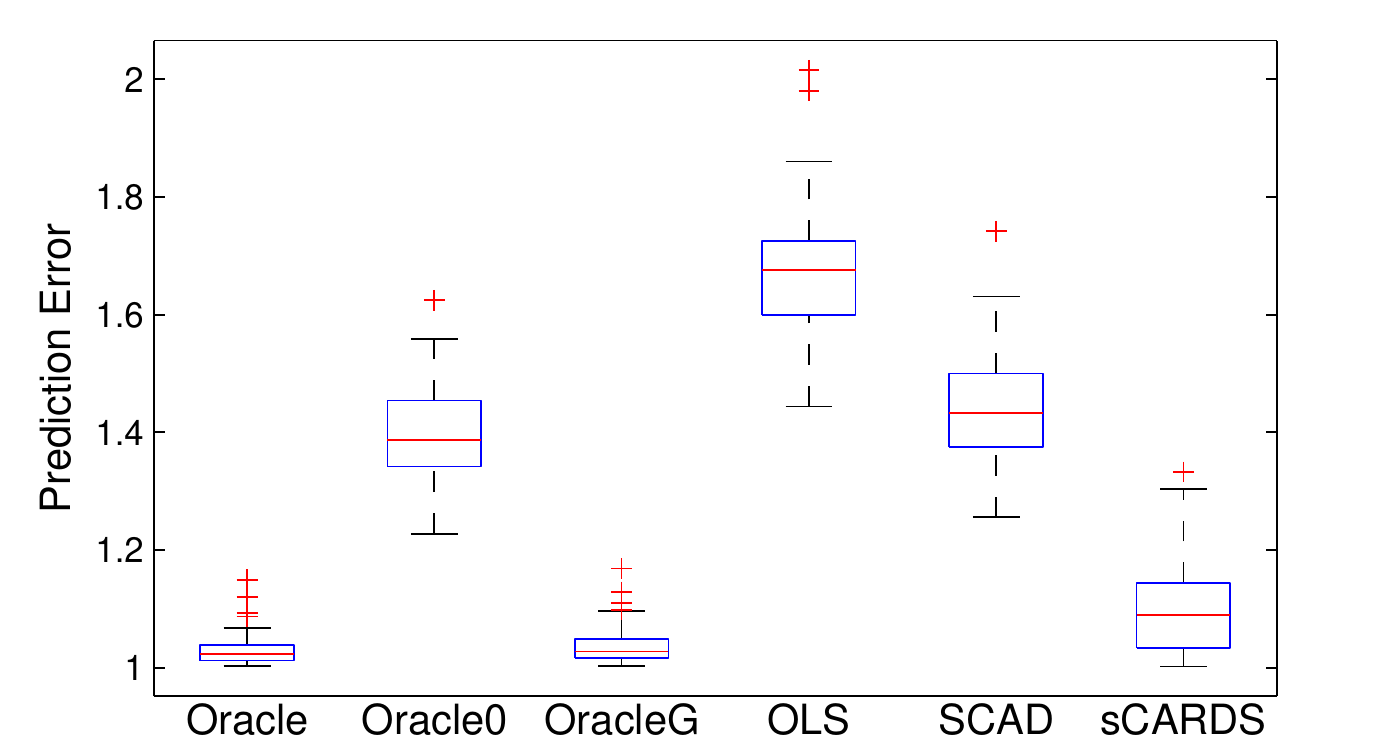}
\caption{r=1}
\end{subfigure}
\begin{subfigure}{0.48 \textwidth}
\centering
\includegraphics[width= 0.99\textwidth]{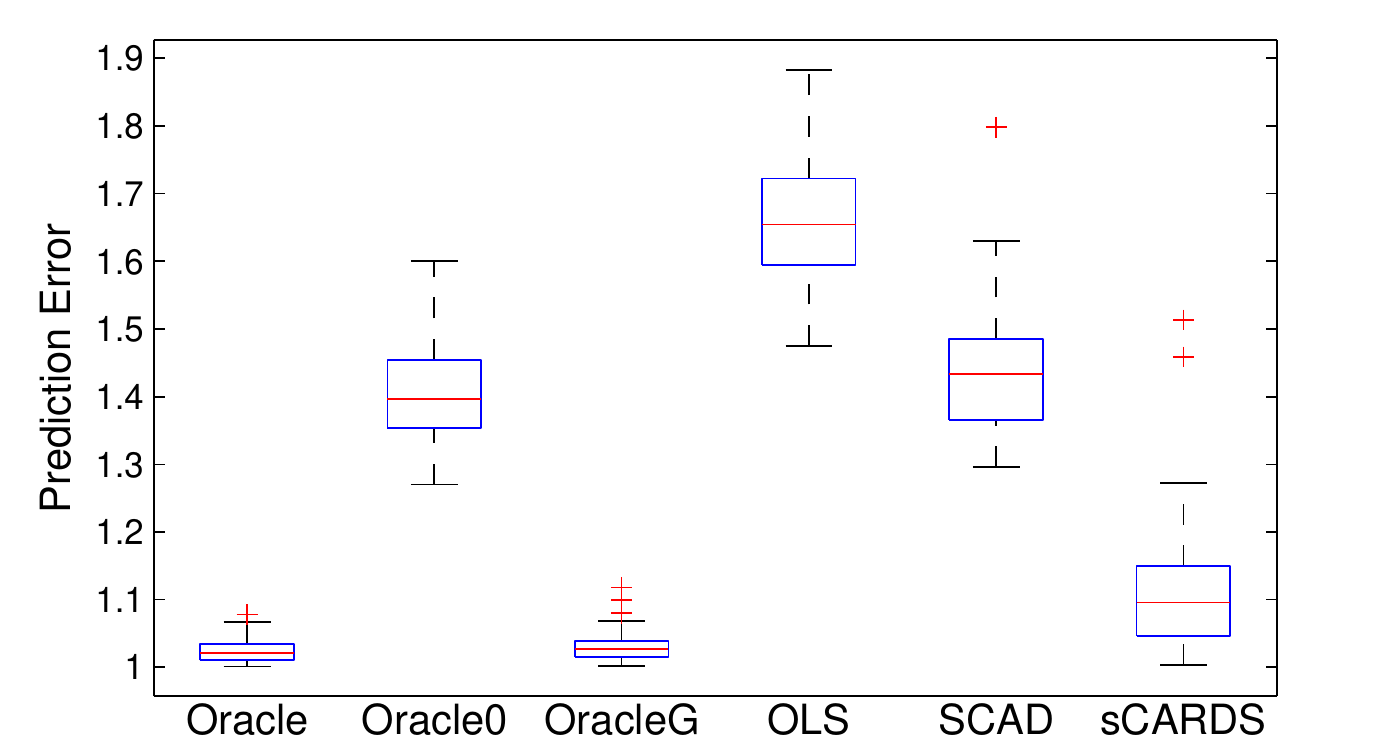}
\caption{r=0.7}
\end{subfigure}
\caption{Boxplots of the average prediction errors over $100$ repetitions in Experiment 2.} \label{fig:simu2box}
\end{figure}

Table \ref{tb:simu2} shows medians of the average prediction error, number of false positives and normalized mutual information on grouping important predictors. Figure \ref{fig:simu2box} displays the boxplots of average prediction errors under different values of $r$.  First, by comparing prediction errors of the three oracles, we see a significant advantage of taking into account both homogeneity and sparsity over pure sparsity. Moreover, the results of Oracle0 and OracleG show that exploring group structure is more important than sparsity. Second, sCARDS achieves a much smaller prediction error than that of OLS and SCAD. Third, compared to the preliminary estimator SCAD, sCARDS can further filter out falsely selected unimportant variables. Fourth, sCARDS successfully recovers the grouping structure on important variables in most cases ($\text{NMI}=1$ means the estimated groups exactly overlap with the true ones).

\begin{table} [tb]
\begin{center}
\caption{Medians of the average prediction error (PE), number of false positives (FP) and NMI on the important variables,  over 100 repetitions for Experiment 2.}\label{tb:simu2}
\begin{tabular}{c|c|cccccc}
\hline
\multicolumn{2}{c|}{} & Oracle & Oracle0 & OracleG & OLS & SCAD & sCARDS \\
\hline
\multirow{2}{*}{PE} & r=1 &   1.0234  &  1.3869  &  1.0273  &  1.6758  &  1.4333  &  1.0895\\
& r=0.7 &  1.0204  &  1.3961  &  1.0274  &  1.6544  &  1.4330  &  1.0960\\
\hline
\multirow{2}{*}{FP} & r=1 & 0  &   0  &  40  &  40  &   5  &   1 \\
& r=0.7 &  0  & 0  & 40 &  40  &  4  &  2.5\\
\hline
\multirow{2}{*}{NMI} & r=1 &  1.0000  &  0.5059  &  1.0000  &  0.5059  &  0.5059  &  1.0000\\
& r=0.7 & 1.0000  &  0.5059  &  1.0000  &  0.5059  &  0.5059  &  1.0000\\
\hline
\end{tabular}
\end{center}
\end{table}

{\bf Experiment 3}: We consider a special case of the spacial-temporal model, where $\bX_{it}=\bX_t$ for $i=1,\cdots, p$, i.e., the predictors are common for all spacial locations. $p = 100$ is the total number of locations. Each $\bbeta_i$ is a $5$-dimensional vector. In each coordinate $j=1,\cdots, 5$, the coefficients $\{\beta_{ij}, 1\leq i\leq 100\}$ are divided into four groups of equal size $25$, with coefficients in the same group sharing a same value. In coordinate $1$, the four true coefficients are $[-2, -1, 1, 2]$; in coordinate $j=2,\cdots, 5$, they are $[-2, -1, 1, 2]+0.1\times(j-1)$.

We extend aCARDS (bCARDS) to this model: given a preliminary estimator, for each coordinate $j=1,\cdots, k$, extract the data-driven segments (ordering) and build the cross-sectional hybrid (fused) penalty $P_{j}(\cdot)$, then sum them up to build the penalty term, and finally solve a penalized maximum likelihood:
\[
\min_{\bbeta=(\bbeta_1^T,\cdots, \bbeta_p^T)^T = (\bb_1,\cdots,\bb_k)} \Big\{  \frac{1}{2T}\sum_{i=1}^p \sum_{t=1}^T (Y_{it} -\bX_t^T\bbeta_i ) + \sum_{j=1}^k P_j(\bb_j)  \Big\}.
\]
We still call the method aCARDS (bCARDS). The Oracle is the maximum likelihood estimator knowing the true groups in each coordinate. We aim to compare the performance of Oracle, OLS and aCARDS.

\begin{figure}
\begin{subfigure}{0.48 \textwidth}
\centering
\includegraphics[width= 0.9\textwidth]{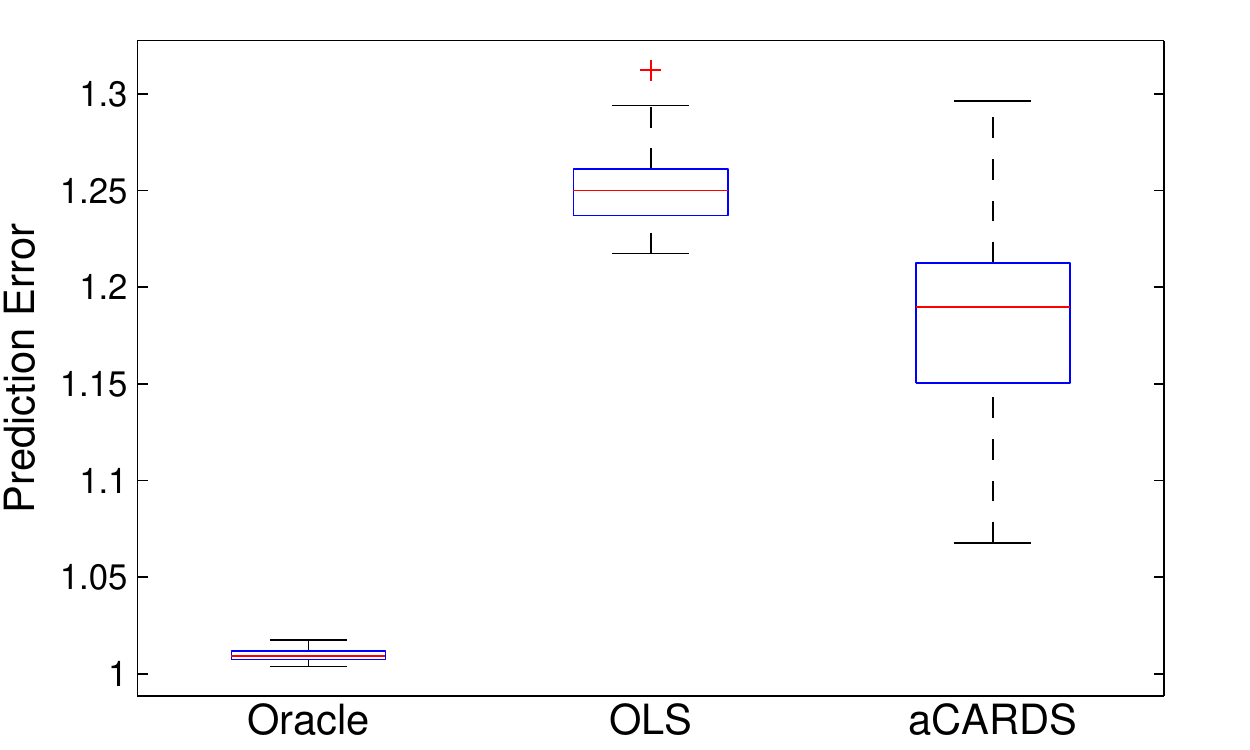}
\caption{T=20}
\end{subfigure}
\begin{subfigure}{0.48 \textwidth}
\centering
\includegraphics[width= 0.9\textwidth]{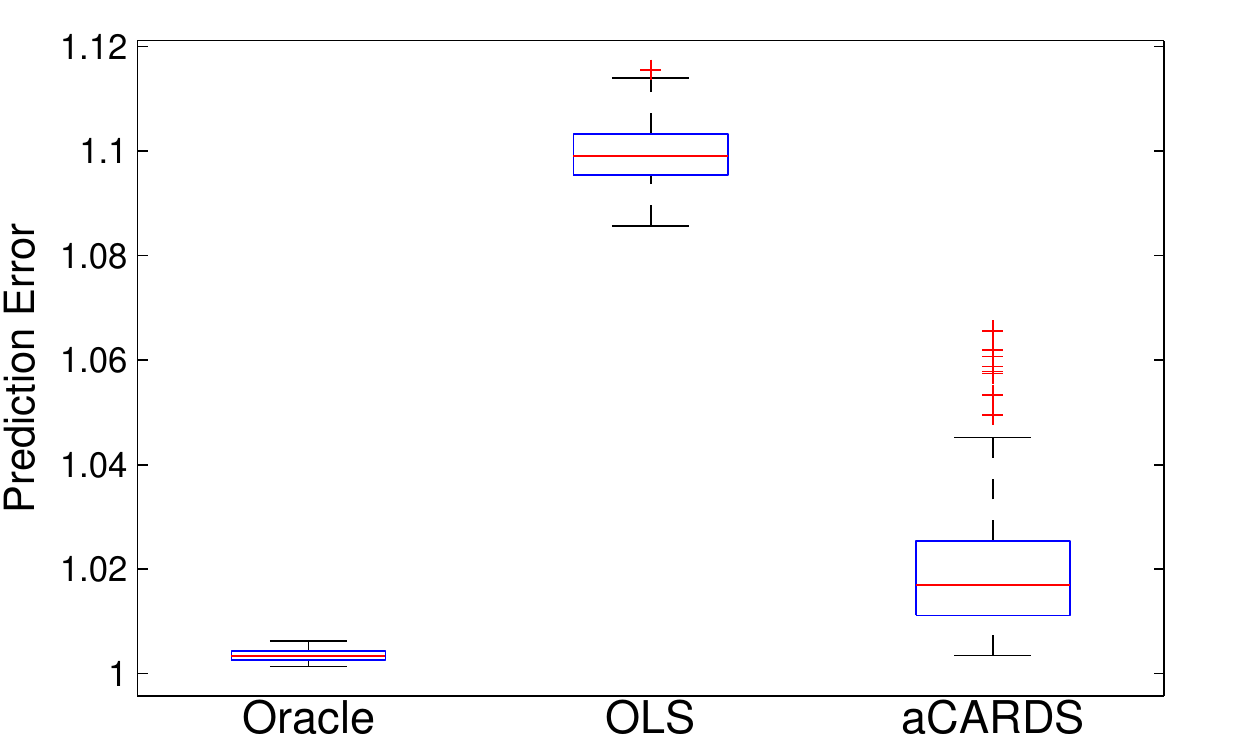}
\caption{T=50}
\end{subfigure}\\
\begin{subfigure}{0.48 \textwidth}
\centering
\includegraphics[width= 0.9\textwidth]{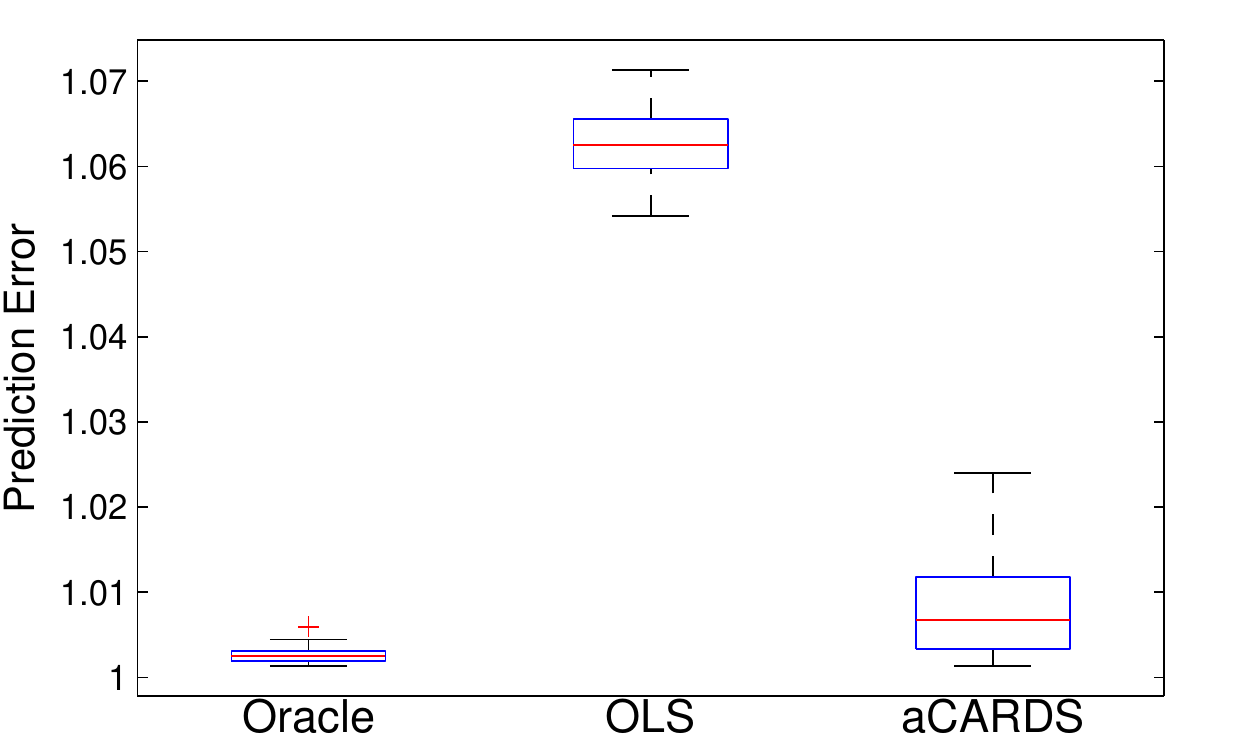}
\caption{T=80}
\end{subfigure}
\begin{subfigure}{0.48 \textwidth}
\centering
\includegraphics[width= 0.995\textwidth]{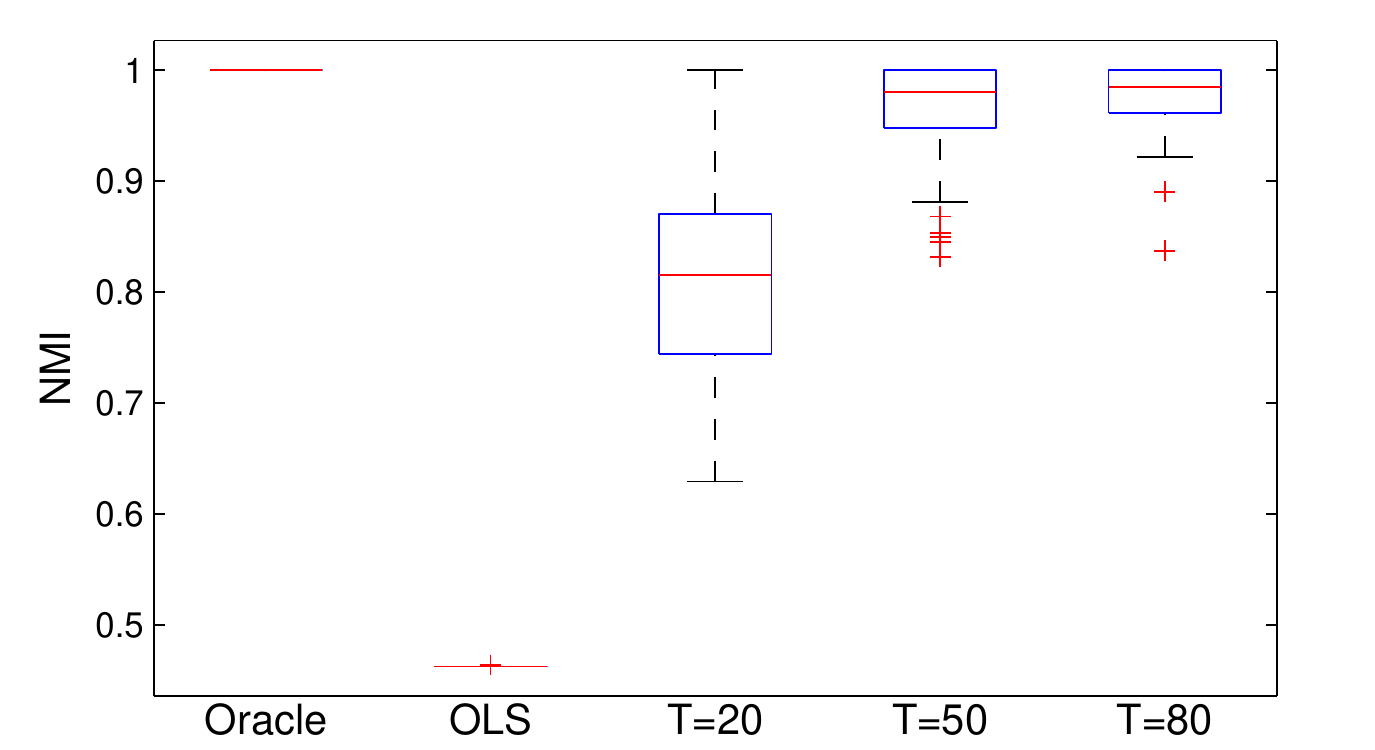}
\caption{NMI}
\end{subfigure}
\caption{Boxplots of the results over $100$ repetitions in Experiment 3.} \label{fig:simu3box}
\end{figure}

Table \ref{tb:simu3} shows medians of the average prediction error and normalized mutual information(averaged over $5$ coordinates). Instead of varying the signal-to-noise ratio directly, we equivalently change $T$, the total number of time points. Figure \ref{fig:simu3box} contains the boxplots. We see that aCARDS achieves significantly lower prediction errors in all cases. Moreover, aCARDS estimates well the true grouping structure; in particular, when $T=50,80$, $\text{NMI}>0.95$ in most repetitions.

\begin{table} [bt]
\begin{center}
\caption{Medians of the average prediction error and NMI over 100 repetitions for Experiment 3.}\label{tb:simu3}
\begin{tabular}{c|ccc|ccc}
\hline
& \multicolumn{3}{c|}{Prediction Error} & \multicolumn{3}{c}{NMI}\\
\cline{2-7}
 & Oracle & OLS & aCARDS & Oracle & OLS & aCARDS \\
\hline
T=20  &  1.0095  &  1.2501  &  1.1898  &  1.0000  &  0.4628  &  0.8154\\
T=50  &  1.0034  &  1.0990  &  1.0170  &  1.0000  &  0.4628  &  0.9803\\
T=80  &  1.0025  &  1.0625  &  1.0067  &  1.0000  &  0.4628  &  0.9851\\
\hline
\end{tabular}
\end{center}
\end{table}

\section{Real data analysis}

\subsection{S\&P500 returns}

\begin{figure}
\centering
\includegraphics[width= 0.7\textwidth]{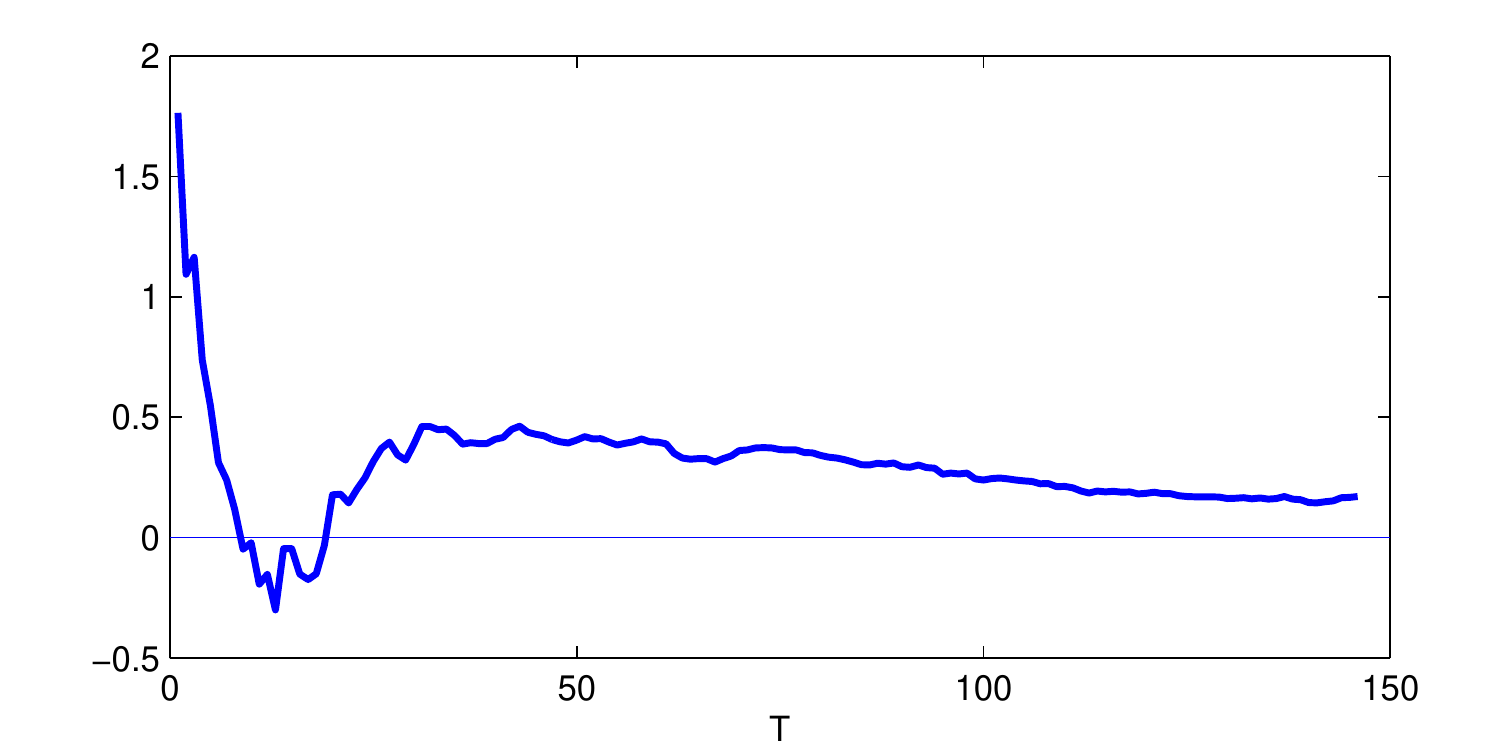}
\caption{Cumulative sum of squared estimation errors of the S\&P500 data from December 1, 2011 to July 2, 2012. The vertical axix is $100(\text{cRSS}_t^{OLS}-\text{cRSS}_t^{bCARDS})/\text{cRSS}_t^{OLS}$.} \label{fig:stockcrss}
\end{figure}

\begin{figure}
\begin{subfigure}{0.495 \textwidth}
\centering
\includegraphics[width= 0.92\textwidth]{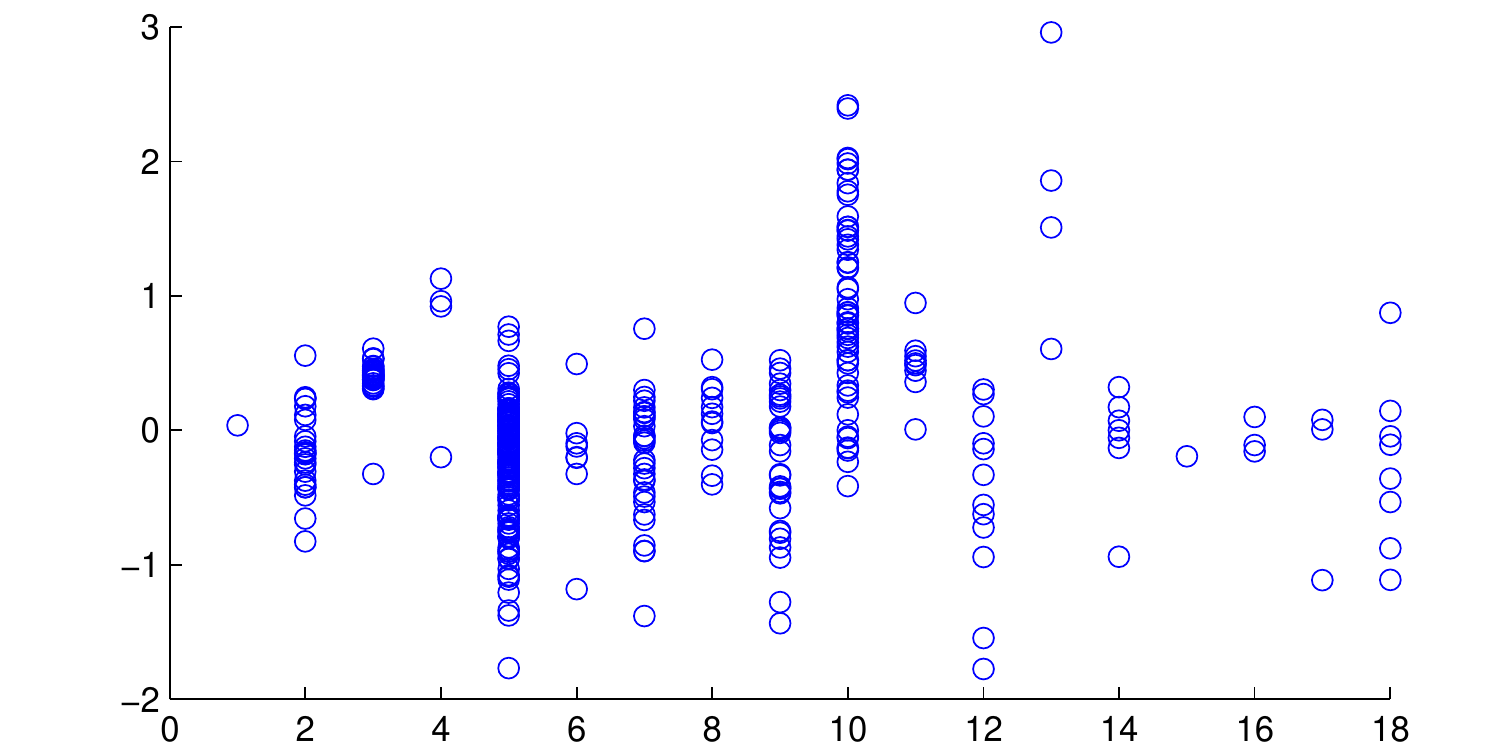}
\caption{}
\end{subfigure}
\begin{subfigure}{0.49 \textwidth}
\centering
\includegraphics[width= 1\textwidth]{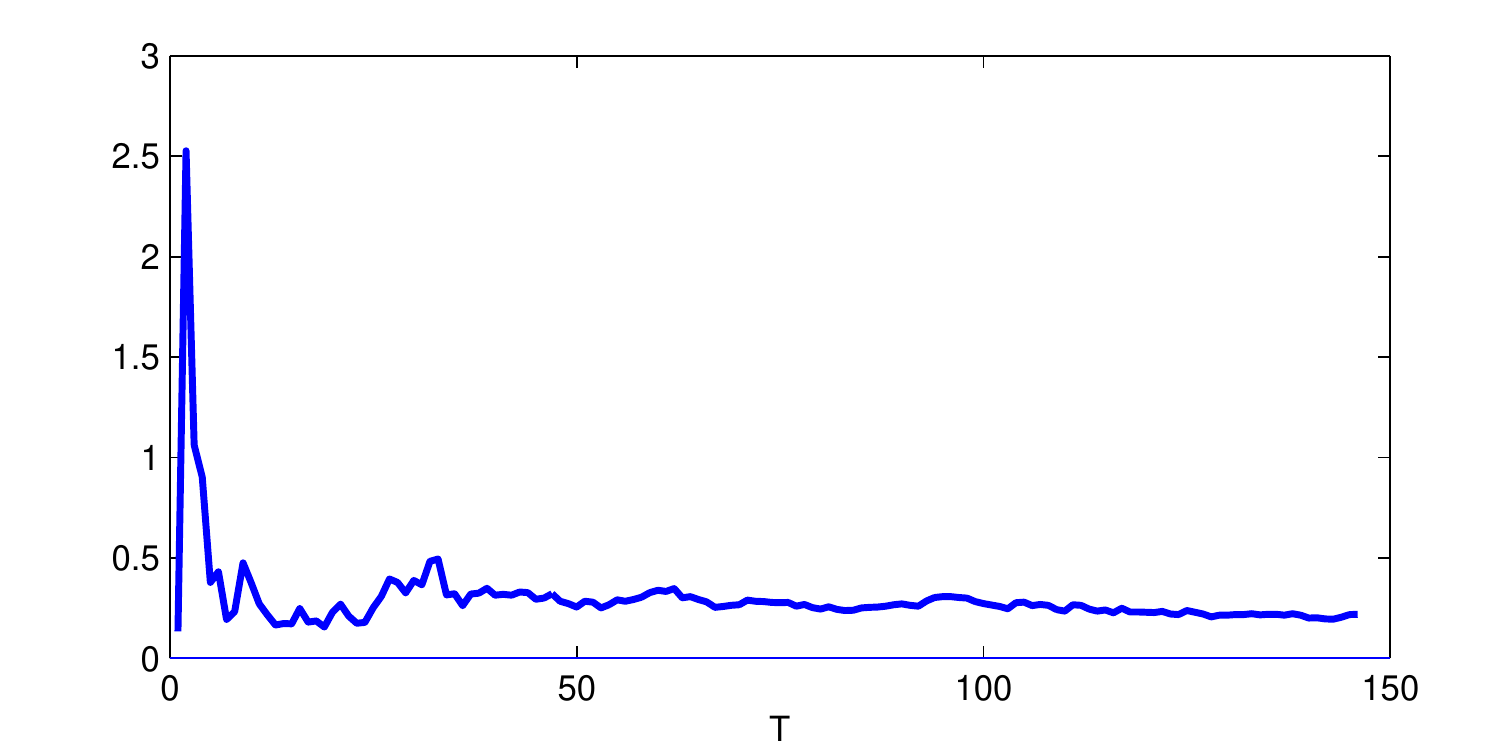}
\caption{}
\end{subfigure}
\caption{(a)OLS coefficients on the ``book-to-market ratio" factor. The x axis represents different sectors. (b)Percentage improvement of the cumulative sum of squared estimation errors for stocks in Sector 2 ``Utilities".} \label{fig:stocksec2}
\end{figure}

In this study, we fit a homogeneous Fama-French model for stock returns: $Y_{it}=\alpha_i + \bX_t^T\bbeta_i^0$, where $\bX_t$ contains three Fama-French factors at time $t$ and $Y_{it}$ is the excess return of stocks. We collected daily returns of $410$ stocks, which were in the components of the S\&P500 index in the period December 1, 2010 to December 1, 2011 ($T=254$). We applied bCARDS as in Experiment 3, except that the intercepts $\alpha_j$'s were also penalized. The tuning parameters were chosen via generalized cross validation (GCV). Table \ref{stock:tb1} shows the number of fitted coefficient groups on three factors and the number of non-zero intercepts. We then used the daily returns of those stocks in the period December 1, 2011 to July 2, 2012 ($T=146$) to evaluate the estimation error. Let $\widehat{y}_{it}$ and $y_{it}$ be the fitted and observed excess returns of stock $i$ at time $t=1,\cdots, 146$, respectively. Define the cumulative sum of squared estimation errors at time $t$ as $\text{cRSS}_t = \sum_{s=1}^t \rho^{\lfloor s/10\rfloor} \sum_{i}(\widehat{y}_{it}-y_{it})^2$, where $\rho$ is a chosen constant between $0$ and $1$. Here we take $\rho=0.95$. Figure \ref{fig:stockcrss} shows the percentage improvement in $\text{cRSS}_t$ of the CARDS estimator over the OLS estimator. We see that CARDS achieves a smaller cumulative sum of squared estimation errors compared to OLS at most time points, especially in the ``very-close" and ``far-away" future. The North American Industry Classification System (NAICS) classifies these 410 companies into $18$ different industry sectors. Figure \ref{fig:stocksec2}(a) shows the OLS coefficients on the ``book-to-market ratio" factor. We can see that stocks belonging to Sector 2 ``Utilities" (29 stocks in total) have very close OLS coefficients, and 17 stocks in this sector were clustered into one group in CARDS estimator. Figure \ref{fig:stocksec2} (b) shows the percentage improvement in $\text{cRSS}_t$ only for stocks in this sector, where the improvement is more significant.

\begin{table}[tb]
\centering
\begin{tabular}{c|c}
\hline
Fama-French factors & No. of coef. groups\\
\hline
 ``market return" & 41 \\
\hline
 ``market capitalization" & 32\\
\hline
 ``book-to-market ratio" & 56\\
\hline
 intercept & 60\\
\hline
\end{tabular}
\caption{Number of groups in fitting the S\&P500 data.}\label{stock:tb1}
\end{table}

\subsection{Polyadenylation signals}

The proposed method can be easily extended to more general settings such as generalized linear models although we have focused on the linear regression setting so far. In this subsection, we will apply the proposed method to a logistic regression example.
This study tried to predict polyadenylation signals (PASes) in human DNA and mRNA sequences by analyzing features around them. The data set was first used in \cite{legendre03} and later analyzed by \cite{Liu2003in-silico}, and it is available at \url{http://datam.i2r.a-star.edu.sg/datasets/krbd/SequenceData/Polya.html}. There is one training data set and five testing data sets. To avoid any platform bias, we use the training data set only. It has 4418 observations each  with 170 predictors and a binary response. The binary response indicates whether a terminal sequence is classified as a ``strong" or ``weak" polyA site, and the predictors are features from the upstream (USE) and downstream (DSE) sequence elements.
We randomly select 2000 observations to perform model estimation and use the rest to evaluate performance. Our numerical analysis consists the following steps. Step 1 is to apply the lasso penalized logistic regression to these 2000 observations with all 170 predictors and to use AIC to select an appropriate regularization parameter. In step 2, we use the logistic regression coefficients obtained in step 1 as our preliminary estimate and apply CARDS accordingly.  Average prediction error (and standard error in parentheses) over 40 random splitting are reported in Table \ref{logistic:tb1}. We also report the average number of non-zero coefficient groups and the average number of selected features. It shows that two versions of CARDS lead to a smaller prediction error when compared with the total variation penalty. In addition, the aCARDS has fewer groups of non-zero coefficients but more selected features.

\begin{table}
\begin{tabular}{c|c|c|c}
\hline
&     aCARDS &  bCARDS &   TV\\
\hline
Prediction Error &    0.2449 (.0015)  &   0.2485 (.0014)  &   0.2757 (.0026)\\
\hline
No. of non-zero coef. groups &    5.5000  &   21.6250  &  5.7500\\
\hline
No. of selected features &   73.2750  &   21.6250  & 40.3500\\
\hline
\end{tabular}
\caption{Results of the PASes data.}\label{logistic:tb1}
\end{table}

\section{Conclusion}

In this paper, we explored homogeneity of coefficients in high-dimensional regression. We proposed a new method called clustering algorithm in regression via data-driven segmentation (CARDS) to estimate regression coefficients and to detect homogeneous groups. The implementation of CARDS does not need any geographical information (neighborhoods, distance, graphs, etc.) as a priori, which differs it from other methods in similar settings and makes it more general to applications. A modification of CARDS, sCARDS, can be used to explore homogeneity and sparsity simultaneously. Our theoretical results show that by exploring homogeneity better estimation accuracy can be achieved. In particular, when the number of homogeneous groups is small, the power of exploring homogeneity and sparsity simuntaneously is much larger than that of exploring sparsity only, which is justified in our simulation studies.

To promote homogeneity, the CARDS uses a preliminary estimate to construct data-driven penalties. This so-called ``hybrid pairwise penalty" is built through a preliminary ranking $\tau$ and a parameter $\delta$ for segmentation. Such idea of taking advantage of a preliminary estimate can be generalized. For example, we may apply clustering methods to these preliminary coefficients, such as $k$-mean algorithm or hierarchical clustering algorithm, to help construct penalties and further promote homogeneity.

This paper only considers the case where predictors in one homogeneous group have equal coefficients. In a more general situation, coefficients of predictors in the same group are close but not exactly equal. The idea of data-driven pairwise penalties still applies, but instead of using the class of folded concave penalty functions, we may need to use penalty functions which are smooth at the origin, e.g., the $L_2$ penalty function. Another possible approach is to use posterior-type estimators combined with, say, a Gaussian prior on the coefficients. These are beyond the scope of this paper and we leave them as future work.

\section{Proofs}  \label{sec:proof}

\subsection{Proof of Theorem \ref{thm:fuse}}

Introduce the mapping $T: \cM_A \to \mathbb{R}^K$, where $T(\bbeta)$ is the $K$-dimensional vector whose $k$-th coordinate equals to the common value of $\beta_j$ for $j\in A_k$. Note that $T$ is a bijection and $T^{-1}$ is well-defined for any $\bmu\in\mathbb{R}^K$. Also, introduce the mapping $T^*: \mathbb{R}^p\to\mathbb{R}^K$, where $T^*(\bbeta)_k = \tfrac{1}{|A_k|}\sum_{j\in A_k}\beta_j$. We see that $T^*=T$ on $\cM_A$, and $T^{-1}\circ T^*$ is the orthogonal projection from $\mathbb{R}^p$ to $\cM_A$. Denote $\bmu^0=T(\bbeta^0)$ and $\hbmu^{oracle}=T(\hbbeta^{oracle})$.

Denote $L_n(\bbeta)= \tfrac{1}{2n}\Vert \by-\bX\bbeta\Vert^2$ and $P_n(\bbeta)= \lambda_n\sum_{j=1}^{p-1}\rho(\beta_{\tau(j+1)}-\beta_{\tau(j)})$, so that we can write $Q_n(\bbeta)= L_n(\bbeta)+ P_n(\bbeta)$. For any $\bmu\in\mathbb{R}^K$, let
\[
L_n^A(\bmu)= \frac{1}{2n}\Vert \by-\bX_A\bmu\Vert^2, \qquad P_n^A(\bmu)= \lambda_n\sum_{k=1}^{K-1}\rho(\mu_{k+1}-\mu_k),
\]
and define $Q_n^A(\bmu)= L_n^A(\bmu)+ P_n^A(\bmu)$. Note that when $\tau$ preserves the order of $\bbeta^0$, there exist $1=j_1<j_2<\cdots <j_K< j_{K+1}=p+1$ such that $A_k=\{\tau(j_k),\tau(j_k+1),\cdots,\tau(j_{k+1}-1)\}$ for $1\leq k\leq K$. Then $Q_n(\bbeta)=Q_n^A(T(\bbeta))$ and $Q_n^A(\bmu) = Q_n(T^{-1}(\bmu))$ for any $\bbeta\in\cM_A$ and $\bmu\in \mathbb{R}^K$.

In the first part of the proof, we show $\Vert \hbbeta^{oracle} - \bbeta^0 \Vert= O_p (\sqrt{K/n})$. By definition and direct calculations,
\[
\Vert \hbbeta^{oracle}-\bbeta^0\Vert = \Vert \bD(\hbmu^{oracle}-\bmu^0)\Vert, \qquad \hbmu^{oracle}-\bmu^0 = (\bX_A^T\bX_A)^{-1}\bX_A^T\bepsilon.
\]
Therefore, we can write
\[
\Vert \hbbeta^{oracle}-\bbeta^0\Vert =\Vert (\bD^{-1}\bX_A^T\bX_A\bD^{-1})^{-1}\bD^{-1}\bX_A^T\bepsilon\Vert.
\]
From Condition 3.1, $\Vert (\bD^{-1}\bX_A^T\bX_A\bD^{-1})^{-1}\Vert \leq (c_1n)^{-1}$ and $\text{tr}(\bD^{-1}\bX_A^T\bX_A\bD^{-1})\leq c_2nK$. By the Markov inequality, for any $\delta>0$,
\[
P\left( \Vert \bD^{-1}\bX_A^T\bepsilon\Vert > \sqrt{\frac{c_2 nK}{\delta}} \right) \leq \frac{E\Vert \bD^{-1}\bX_A^T\bepsilon\Vert^2}{c_2nK/\delta}  = \frac{\text{tr}(\bD^{-1}\bX_A^T\bX_A\bD^{-1})}{c_2nK/\delta} \leq \delta.
\]
Combining the above, we have shown that with probability at least $1-\delta$, $\Vert \hbbeta^{oracle}-\bbeta^0\Vert\leq C\delta^{-1/2}\sqrt{K/n}$. This proves $\Vert \hbbeta-\bbeta^0\Vert= O_p(\sqrt{K/n})$.

Furthermore, we can write $\bD^{-1}\bX_A^T\bepsilon=(\bv_1^T\bepsilon, \cdots, \bv_k^T\bepsilon)^T$,  where $\bv_k = \bX_A^{-1}\bD\be_k$ and $\be_k$ is the unit vector with $1$ on the $k$-th coordinate and $0$ elsewhere. Note that $\Vert \bv_k\Vert\leq \Vert \bD^{-1}\bX_A^T\bX_A\bD^{-1}\Vert\leq c_2n$. It follows from Condition 3.3 and the union bound that \beq \label{noise}
P\left( \Vert \bD^{-1}\bX_A^T\bepsilon\Vert_{\infty} > \sqrt{c_2c_3^{-1}n\log(2n)} \right)
\leq  \sum_{k=1}^K P\left( \Vert \bv_k^T\bepsilon\Vert > \Vert \bv_k\Vert \sqrt{c_3^{-1}\log(2n)}  \right)
\leq  n^{-1}K.
\eeq
Since $\Vert \bD^{-1}\bX_A^T\bepsilon\Vert\leq K^{1/2}\Vert \bD^{-1}\bX_A^T\bepsilon\Vert_{\infty}$, we have
\beq \label{oracle_L2bound}
\Vert \hbbeta^{oracle} - \bbeta^0\Vert \leq C\sqrt{K\log(n)/n}, \qquad \text{with probability }\geq 1-n^{-1}K.
\eeq

In the second part of the proof, we show that $\hbbeta^{oracle}$ is a strictly local minimum of $Q_n(\bbeta)$ with probability at least $1-\epsilon_0- n^{-1}K- 2p^{-1}$. By assumption, there is an event $E_1$ such that $P(E_1^c)\leq \epsilon_0$ and over the event $E_1$, $\tau$ preserves the order of $\bbeta^0$. Consider the neighborhood of $\bbeta^0$:
\[
\cB = \left\lbrace \bbeta\in\mathbb{R}^p:\ \Vert \bbeta-\bbeta^0 \Vert < 2C\sqrt{K\log(n)/n}  \right\rbrace.
\]
By \eqref{oracle_L2bound}, there is an event $E_2$ such that $P(E_2^c)\leq n^{-1}K$ and over the event $E_2$, $\Vert \hbbeta^{oracle}-\bbeta^0\Vert\leq C\sqrt{K\log(n)/n}$. Hence, $\hbbeta^{oracle}\in\cB$ over the event $E_2$.
For any $\bbeta\in \cB$, write $\bbeta^*$ as its orthogonal projection to $\cM_A$. We aim to show
\begin{enumerate}
\item[(a)] Over the event $E_1\cap E_2$,
\beq \label{key1}
Q_n(\bbeta^*)\geq Q_n(\hbbeta^{oracle}), \qquad \text{ for any } \bbeta\in\cB,
\eeq
and the inequality is strict whenever $\bbeta^* \neq \hbbeta^{oracle}$.
\item[(b)] There is an event $E_3$ such that $P(E_3^c)\leq 2p^{-1}$. Over the event $E_1\cap E_2\cap E_3$, there exists $\cB_n$, a neighborhood of $\hbbeta^{oracle}$, such that
\beq \label{key2}
Q_n(\bbeta) \geq Q_n( \bbeta^*),\qquad  \text{ for any } \bbeta\in \cB_n,
\eeq
and the inequality is strict whenever $\bbeta\neq\bbeta^*$.
\end{enumerate}
Combining (a) and (b), $Q_n(\bbeta)\geq Q_n(\hbbeta^{oracle})$ for any $\bbeta\in \cB_n$, a neighborhood of $\hbbeta^{oracle}$, and the inequality is strict whenever $\bbeta\neq\hbbeta^{oracle}$. This proves that $\hbbeta^{oracle}$ is a strictly local minimum of $Q_n$ over the event $E_1\cap E_2\cap E_3$, and the claim follows immediately. Below we show (a) and (b).

Consider (a) first. We claim
\beq \label{P=0}
P_n^A(T^*(\bbeta))=0 \qquad \text{for any } \bbeta\in\cB.
\eeq
To see this, for a given $\bbeta\in\cB$, write $\bmu = T^*(\bbeta)$. It suffices to check $|\mu_{k+1}-\mu_k|> a\lambda_n$ for $k=1,\cdots,K-1$. Note that $|\mu_{k+1}-\mu_k|\geq \min_{i\in A_k, j\in A_{k+1}}|\beta_i-\beta_j|\geq \min_{i,j}|\beta^0_i-\beta^0_j|- 2\Vert \bbeta-\bbeta^0\Vert_{\infty}\geq 2b_n - 2C\sqrt{K\log(n)/n}$. Since $b_n>a\lambda_n\gg \sqrt{K\log(n)/n}$, it is easy to see that $|\mu_{k+1}-\mu_k|>a\lambda_n$.

Using \eqref{P=0}, we see that $Q_n^A(T^*(\bbeta))= L_n^A(T^*(\bbeta))$, for all $\bbeta\in\cB$. By definition and the fact that $\tfrac{\partial^2 L_n^A(\bmu)}{\partial\bmu\partial\bmu^T}=\tfrac{1}{2n} \bX_A^T\bX_A$ is positive definite, $\hbmu^{oracle}$ is the unique global minimum of $L_n^A(\bmu)$. As a result, $L_n^A(T^*(\bbeta))\geq L_n^A(\hbmu^{oracle})= L_n(\hbbeta^{oracle})$, and the inequality is strict for any $T^*(\bbeta)\neq\hbmu^{oracle}$. Note that $Q_n^A=Q_n\circ T^{-1}$ and $T^{-1}\circ T^*$ is the orthogonal projection from $\mathbb{R}^p$ to $\cM_A$. Combining the above, for any $\bbeta\in\cB$,
\[
Q_n(\bbeta^*)= Q_n( T^{-1}\circ T^*(\bbeta)) = Q_n^A(T^*(\bbeta))= L_n^A(T^*(\bbeta))\geq L_n(\hbbeta^{oracle}),
\]
and the inequality is strict whenever $T^*(\bbeta)\neq \hbmu^{oracle}$, i.e., $\bbeta^*\neq T^{-1}(\hbmu^{oracle})=\hbbeta^{oracle}$. This proves \eqref{key1}.

Second, consider (b). For a positive sequence $t_n$ to be determined, let
\[
\cB_n =\cB\cap \{\bbeta: \Vert \bbeta-\hbbeta^{oracle} \Vert\leq t_n\}.
\]
Since $\bbeta^*$ is the orthogonal projection of $\bbeta$ to $\cM_A$, $\Vert \bbeta-\bbeta^*\Vert\leq \Vert \bbeta-\bbeta'\Vert$ for any $\bbeta'\in \cM_A$. In particular, $\Vert \bbeta- \bbeta^* \Vert\leq \Vert \bbeta-\hbbeta^{oracle}\Vert$. As a result, to show \eqref{key2}, it suffices to show
\beq \label{key2suffcond}
Q_n(\bbeta) \geq Q_n(\bbeta^*), \qquad \text{for any } \bbeta \text{ such that } \Vert \bbeta-\bbeta^*\Vert\leq t_n,
\eeq
and the inequality is strict whenever $\bbeta\neq\bbeta^*$.

To show \eqref{key2suffcond}, write $\bmu= T^*(\bbeta)$ so that $\bbeta^*= T^{-1}(\bmu)$. By Taylor expansion,
\begin{eqnarray*}
Q_n(\bbeta) - Q_n(\bbeta^*) &=&  -\frac{1}{n} (\by - \bX\bbeta^m)^T\bX(\bbeta - \bbeta^*) + \sum_{j=1}^p \frac{\partial P_n(\bbeta^m)}{\partial \beta_{\tau(j)}}(\beta_{\tau(j)} - \beta^*_{\tau(j)})\\
&\equiv & I_3 + I_4,
\end{eqnarray*}
where $\bbeta^m$ is in the line between $\bbeta$ and $\bbeta^*$.
Consider $I_4$ first. Direct calculations yield
\[
\frac{\partial P_n(\bbeta)}{\partial \beta_{\tau(j)}} = \left\lbrace \begin{array}{lr}
- \lambda_n \bar{\rho}(\beta_{\tau(2)}-\beta_{\tau(1)}), & j=1\\
\lambda_n \bar{\rho}(\beta_{\tau(j)}-\beta_{\tau(j-1)}) - \lambda_n \bar{\rho}(\beta_{\tau(j+1)}-\beta_{\tau(j)}), & 2\leq j\leq p-1\\
\lambda_n \bar{\rho}(\beta_{\tau(p)}-\beta_{\tau(p-1)}), & j=p,
\end{array} \right.
\]
where $\bar{\rho}(t)=\rho(t)\sgn(t)$ and $\rho(t)=\lambda^{-1}p_{\lambda}(t)$.
Plugging it into $I_4$ and rearranging the sum, we obtain
\beq  \label{I4(original)}
I_4 = \lambda_n \sum_{j=1}^{p-1} \bar{\rho}(\beta^m_{\tau(j+1)}-\beta^m_{\tau(j)}) \big[(\beta_{\tau(j+1)} - \beta_{\tau(j)}) - (\beta^*_{\tau(j+1)}- \beta^*_{\tau(j)}) \big].
\eeq
Note that when $\tau(j)$ and $\tau(j+1)$ belong to the same group, $\beta^*_{\tau(j)}=\beta^*_{\tau(j+1)}$, and hence the sign of $(\beta^m_{\tau(j+1)} - \beta^m_{\tau(j)})$ is the same as the sign of $(\beta_{\tau(j+1)} - \beta_{\tau(j)})$ if neither of them is $0$. In addition, recall that $A_k= \{\tau(j_k),\tau(j_k+1),\cdots,\tau(j_{k+1}-1)\}$ for all $1\leq k\leq K$, for some indices $1=j_1<j_2<\cdots <j_K=p$.
Combining the above, we can rewrite
\begin{eqnarray*}
I_4 &=& \lambda_n \sum_{k=1}^{K}\sum_{j=j_k}^{j_{k+1}-2} \rho'(|\beta^m_{\tau(j+1)}-\beta^m_{\tau(j)}|)|\beta_{\tau(j+1)} - \beta_{\tau(j)} | \notag\\
&& +\lambda_n \sum_{k=2}^{K} \bar{\rho}(|\beta^m_{\tau(j_k)}-\beta^m_{\tau(j_k-1)}|) \big[(\beta_{\tau(j_k)} - \beta_{\tau(j_k-1)}) - (\beta^*_{\tau(j_k)}- \beta^*_{\tau(j_k-1)})\big].
\end{eqnarray*}
First, since $\bbeta^0\in\cM_A$ and $\bbeta^*$ is the orthogonal projection of $\bbeta$ to $\cM_A$, $\Vert \bbeta^*-\bbeta^0\Vert\leq \Vert \bbeta-\bbeta^0\Vert$. Hence, $\bbeta\in\cB$ implies $\bbeta^*,\bbeta^m\in\cB$. By repeating the proof of \eqref{P=0}, we can show  $\bar{\rho}(|\beta^m_{\tau(j_k)}-\beta^m_{\tau(j_k-1)}|)=0$ for $2\leq k\leq K$. So the second term in $I_4$ disappears.
Second, in the first term of $I_4$, since $|\beta^m_{\tau(j+1)}-\beta^m_{\tau(j)}|\leq 2\Vert \bbeta^m-\bbeta^*\Vert_\infty\leq 2\Vert \bbeta-\bbeta^* \Vert_\infty \leq 2t_n$,  it follows by concavity that $\rho'(|\beta^m_{\tau(j+1)}-\beta^m_{\tau(j)}|)\geq \rho'(2t_n)$. Together, we have
\beq  \label{I4}
I_4 \geq \lambda_n  \sum_{k=1}^{K}\sum_{j=j_k}^{j_{k+1}-2} \rho'(2t_n) |\beta_{\tau(j+1)} - \beta_{\tau(j)} |.
\eeq

Next, we simplify $I_3$. Denote $\bz = \bz(\bbeta^m)= \bX^T(\by - \bX\bbeta^m)$ and write $I_3=-\bz^T(\bbeta-\bbeta^*)$. For any fixed $k$ and $l$ such that $\tau(l)\in A_k$ and $l\neq j_{k+1}-1$, let $A_{kl}^1=\{\tau(j)\in A_k: j\leq l\}$ and $A_{kl}^2=\{\tau(j)\in A_k: j>l\}$.
Regarding that $\beta_{\tau(i)}^*=\tfrac{1}{|A_k|}\sum_{j=j_k}^{j_{k+1}-1} \beta_{\tau(j)}$ for $i\in A_k$, we can reexpress $I_3$ as
\begin{eqnarray}  \label{reorder}
I_3 &=& - \frac{1}{2n} \sum_{k=1}^{K}\sum_{i=j_k}^{j_{k+1}-1} \frac{1}{n} z_{\tau(i)} \big[\beta_{\tau(i)} - \beta^*_{\tau(i)}\big] - \frac{1}{2n} \sum_{k=1}^{K}\sum_{j=j_k}^{j_{k+1}-1} \frac{1}{n} z_{\tau(j)} \big[ \beta_{\tau(j)} - \beta^*_{\tau(j)}\big] \notag\\
&=& -  \sum_{k=1}^{K} \frac{1}{2n|A_k|} \sum_{i,j=j_k}^{j_{k+1}-1} z_{\tau(i)}\big[ \beta_{\tau(i)} - \beta_{\tau(j)}\big] - \sum_{k=1}^{K} \frac{1}{2n|A_k|} \sum_{i,j=j_k}^{j_{k+1}-1} z_{\tau(j)}\big[ \beta_{\tau(j)} - \beta_{\tau(i)}\big]\notag\\
&=& - \sum_{k=1}^{K} \frac{1}{2n|A_k|} \sum_{i,j=j_k}^{j_{k+1}-1} \big[ z_{\tau(j)}-z_{\tau(i)}\big]\big[ \beta_{\tau(j)} - \beta_{\tau(i)}\big]\notag\\
&=& - \sum_{k=1}^{K} \frac{1}{n|A_k|} \sum_{j_k\leq i<j=j_{k+1}-1} \big[ z_{\tau(j)}-z_{\tau(i)}\big] \sum_{i\leq l< j}\big[ \beta_{\tau(l+1)} - \beta_{\tau(l)}\big]\notag\\
&=& - \sum_{k=1}^K \frac{1}{n|A_k|} \sum_{l=j_k}^{j_{k+1}-2} \big[ \beta_{\tau(l+1)} - \beta_{\tau(l)}\big]
\bigg[ |A_{kl}^1| \sum_{j\in A_{kl}^2} z_{\tau(j)} - |A_{kl}^2| \sum_{i\in A_{kl}^1} z_{\tau(i)}  \bigg]\notag\\
&\equiv & \sum_{k=1}^K \sum_{l=j_k}^{j_{k+1}-2} w_{\tau(l)}(\bz)\big[ \beta_{\tau(l+1)} - \beta_{\tau(l)}\big],
\end{eqnarray}
where for any vector $\bv\in\mathbb{R}^p$,
\[
w_{\tau(l)}(\bv) = n^{-1}\bigg[ \frac{|A^2_{kl}|}{|A_k|}\sum_{j\in A_{kl}^1} v_{\tau(j)} - \frac{|A^1_{kl}|}{|A_k|}\sum_{j\in A_{kl}^2} v_{\tau(j)}  \bigg].
\]
We aim to bound $|w_{\tau(l)}(\bz)|$. Denote $\boldsymbol{\eta}=\bX^T\bX(\bbeta^*-\bbeta^0)$, $\boldsymbol{\eta}^m=\bX^T\bX(\bbeta^m-\bbeta^*)$
and write $\bz= \bX^T\bepsilon + \boldsymbol{\eta} + \boldsymbol{\eta}^m$. First,  $w_{\tau(l)}(\bv)$ is a linear function of $\bv$. Second, since $\bbeta^m$ lies between $\bbeta$ and $\bbeta^*$, we have $\Vert \bbeta^*-\bbeta^m\Vert\leq \Vert \bbeta^*-\bbeta\Vert\leq t_n$. It follows that $\Vert \boldsymbol{\eta}^m \Vert\leq \lambda_{\max}(\bX^T\bX)t_n$. Moreover, $|w_{\tau(l)}(\bv)|\leq (|A_k|/n)\Vert \bv\Vert_\infty\leq (p/n)\Vert \bv\Vert$ for all $\bv$. Combining the above yields
\begin{eqnarray}  \label{w_decomp}
|w_{\tau(l)}(\bz)| &\leq & |w_{\tau(l)}(\bX^T\bepsilon)| + |w_{\tau(l)}(\boldsymbol{\eta})| + \sup_{\bv: \Vert\bv\Vert\leq \lambda_{\max}(\bX^T\bX)t_n} |w_{\tau(l)}(\bv)| \notag\\
&\leq & |w_{\tau(l)}(\bX^T\bepsilon)| + |w_{\tau(l)}(\boldsymbol{\eta})| + (p/n)\lambda_{\max}(\bX^T\bX)\cdot t_n.
\end{eqnarray}

First, we bound the term $w_{\tau(l)}(\bX^T\bepsilon)$. Let $E_3$ be the event that
\beq  \label{w_term1}
\max_{\tau(l)\in A_k} |w_{\tau(l)}(\bX^T\bepsilon)|\leq  n^{-1/2}\sqrt{2\sigma_k|A_k|\log(p)/c_3}, \qquad k=1,\cdots, K,
\eeq
where we recall $\sigma_k$ is the maximum eigenvalue of $n^{-1}\bX^T\bX$ restricted to the $(A_k, A_k)$-block. Given $\tau(l)$,
we can express $w_{\tau(l)}(\bX^T\bepsilon)$ as
\[
w_{\tau(l)}(\bX^T\bepsilon)=\ba_{\tau(l)}^T\bepsilon, \quad \text{where } \ba_{\tau(l)} = n^{-1}\bigg( \frac{|A^2_{kl}|}{|A_k|}\bX_{A^1_{kl}}\boldsymbol{1}_{A^1_{kl}} - \frac{|A^1_{kl}|}{|A_k|}\bX_{A^2_{kl}}\boldsymbol{1}_{A^2_{kl}}\bigg).
\]
Write $L_1=|A^1_{kl}|$ and $L_2=|A^2_{kl}|$, so that $|A_k|=L_1+L_2$. It is observed that $\Vert \bX_{A^1_{kl}}{\boldsymbol 1}_{A^1_{kl}}\Vert^2\leq n\sigma_k \Vert {\boldsymbol 1}_{A^1_{kl}}\Vert^2\leq n\sigma_k L_1$. Using the fact that $(a+b)^2\leq 2(a^2+b^2)$ for any real values $a,b$, we have $\Vert \ba_{\tau(l)}\Vert^2 \leq 2n^{-1}\sigma_k(L_2^2L_1/|A_k|^2 + L_1^2L_2/|A_k|^2) = 2\sigma_k L_1L_2/(n|A_k|)\leq \sigma_k|A_k|/(2n)$.
Applying Condition 3.3 and the probability union bound,
\begin{eqnarray} \label{prob_E3}
P(E_3^c)&\leq & \sum_{k=1}^K \sum_{\tau(l)\in A_k} P\left( |w_{\tau(l)}(\bX^T\bepsilon)|> n^{-1/2}\sqrt{\sigma_k|A_k|\log(2p)/c_3} \right) \nonumber\\
&\leq &  \sum_{1\leq j\leq p} P\left( |\ba_j^T\bepsilon|> \Vert \ba_j\Vert \sqrt{2\log(p)/c_3} \right)\ \leq\ 2p^{-1}.
\end{eqnarray}
Second, we bound the term $w_{\tau(l)}(\boldsymbol{\eta})$. Observing that for any vector $\bv$, $w_{\tau(l)}(\bv)= w_{\tau(l)}(\bv - \bar{v}_k\boldsymbol{1})$, where $\bar{v}_k$ is the mean of $\{v_j, j\in A_k\}$, we have
\[
|w_{\tau(l)}(\bv)|\leq \frac{2|A^1_{kl}||A^2_{kl}|}{n|A_k|}\max_{j\in A_k}|v_j-\bar{v}_k|\leq \frac{|A_k|}{2n} \max_{j\in A_k} |v_j-\bar{v}_k|,
\]
Since $\boldsymbol{\eta}=\bX^T\bX(\bbeta^*-\bbeta^0)$ and $\bbeta^*-\bbeta^0\in \cM_A$, we have $\max_{j\in A_k}|\eta_j-\bar{\eta}_k|\leq n\nu_k\Vert \bbeta^*-\bbeta^0\Vert$. As a result,
\beq  \label{w_term2}
\max_{\tau(l)\in A_k} |w_{\tau(l)}(\boldsymbol{\eta})| \leq (\nu_k/2)|A_k|\cdot \Vert \bbeta^*-\bbeta^0\Vert \leq C\nu_k|A_k|\sqrt{K\log(n)/n}.
\eeq

Combining \eqref{reorder}-\eqref{w_term2}, we find that over the event $E_1\cap E_2\cap E_3$,
\begin{eqnarray}  \label{I3}
&& |I_3| \nonumber\\
&\leq &  \sum_{k=1}^K  \sum_{l=j_k}^{j_{k+1}-2}  \left[ C\bigg( \sqrt{\frac{\sigma_k|A_k|\log(p)}{n}} + \nu_k|A_k|\sqrt{\frac{K\log(n)}{n}} \bigg) + \frac{p\lambda_{\max}(\bX^T\bX)}{n}t_n \right] | \beta_{\tau(l+1)} - \beta_{\tau(l)}|\nonumber\\
&\leq & \sum_{k=1}^K  \sum_{l=j_k}^{j_{k+1}-2} \left( \frac{\lambda_n}{2} + \frac{p\lambda_{\max}(\bX^T\bX)}{n}t_n \right) | \beta_{\tau(l+1)} - \beta_{\tau(l)}|,
\end{eqnarray}
where we have used the fact $\lambda_n\gg \max_k \{ \sqrt{\sigma_k|A_k|\log(p)/n} + \nu_k|A_k|\sqrt{K\log(n)/n}\}$.

From \eqref{I4} and \eqref{I3}, over the event $E_1\cap E_2\cap E_3$,
\[
\inf_{\bbeta\in\cB: \Vert \bbeta - \bbeta^* \Vert \leq t_n} \big[ Q_n(\bbeta) - Q_n(\bbeta^*)
 \big] \geq \sum_{k=1}^K \sum_{l=j_k}^{j_{k+1}-2} \Big[\frac{\lambda_n}{2} - g_n(t_n) \Big] | \beta_{\tau(l+1)} - \beta_{\tau(l)} | ,
\]
where $g_n(t_n)= n^{-1}p\lambda_{\max}(\bX^T\bX)t_n - \lambda_n[1-\rho'(2t_n)]$. Since $\rho'(0+)=1$, $g_n(0+)=0$. So we can always choose $t_n$ sufficiently small to make sure $|g_n(t_n)|< \lambda_n/2$; consequently, the right hand side is non-negative, and strictly positive when $\sum_{k=1}^K \sum_{l=j_k}^{j_{k+1}-2}|\beta_{\tau(l+1)} - \beta_{\tau(l)} |>0$, i.e., $\bbeta\neq\bbeta^*$. This proves \eqref{key2}.
\qed

\subsection{Proof of Theorem \ref{thm:LLA}}

First, we show that the LLA algorithm yields $\hbbeta^{oracle}$ after one iteration. Let $E_1$ be the event that $\tau$ preserves the order of $\bbeta^0$, $E_2$ the event that $\Vert \hbbeta-\bbeta^0\Vert\leq C\sqrt{K\log(n)/n}$ and $E_3$ the event that \eqref{w_term1} holds. We have shown that $P(E_1\cap E_2\cap E_3)\geq 1-\epsilon_0-n^{-1}K - 2p^{-1}$. It suffices to show that over the event $E_1\cap E_2\cap E_3$, the LLA algorithm gives $\hbbeta^{oracle}$ after the first iteration.

Let $w_j=\rho'(|\widehat{\beta}^{initial}_{\tau(j+1)}-\widehat{\beta}^{initial}_{\tau(j)}|)$. At the first iteration, the algorithm minimizes
\[
Q_n^{initial}(\bbeta) \equiv \frac{1}{2n}\Vert \by-\bX\bbeta\Vert^2 + \lambda_n \sum_{j=1}^{p-1} w_j|\beta_{\tau(j+1)}-\beta_{\tau(j)}|.
\]
This is a convex function, hence it suffices to show that $\hbbeta^{oracle}$ is a strictly local minimum of $Q_n^{initial}$. Using the same notations as in the proof of Theorem \ref{thm:fuse}, for any $\bbeta\in\mathbb{R}^p$, write $\bbeta^*=T^{-1}\circ T^*(\bbeta)$ as its orthogonal projection to $\cM_A$. Let $\cB=\{\bbeta\in\mathbb{R}^p: \Vert \bbeta-\bbeta^0\Vert\leq C\sqrt{K\log(n)/n}\}$, and for a sequence $\{t_n\}$ to be determined, consider the neighborhood of $\hbbeta^{oralce}$ defined by $\cB_n=\{\bbeta\in\cB: \Vert \bbeta-\hbbeta^{oracle}\Vert\leq t_n \}$.
It suffices to show
\beq  \label{keyobj_LLA}
Q_n^{initial}(\bbeta)\geq Q_n^{initial}(\bbeta^*) \geq Q_n^{initial}(\hbbeta^{oracle}), \qquad \text{for any } \bbeta\in\cB_n,
\eeq
and the first inequality is strict whenever $\bbeta\neq\bbeta^*$, and the second inequality is also strict whenever $\bbeta\neq\hbbeta^{oracle}$.

We first show the second inequality in \eqref{keyobj_LLA}. For $\tau(j)$ and $\tau(j+1)$ in different groups, $|\beta^0_{\tau(j+1)}-\beta^0_{\tau(j)}|> 2b_n$; also, $\Vert \hbbeta^{initial}-\bbeta^0\Vert_\infty\leq\lambda_n/2< b_n$. Hence, $|\widehat{\beta}^{initial}_{\tau(j+1)}-\widehat{\beta}^{initial}_{\tau(j)}|\geq 2b_n-\lambda_n>a\lambda_n$, and it follows that $w_j=0$. On the other hand, for $\tau(j)$ and $\tau(j+1)$ in the same group, $\beta_{\tau(j+1)}-\beta_{\tau(j)}=0$ whenever $\bbeta\in \cM_A$.
Consequently,
\[
Q_n^{initial}(\bbeta) = \frac{1}{2n}\Vert\by - \bX\bbeta \Vert^2 = L_n(\bbeta), \qquad \text{for }\ \bbeta\in\cM_A.
\]
We have seen in the proof of Theorem \ref{thm:fuse} that $\hbbeta^{oracle}$ is the unique global minimum of $L_n$ constrained on $\cM_A$. So the second inequality in \eqref{keyobj_LLA} holds.

Next, consider the first inequality in \eqref{keyobj_LLA}. By applying Talylor expansion and rearranging terms, for some $\bbeta^m$ that lies in the line between $\bbeta$ and $\bbeta^*$,
\begin{eqnarray*}
&& Q_n^{initial}(\bbeta) - Q_n^{initial}(\bbeta^*) \\
&=& \lambda_n \sum_{j=1}^{p-1} w_j \cdot \sgn(\beta^m_{\tau(j+1)}-\beta^m_{\tau(j)}) \big[(\beta_{\tau(j+1)} - \beta_{\tau(j)}) - (\beta^*_{\tau(j+1)}- \beta^*_{\tau(j)}) \big]\\
&& -\frac{1}{n} (\by - \bX\bbeta^m)^T\bX(\bbeta - \bbeta^*) \equiv  J_1 + J_2.
\end{eqnarray*}
We first simplify $J_1$. Note that $w_j=0$ when $\tau(j)$ and $\tau(j+1)$ are in different groups. When $\tau(j)$ and $\tau(j+1)$ are in the same $A_k$, first, $\beta^*_{\tau(j+1)}=\beta^*_{\tau(j)}$, and $[\beta^m_{\tau(j+1)}-\beta^m_{\tau(j)}]$ has the same sign as  $[\beta_{\tau(j+1)}-\beta_{\tau(j)}]$; second, $|\widehat{\beta}^{initial}_{\tau(j+1)}-\widehat{\beta}^{initial}_{\tau(j)}|\leq 2\Vert \hbbeta^{initial}-\bbeta^0\Vert_{\infty}\leq \lambda_n$, and hence $w_j\geq \rho'(\lambda_n)\geq a_0$. Combining the above yields
\beq \label{J1}
J_1 = \lambda_n \sum_{k=1}^K \sum_{j=j_k}^{j_{k+1}-2} w_j |\beta_{\tau(j+1)} - \beta_{\tau(j)}|
\geq   a_0 \lambda_n \sum_{k=1}^K \sum_{j=j_k}^{j_{k+1}-2} |\beta_{\tau(j+1)} - \beta_{\tau(j)}|
\eeq
Next, we simplify $J_2$. Denote $\bz=\bX^T(\by - \bX\bbeta^m)$. Similarly to \eqref{reorder}-\eqref{I3}, we find that
\[
J_2 = - \sum_{k=1}^K \sum_{l=j_k}^{j_{k+1}-2} w_{\tau(l)}(\bz)\big[ \beta_{\tau(l+1)} - \beta_{\tau(l)} \big],
\]
where over the event $E_3$, for any $j_k\leq l\leq j_{k+1}-2$,
\[
|w_{\tau(l)}(\bz)| \leq \sqrt{\frac{2\sigma_k|A_k|\log(p)}{c_3n}} + \frac{\nu_k|A_k|}{2}\sqrt{\frac{K\log(n)}{n}} + \frac{p\lambda_{\max}(\bX^T\bX)}{n}t_n.
\]
By the choice of $\lambda_n$, the sum of the first two terms is upper bounded by $a_0\lambda_n/3$ for large $n$; in addition, we choose $t_n=a_0n\lambda_n/(3p\lambda_{\max}(\bX^T\bX))$. It follows that
\beq \label{J2}
|J_2|\leq \sum_{k=1}^K \sum_{l=j_k}^{j_{k+1}-2} \frac{2a_0\lambda_n}{3} | \beta_{\tau(l+1)} - \beta_{\tau(l)} |.
\eeq
Combining \eqref{J1} and \eqref{J2}, over the event $E_1\cap E_2\cap E_3$,
\[
Q_n^{initial}(\bbeta) - Q_n^{initial}(\bbeta^*) \geq \frac{a_0\lambda_n}{3}\sum_{k=1}^K \sum_{l=j_k}^{j_{k+1}-2} | \beta_{\tau(l+1)} - \beta_{\tau(l)} |\geq 0.
\]
This proves the first inequality in \eqref{keyobj_LLA}.

Second, we show that over the event $E_1\cap E_2\cap E_3$, at the second iteration, the LLA algorithm still yields $\hbbeta^{oracle}$ and therefore it converges to $\hbbeta^{oracle}$. We have shown that after the first iteration, the algorithm outputs $\hbbeta^{oracle}$. It then treats $\hbbeta^{oracle}$ as the initial solution for the second iteration. So it suffices to check
\[
\Vert \hbbeta^{oracle} - \bbeta^0\Vert_\infty \leq \lambda_n/2.
\]
This is true because over the event $E_1$, $\Vert \hbbeta^{oracle} - \bbeta^0\Vert\leq C\sqrt{K\log(n)/n}\ll \lambda_n$. \qed

\subsection{Proof of Theorem \ref{thm:fuseL1}}

Since $\tau$ is consistent with $\bbeta^0$, there exists $1=j_1<j_2<\cdots <j_{K+1}=p+1$ such that $A_k=\{\tau(j_k),\tau(j_k+1), \cdots, \tau(j_{k+1}-1)\}$ for all $k$. We shall write $\tau(j)=j$ without loss of generality.

In the first part of the proof, we show that $\hbbeta\in\cM_A$, and it satisfies the sign restrictions $\sgn(\hat{\beta}_{A,k+1}-\hat{\beta}_{A,k})=\sgn(\beta^0_{A,k+1}-\beta^0_{A,k})$, $k=1,\cdots, K-1$.

When $\rho(t)=|t|$, $Q_n(\bbeta)$ is strictly convex. So $\hbbeta$ is the unique global minimum if and only if it satisfies the first-order conditions:
\[
0 =
\left\lbrace
\begin{array}{lr}
- \tfrac{1}{n}\bx_1^T\bepsilon + \tfrac{1}{n}\bx_1^T \bX(\hbbeta - \bbeta^0) - \lambda_n \sgn(\hat{\beta}_{2}- \hat{\beta}_j), \\
- \tfrac{1}{n}\bx_j^T\bepsilon + \tfrac{1}{n}\bx_j^T \bX(\hbbeta - \bbeta^0) + \lambda_n \sgn(\hat{\beta}_{j}-\hat{\beta}_{j-1}) - \lambda_n \sgn(\hat{\beta}_{j+1}- \hat{\beta}_j), & 2\leq j\leq p\\
- \tfrac{1}{n}\bx_p^T\bepsilon + \tfrac{1}{n}\bx_p^T \bX(\hbbeta - \bbeta^0) + \lambda_n \sgn(\hat{\beta}_{p}-\hat{\beta}_{p-1}),
\end{array} \right.
\]
where $\sgn(t)=1$ when $t>0$, $-1$ when $t<0$, and any value in $[-1,1]$ when $t=0$. Therefore, it suffices to show there exists $\hbbeta\in \cM_A$ that satisfy the sign restrictions and the first-order conditions simultaneously.

For $\hbbeta\in\cM_A$, we write $\hbmu=T(\hbbeta)$ and $\bmu^0=T(\bbeta^0)$, where the mapping $T$ is the same as that in the proof of Theorem \ref{thm:fuse}. The sign restrictions now become $\sgn(\hat{\mu}_{k+1}-\hat{\mu}_{k})=\sgn(\mu^0_{k+1}-\mu^0_{k})$ for all $k=1,\cdots, K-1$.  Note that $\hat{\beta}_{j}=\hat{\beta}_{j+1}$ when predictors $j$ and $(j+1)$ belong to the same group in $\cA$. The first-order conditions can be re-expressed as
\beq  \label{L1_foc}
0 = \left\lbrace
\begin{array}{ll}
- \tfrac{1}{n}\bx_{j}^T\bepsilon + \tfrac{1}{n}\bx_{j}^T\bX_A(\hbmu - \bmu^0) + \lambda_n \sgn(\hat{\mu}_{k}-\hat{\mu}_{k-1}) - \lambda_n r_{j}, & j=j_k \\
- \tfrac{1}{n}\bx_{j}^T\bepsilon + \tfrac{1}{n}\bx^T_{j}\bX_A(\hbmu - \bmu^0) + \lambda_n r_{j-1} - \lambda_n \sgn(\hat{\mu}_{k}- \hat{\mu}_{k-1}),  & j=j_k-1 \\
- \tfrac{1}{n}\bx_j^T\bepsilon + \tfrac{1}{n}\bx_{j}^T\bX_A(\hbmu - \bmu^0) + \lambda_n r_{j+1} - \lambda_n r_{j}, & \text{elsewhere},
\end{array} \right.
\eeq
where $r_j$'s take any values on $[-1,1]$ and we set $\sgn(\hat{\mu}_{1}-\hat{\mu}_{0})=\sgn(\hat{\mu}_{K+1}-\hat{\mu}_{K})=0$ by default. Denote by $\delta^0_k=\sgn(\mu^0_{k+1}-\mu^0_k)$ when $1\leq k\leq K-1$ and $\delta^0_k= 0$ when $k=0, K$; similarly, $\hat{\delta}_k$ for $1\leq k\leq K$.
In \eqref{L1_foc}, we first remove $r_j$'s by summing up the equations corresponding to indices in each $A_k$. Using the fact that $\bx_{A,k}=\sum_{j\in A_k}\bx_j$, we obtain
\[
- \tfrac{1}{n} \bx_{A,k}^T\bepsilon + \tfrac{1}{n}\bx_{A,k}\bX_A(\hbmu - \bmu^0) + \lambda_n \hat{\delta}_{k-1} - \lambda_n \hat{\delta}_k = 0, \quad k=1,\cdots, K.
\]
Under the sign restrictions $\hat{\delta}_k=\delta^0_k$, $k=1,\cdots, K-1$, it becomes a pure linear equation of $(\hbmu-\bmu^0)$:
\[
- \tfrac{1}{n} \bX_{A}^T\bepsilon + \tfrac{1}{n}\bX_{A}\bX_A(\hbmu - \bmu^0) + \lambda_n \bd^0 = 0,
\]
where $\bd^0$ is the $K$-dimensional vector with $d^0_k=\delta^0_{k}-\delta^0_{k-1}$, as defined in Section \ref{subsec:simple_L2}.
It follows immediately that
\beq \label{L1_mu}
\hbmu - \bmu^0 = n\lambda_n (\bX_A^T\bX_A)^{-1} \bd^0 + (\bX_A^T\bX_A)^{-1}\bX_A^T\bepsilon.
\eeq
Second, given $(\hbmu-\bmu^0)$, \eqref{L1_foc} can be viewed as equations of $r_j$'s and we can solve them directly. Denote $\btheta = \tfrac{1}{n}\bX^T\bX_A(\hbmu - \bmu^0)-\tfrac{1}{n}\bX^T\bepsilon$. For each $j\in A_k$, define $A^1_{kj}=\{j_k,\cdots, j\}$ and $A^2_{kj}=\{j+1,\cdots, j_{k+1}-1\}$. The solutions of \eqref{L1_foc} are
\[
r_{j} = \hat{\delta}_{k-1} + \lambda_n^{-1} \sum_{i\in A^1_{kj}} \theta_{i} = \hat{\delta}_k - \lambda_n^{-1}\sum_{i\in A^2_{kj}}\theta_i, \qquad j\in A_k.
\]
Here the two expressions of $r_j$ are equivalent because $\lambda_n\sum_{i\in A_k}\theta_i=\hat{\delta}_k-\hat{\delta}_{k-1}$ from \eqref{L1_foc}. It follows that any convex combination of the two expressions is also an equivalent expression of $r_j$. Taking the combination coefficients as $|A^1_{kj}|/|A_k|$ and $|A^2_{kj}|/|A_k|$, and plugging in the sign restrictions $\hat{\delta}_k=\delta^0_k$, $k=1,\cdots, K-1$, we obtain
\begin{align*}
r_j &= \lambda_n^{-1} \Big(\frac{|A^2_{kj}|}{|A_k|}\sum_{i\in A^1_{kj}}\theta_i - \frac{|A^1_{kj}|}{|A_k|}\sum_{i\in A^2_{kj}}\theta_i \Big) + \Big( \frac{|A^2_{kj}|}{|A_k|}\delta^0_{k-1} + \frac{|A^1_{kj}|}{|A_k|}\delta^0_{k}\Big)\cr
&= n\lambda_n^{-1} w_j(\btheta) + \Big( \frac{|A^2_{kj}|}{|A_k|}\delta^0_{k-1} + \frac{|A^1_{kj}|}{|A_k|}\delta^0_{k}\Big),
\end{align*}
where the function $w_j(\cdot)$ is defined as in \eqref{reorder}. Here $r_j$'s still depend on $(\hbmu-\bmu^0)$ through $\btheta$. Combining \eqref{L1_mu} to the definition of $\btheta$ gives
\begin{align*}
\btheta  &= -\tfrac{1}{n}\bX^T\left[\bI - \bX_A(\bX_A^T\bX_A)^{-1}\bX_A^T\right] \bepsilon + \lambda_n \bX^T\bX_A(\bX_A^T\bX_A)^{-1}\bd^0 \\
&\equiv -\tfrac{1}{n}\bX^T\bar{\bP}_A\bepsilon + \lambda_n \bb^0,
\end{align*}
where $\bar{\bP}_A=\bI - \bX_A(\bX_A^T\bX_A)^{-1}\bX_A^T$ and $\bb^0$ is defined as in Section \ref{subsec:simple_L2}. By plugging in the expression of $\btheta$, we can remove the dependence on $(\hbmu-\bmu^0)$ of the solutions of $r_j$'s:
\beq  \label{L1_r}
r_j = - \lambda_n^{-1} w_j(\bX^T\bar{\bP}_A\bepsilon) + n w_j(\bb^0) + \Big( \frac{|A^2_{kj}|}{|A_k|}\delta^0_{k-1} + \frac{|A^1_{kj}|}{|A_k|}\delta^0_k \Big).
\eeq

Now, to show the the existence of $\hbbeta\in\cM_A$ that satisfies both the sign restrictions and first-order conditions, it suffices to show with probability at least $1-\epsilon_0-n^{-1}K-2p^{-1}$,
\begin{itemize}
\item[(a)] the $r_j$'s in \eqref{L1_r} take values on $[-1,1]$;
\item[(b)] the $\hbmu$ in \eqref{L1_foc} satisfy the sign restrictions, i.e.,  $\sgn(\hat{\mu}_{k+1}-\hat{\mu}_k)=\sgn(\mu^0_{k+1}-\mu^0_k)$ for all $k=1,\cdots, K-1$.
\end{itemize}

Consider (a) first. In \eqref{L1_r}, under the ``irrepresentability" condition, the sum of the last two terms is bounded by $(1-\omega_n)$ in magnitude. To deal with the first term, recall that in deriving \eqref{w_term1}, we write $w_j(\bX^T\bepsilon)=\ba_j^T\bepsilon$. It follows immediately that $w_j(\bX^T\bar{\bP}_A\bepsilon)=\ba_j^T\bar{\bP}_A\bepsilon=(\bar{\bP}_A\ba_j)^T\bepsilon$. Since $\Vert \bar{\bP}_A\ba_j\Vert\leq \Vert \ba_j \Vert$, similarly to \eqref{w_term1}, we obtain
\[
\max_{j\in A_k}|w_j(\bX^T\bar{\bP}_A\bepsilon)| \leq C\sqrt{\sigma_k|A_k|\log(p)/n}, \qquad 1\leq k\leq K,
\]
except for a probability at most $2p^{-1}$.
Therefore, by the choice of $\lambda_n$, the absolute value of the first term is much smaller than $\omega_n$. So $\max_{j}|r_j|\leq 1$ except for a probability at most $2p^{-1}$, i.e., (a) holds.

Next, consider (b). Since $|\mu^0_{k+1}-\mu^0_k|\geq 2b_n$, it suffices to show that $\Vert \hbmu-\bmu^0\Vert_\infty < b_n$. Note that \eqref{L1_mu} can be rewritten as
\[
\hbmu - \bmu^0  = \lambda_n \bD^{-1}(\tfrac{1}{n}\bD^{-1}\bX_A^T\bX_A\bD^{-1})^{-1}(\lambda_n \bD^{-1}\bd^0 + \bD^{-1}\bX_A^T\bepsilon).
\]
It follows from Condition 3.1 that $\Vert \bmu-\bmu^0\Vert\leq c_1^{-1}(\lambda_n\Vert \bD^{-2}\bd^0\Vert + \Vert \bD^{-1}\Vert\Vert \bD^{-1}\bX_A^T\bepsilon\Vert)$. First, note that $\Vert \bD^{-2}\bd^0\Vert^2\leq 4\sum_{k=1}^K\tfrac{1}{|A_k|^2}$. Second, from \eqref{oracle_L2bound}, $\Vert \bD^{-1}\bX_A^T\bepsilon\Vert\leq C\sqrt{nK\log(n)}$, except a probability of at most $n^{-1}K$. Moreover, $\Vert \bD^{-1}\Vert=(\min_{k}|A_k|)^{-1}\leq 1$.  These together imply
\[
\Vert \hbmu - \bmu^0 \Vert \leq C\lambda_n \Big( \sum_{k=1}^K \frac{1}{|A_k|^2}\Big)^{1/2} + C\sqrt{\frac{K\log(n)}{n}}.
\]
From the conditions on $b_n$, the right hand side is much smaller than $b_n$. It follows that $\Vert \hbmu-\bmu^0\Vert_\infty \ll b_n$. This proves (b).

In the second part of the proof, we derive the convergence rate of $\Vert \hbbeta-\bbeta^0\Vert$.
Note that $\Vert \hbbeta - \bbeta^0\Vert= \Vert\bD(\hbmu - \bmu^0 )\Vert$, and from \eqref{L1_mu},
\[
 \bD(\hbmu - \bmu^0) = (\tfrac{1}{n}\bD^{-1}\bX_A^T\bX_A\bD^{-1})^{-1} \big( \lambda_n \bD^{-1}\bd^0 + n^{-1} \bD^{-1}\bX_A^T\bepsilon \big).
\]
Therefore, $\Vert  \hbbeta-\bbeta^0\Vert\leq c_1^{-1}(\lambda_n \Vert \bD^{-1}\bd^0\Vert + n^{-1} \bD^{-1}\bX_A^T\bepsilon )$, where $\Vert \bD^{-1}\bd^0\Vert^2\leq 4\sum_{k=1}^K \tfrac{1}{|A_k|}$ and $\Vert  \bD^{-1}\bX_A^T\bepsilon \Vert= O_p (\sqrt{nK})$ by \eqref{noise}. Combining these gives
\[
\Vert \hbbeta - \bbeta^0\Vert = O_p\Big(\sqrt{K/n} + \lambda_n\big(\sum_k \tfrac{1}{|A_k|}\big)^{1/2} \Big).
\]

\subsection{Proof of Theorem \ref{thm:ols}}

The order generated by $\hbbeta^{ols}$ preserves the order of $\bbeta^0$, if and only if, $\beta^0_i<\beta^0_j$ implies $\hat{\beta}^{ols}_i\leq \hat{\beta}^{ols}_j$ for any pair $1\leq i,j\leq p$. Note that when $\beta^0_i<\beta^0_j$, necessarily $\beta^0_j-\beta^0_i\geq 2b_n$. Moreover, $\hat{\beta}_j^{ols}- \hat{\beta}_i^{ols}\geq (\beta^0_j-\beta^0_i) - 2\Vert \hbbeta^{ols}-\bbeta^0\Vert_\infty$. So
it suffices to show that
$\Vert \hbbeta^{ols}- \bbeta^0\Vert_\infty \leq b_n$ with probability at least $1-2p^{-1}$.

From direct calculations, $\bbeta^{ols}=\bbeta^0 + (\bX^T\bX)^{-1}\bX^T\bepsilon$. Let $\ba_j=\bX(\bX^T\bX)^{-1}\be_j$, $j=1,\cdots, p$. Then $\Vert \ba_j\Vert^2 = \be_j^T(\bX^T\bX)^{-1}\be_j\leq c_4n^{-1}$. Note that $\hat{\beta}^{ols}_j-\beta^0_j=\ba_j^T\bepsilon$. By Condition 3.3 and applying the union bound,
\begin{eqnarray*}
&& P\left( \Vert \hbbeta^{ols} - \bbeta^0 \Vert_\infty > \sqrt{(2c_4/c_3)\log(p)/n} \right) \\
&\leq & \sum_{j=1}^p P\left( | \ba_j^T\bepsilon| > \Vert \ba\Vert \sqrt{2\log(p)/c_3} \right)\ \leq \sum_{j=1}^p 2p^{-2}.
\end{eqnarray*}
So, with probability at least $1-2p^{-1}$, $\Vert \hbbeta^{ols} - \bbeta^0 \Vert_\infty < b_n$. This completes the proof.
\qed

\subsection{Proof of Proposition \ref{prop:A&B}}

Consider the first claim. Given $k$, let $d_k=\min\{l: V_{kl}\neq\emptyset\}$ and $u_k=\max\{l: V_{kl}\neq\emptyset\}$. Then $A_k=\cup_{l=d_k}^{u_k}V_{kl}$. Moreover, for any $d_k<l<u_k$,
\[
\beta^0_{A,k}\leq \max_{i\in B_{d_k}} \beta^0_i \leq \min_{j\in B_l} \beta^0_j \leq \max_{j\in B_l}\leq \min_{i\in B_{u_k}} \beta^0_i \leq \beta^0_{(k)},
\]
where the first and last inequalities are because $A_k\cap B_{d_k}\neq\emptyset$ and $A_k\cap B_{u_k}\neq \emptyset$, and the inequalities between come from Definition \ref{def:cond_seg}. It follows that
$\beta^0_j=\beta^0_{A,k}$ for all $j\in B_l$. This means $B_l\subset A_k$, and hence $V_{kl}= B_l$.

Consider the second claim. Given $l$, let $a_l=\min\{k: V_{kl}\neq\emptyset\}$ and $b_l=\max\{k: V_{kl}\neq\emptyset\}$, and so $B_l=\cup_{k=a_l}^{b_l}V_{kl}$. For any $a_l < k < b_l$ and $l'< l$,
\[
\max_{i\in B_{l'}} \beta^0_i \leq \min_{i\in B_l} \beta^0_i \leq \beta^0_{A,a_l}< \beta^0_{A,k},
\]
where the first inequality comes from Definition \ref{def:cond_seg}, the second inequality is because $A_{a_l}\cap B_l\neq\emptyset$ and the last inequality is from the labelling of groups and the fact that $a_l<k$. It follows that
$B_{l'}\cap A_k=\emptyset$. Similarly, for any $l'> l$, $B_{l'}\cap A_k=\emptyset$. As a result, $A_k\subset B_l$ and $V_{kl}=A_k$.
\qed

\subsection{Proof of Theorem \ref{thm:rate}}

Recall the mappings $T$, $T^{-1}$ and $T^*$ defined in the proof of Theorem \ref{thm:fuse}. Write $Q_n(\bbeta)= L_n(\bbeta)+ P_n(\bbeta)$, where $L_n(\bbeta)=\tfrac{1}{2n}\Vert \by - \bX\bbeta\Vert^2$ and $P_n(\bbeta)= P_{\Upsilon, \lambda_1,\lambda_2}(\bbeta)$.
For any $\bmu\in\mathbb{R}^K$, let
\[
L_n^A(\bmu)= L_n(T^{-1}(\bmu)), \qquad P_n^A(\bmu)= P_n(T^{-1}(\bmu)),
\]
and define $Q_n^A(\bmu)= L_n^A(\bmu)+ P_n^A(\bmu)$.

We only need to show that $\hbbeta^{oracle}$ is a strictly local minimum of $Q_n$ with probability at least $1-\epsilon_0-n^{-1}K-5p^{-1}$. Let $E_1'$ be the event that $\cB$ preserves the order of $\bbeta^0$, and define the event $E_2$ and the set $\cB$ the same as in the proof of Theorem \ref{thm:fuse}. For an event $E_3'$ to be defined such that $P((E_3')^c)\leq 5p^{-1}$, we shall show that \eqref{key1} and \eqref{key2} hold on the event $E_1'\cap E_2\cap E_3'$. The claim then follows immediately. Similar to the proof of Theorem \ref{thm:fuse}, it suffices to show \eqref{P=0} and \eqref{key2suffcond}.

Consider \eqref{P=0} first. Recall that $V_{kl}=A_k\cap B_l$. Define $m_{1,kk'}=\sum_{l=1}^{L-1}(|V_{kl}||V_{k'(l+1)}|+ |V_{k'l}||V_{k(l+1)}|)$ and $m_{2,kk'}=\sum_{l=1}^L|V_{kl}||V_{k'l}|$, for $1\leq k<k'\leq K$. Write for short $\rho_1=\rho_{\lambda_1}$ and $\rho_2=\rho_{\lambda_2}$. It follows that
\[
P^A_n(\bmu) = \lambda_1 \sum_{1\leq k<k'\leq K}  m_{1,kk'}\rho_1(|\mu_{k}-\mu_{k'}|) + \lambda_2 \sum_{1\leq k<k'\leq K} m_{2,kk'} \rho_2(|\mu_{k}-\mu_{k'}|).
\]
Therefore, it suffices to check $\min_{k\neq k'}|\mu_k - \mu_{k'}|> a\max\{ \lambda_{1n}, \lambda_{2n}\}$. Note that the left hand side is lower bounded by $2b_n - \Vert \bbeta- \bbeta^0\Vert_{\infty}\geq 2b_n - C\sqrt{K\log(n)/n}\gg b_n> a\max\{ \lambda_{1n}, \lambda_{2n}\}$, which proves \eqref{P=0}.

Next, consider \eqref{key2suffcond}. For $\bbeta\in\cB$, write $\bmu=T^*(\bbeta)$, $\bbeta^*=T^{-1}(\bmu)$. By Taylor expansion,
\[
Q_n(\bbeta) - Q_n(\bbeta^*) =  -\frac{1}{n} (\by - \bX\bbeta^m)^T\bX(\bbeta - \bbeta^*) + \sum_{j=1}^p \frac{\partial P_n(\bbeta^m)}{\partial \beta_j}(\beta_j - \beta^*_j)\equiv K_1 + K_2,
\]
where $\bbeta^m$ is in the line between $\bbeta$ and $\bbeta^*$. Let $\bar{\rho}_i(t)=\rho'_i(t)\sgn(t)$, $i=1,2$. By rearranging terms in $K_2$, we can write
\begin{eqnarray*}
K_2 &=& \lambda_1 \sum_{l=1}^{L-1}\sum_{i\in B_l, j\in B_{l+1}}\bar{\rho}_1(\beta^m_i-\beta^m_j)\big[ (\beta_i-\beta_j) - (\beta^*_i - \beta^*_j) \big] \\
&& + \lambda_2 \sum_{l=1}^L \sum_{i,j\in B_l} \bar{\rho}_2(\beta^m_i-\beta^m_j) \big[ (\beta_i-\beta_j) - (\beta^*_i - \beta^*_j) \big].
\end{eqnarray*}
For those $i,j$ not belonging the same true group, $|\beta^m_i-\beta^m_j|\geq 2b_n - 2\Vert\bbeta^m-\bbeta^0 \Vert_\infty\geq 2b_n - \Vert \bbeta^* - \bbeta^0\Vert_{\infty}$. Similarly as before, we obtain $\rho(|\beta^m_i-\beta^m_j|)=0$.
On the other hand, for those $i,j$ belonging to the same true group, $\beta^*_i=\beta^*_j$ and hence $\sgn(\beta^m_i-\beta^m_j)= \sgn(\beta_i-\beta_j)$. Together, we find that
\begin{eqnarray}  \label{K1}
K_2 &=& \lambda_1 \sum_{l=1}^{L-1}\sum_{i\in B_l, j\in B_{l+1}, i\sameA j}\rho'_1(|\beta^m_i-\beta^m_j|) |\beta_i-\beta_j|  + \lambda_2 \sum_{l=1}^L \sum_{i,j\in B_l, i\sameA j} \rho'_2(|\beta^m_i-\beta^m_j|)|\beta_i-\beta_j|  \notag\\
&\geq & \lambda_1  \sum_{l=1}^{L-1}\sum_{i\in B_l, j\in B_{l+1}, i\sameA j} \rho_1'(2t_n)|\beta_i-\beta_j| +  \lambda_2 \sum_{l=1}^L \sum_{i,j\in B_l, i\sameA j}\rho_2'(2t_n) |\beta_i-\beta_j|,
\end{eqnarray}
where $i\sameA j$ means $i$ and $j$ are in the same true group, and the last inequality comes from the concavity of $\rho$ and the fact that $|\beta^m_i-\beta^m_j|\leq 2\Vert \bbeta-\bbeta^*\Vert_{\infty}\leq 2t_n$.

Now, we simplify $K_1$. Let $\bz=\bz(\bbeta^m)=\bX^T(\by-\bX\bbeta^m)$ and write $K_1=-\tfrac{1}{n}\bz^T(\bbeta-\bbeta^*)$. Note that for each $j\in A_k$, $\beta^*_j=\tfrac{1}{|A_k|}\sum_{i\in A_k}\beta_i=\tfrac{1}{|A_k|}\sum_{l=d_k}^{u_k}\sum_{i\in V_{kl}}\beta_i$, where $V_{kl}$, $d_k$ and $u_k$ are as in Proposition \ref{prop:A&B}.
\begin{eqnarray*}
K_1 &=& - \frac{1}{n}\sum_{k=1}^K \sum_{l=d_k}^{u_k} \sum_{j\in V_{kl}} z_j (\beta_j-\beta^*_j)\\
&=& - \frac{1}{n}\sum_{k=1}^K \sum_{l=d_k}^{u_k} \sum_{j\in V_{kl}} z_j \frac{1}{|A_k|}\sum_{l'=d_k}^{u_k}\sum_{j'\in V_{kl'}}(\beta_j-\beta_{j'})\\
&=& -\frac{1}{2n}\sum_{k=1}^K \frac{1}{|A_k|} \sum_{l=d_k}^{u_k}\sum_{l'=d_k}^{u_k}\sum_{j\in V_{kl}}\sum_{j'\in V_{kl'}} (z_j-z_{j'})(\beta_j-\beta_{j'})\\
&=& -\frac{1}{2n}\sum_{k=1}^K \frac{1}{|A_k|} \sum_{l=d_k}^{u_k} \sum_{j,j'\in V_{kl}}(z_j-z_{j'})(\beta_j-\beta_{j'})\\
&& -\frac{1}{n}\sum_{k=1}^K \frac{1}{|A_k|}\sum_{d_k\leq l<l'\leq u_k} \sum_{j\in V_{kl}, j'\in V_{kl'}}(z_j-z_{j'})(\beta_j-\beta_{j'})\\
& \equiv & K_{11}+ K_{12}.
\end{eqnarray*}
Using notations in Proposition \ref{prop:A&B}, $\sum_{k=1}^K\sum_{l=d_k}^{u_k}=\sum_{l=1}^L\sum_{k=a_l}^{b_l}$. Therefore,
\begin{eqnarray}   \label{K21}
K_{11} &=& -\frac{1}{2n} \sum_{l=1}^L \sum_{k=a_l}^{b_l} \sum_{j,j'\in V_{kl}} \frac{1}{|A_k|}(z_j-z_{j'})(\mu_j-\mu_{j'})\notag\\
&=& -\frac{1}{2n}\sum_{l=1}^L \sum_{j,j'\in B_l, j\sameA j'} \theta_{jj'}(\bz) (\mu_j-\mu_{j'}),
\end{eqnarray}
where $\theta_{jj'}(\bz)\equiv \tfrac{1}{|A_k|}(z_j-z_{j'})$ for $j,j'\in A_k$.
To simplify $K_{12}$, note that given any $(j,j')$ such that $j\in V_{kl}$ and $j'\in V_{kl'}$, for some $k$ and $l<l'$, we have
\[
\beta_{j}-\beta_{j'} = \frac{1}{\prod_{h=l+1}^{l'-1} |V_{kh}|} \sum_{ \big\{\substack{(i_l,i_{l+1},\cdots, i_{l'}):\ i_l=j,\ i_{l'}=j';\\ i_h\in V_{kh}, h=l+1,\cdots, l'-1} \big\}} \sum_{h=l}^{l'-1} (\beta_{i_h}-\beta_{i_{h+1}}).
\]
Plugging this into the expression $K_{12}$, we obtain
\begin{eqnarray*}
K_{12} &=& -\frac{1}{n}\sum_{k=1}^K \frac{1}{|A_k|}\sum_{d_k\leq l<l'\leq u_k}  \sum_{ \{(i_l,i_{l+1},\cdots, i_{l'}):\ i_h\in V_{kh}\}} \frac{(z_{i_l}-z_{i_{l'}})}{\prod_{h=l+1}^{l'-1} |V_{kh}|}\sum_{h=l}^{l'-1} (\beta_{i_h}-\beta_{i_{h+1}})\\
&=& -\frac{1}{n}\sum_{k=1}^K \frac{1}{|A_k|}\sum_{d_k\leq l<l'\leq u_k} \sum_{h=l}^{l'-1}\sum_{j\in V_{kh}, j'\in V_{k(h+1)}} \omega_{jj',ll'h}(\bz)(\beta_j-\beta_{j'}),
\end{eqnarray*}
where for $(j,j',l,l',h)$ such that $j\in V_{kh}$, $j'\in V_{k(h+1)}$ and $l\leq h\leq l'-1$,
\[
\omega_{jj',ll'h}(\bz) = \left\lbrace
\begin{array}{ll}
z_j - z_{j'}, & l=h=l'-1\\
\frac{|V_{kl'}|}{|V_{k(l+1)}|}(z_j - \bar{z}_{kl'}), &  l=h<l'-1\\
\frac{|V_{kl}||V_{kl'}|}{|V_{kh}||V_{k(h+1)}|}(\bar{z}_{kl}-\bar{z}_{kl'}), & l<h<l'-1\\
\frac{|V_{kl}|}{|V_{k(l'-1)}|}(\bar{z}_{kl}- z_{j'}),  & l<h=l'-1
\end{array} \right.,
\]
and $\bar{z}_{kl}$ is the average of $\{z_j: j\in V_{kl}\}$. By rearranging terms, $\sum_{k=1}^K \sum_{d_k\leq l<l'\leq u_k}\sum_{h=l}^{l'-1}= \sum_{h=1}^L\sum_{k=a_h}^{b_h}\sum_{(l,l'): d_k\leq l\leq h<l'\leq u_k}$. Therefore,
\begin{eqnarray}  \label{K22}
K_{12} &=& - \frac{1}{n}\sum_{h=1}^{L-1} \sum_{k=a_h}^{b_h} \frac{1}{|A_k|} \sum_{j\in V_{kh}, j'\in V_{k(h+1)}}
\Big[ \sum_{l=d_k}^h \sum_{l'=h+1}^{u_k} \omega_{jj',ll'h}(\bz)\Big] (\beta_j-\beta_{j'}) \notag\\
&=& - \frac{1}{n}\sum_{h=1}^{L-1} \sum_{j\in B_h, j'\in B_{h+1}, j\sameA j'} \tau_{jj'}(\bz) (\beta_j-\beta_{j'}),
\end{eqnarray}
where
\begin{eqnarray*}
\tau_{jj'}(\bz) &=& \frac{1}{|A_k|} \sum_{l=d_k}^h \sum_{l'=h+1}^{u_k} \omega_{jj',ll'h}(\bz)\\
&=& \frac{1}{|A_k|}\sum_{l=d_k}^{h-1}\sum_{l'=h+2}^{u_k} \frac{|V_{kl}||V_{kl'}|}{|V_{kh}||V_{k(h+1)}|}(\bar{z}_{kl}-\bar{z}_{kl'})
+ \frac{1}{|A_k|}\sum_{l=d_k}^{h-1} \frac{|V_{kl}|}{|V_{kh}|}(\bar{z}_{kl}-z_{j'})\\
&& + \frac{1}{|A_k|}\sum_{l'=h+2}^{u_k} \frac{|V_{kl'}|}{|V_{k(h+1)}|}(z_j - \bar{z}_{kl'}) + \frac{1}{|A_k|}(z_j-z_{j'}) \\
&=& \frac{1}{|A_k|}\sum_{l=d_k}^{h-1}\frac{|V_{kl}|(\sum_{l'=h+1}^{u_k}|V_{kl'}|)}{|V_{kh}||V_{k(h+1)}|}\bar{z}_{kl}
+ \frac{1}{|A_k|}\frac{\sum_{l'=h+1}^{u_k}|V_{kl'}|}{|V_{k(h+1)}|}z_j \\
&&- \frac{1}{|A_k|}\sum_{l'=h+2}^{u_k}\frac{(\sum_{l=d_k}^{h}|V_{kl}|)|V_{kl'}|}{|V_{kh}||V_{k(h+1)}|}\bar{z}_{kl'}
- \frac{1}{|A_k|}\frac{\sum_{l=d_k}^{h}|V_{kl}|}{|V_{kh}|}z_{j'} \\
&=& \frac{1}{|V_{kh}||V_{k(h+1)}|}\Big[ \frac{(\sum_{l'=h+1}^{u_k}|V_{kl'}|)}{|A_k|}\sum_{l=d_k}^{h}\sum_{i\in V_{kl}}z_i - \frac{(\sum_{l=d_k}^{h}|V_{kl}|)}{|A_k|}\sum_{l'=h+1}^{u_k}\sum_{i\in V_{kl'}}z_i   \Big] \\
&& +  \frac{1}{|V_{kh}||V_{k(h+1)}|} \Big[\frac{(\sum_{l'=h+1}^{u_k}|V_{kl'}|)}{|A_k|}\sum_{i\in V_{kh}}(z_j-z_i) + \frac{(\sum_{l=d_k}^{h}|V_{kl}|)}{|A_k|}\sum_{i\in V_{k(h+1)}}(z_i-z_{j'})\Big].
\end{eqnarray*}
Let $A^1_{kh}=\cup_{l\leq h}V_{kl}$ and $A^2_{kh}=\cup_{l>h}V_{kl}$. Then for any $(j,j')$ such that $j\in B_h$, $j'\in B_{h+1}$ and $j,j'\in A_k$, we have the following expression
\begin{eqnarray}  \label{tau(ij)}
\tau_{jj'}(\bz) &=& \frac{1}{|V_{kh}||V_{k(h+1)}|}\Big[ \frac{|A^2_{kh}|}{|A_k|}\sum_{i\in A^1_{kh}}z_i - \frac{|A^1_{kh}|}{|A_k|}\sum_{i\in A^1_{kh}}z_i \Big] \notag\\
&& +  \frac{1}{|V_{kh}||V_{k(h+1)}|} \Big[\frac{|A^2_{kh}|}{|A_k|}\sum_{i\in A^1_{kh}}(z_j-z_i) + \frac{|A^1_{kh}|}{|A_k|}\sum_{i\in A^2_{kh}}(z_i-z_{j'})\Big].
\end{eqnarray}
Combining \eqref{K1}, \eqref{K21} and \eqref{K22} gives
\begin{eqnarray*}
Q_n(\bbeta)-Q_n(\bbeta^*) &=& \sum_{l=1}^{L-1}\sum_{i\in B_l, j\in B_{l+1}, i\sameA j} \big[ \lambda_1  \rho_1'(2t_n)|\beta_i-\beta_j| - n^{-1}\tau_{ij}(\bz)(\beta_i-\beta_j)\big]\\
&& +  \sum_{l=1}^L \sum_{i,j\in B_l, i\sameA j} \big[\lambda_2 \rho_2'(2t_n) |\beta_i-\beta_j| - n^{-1} \theta_{ij}(\bz)(\beta_i-\beta_j)\big].
\end{eqnarray*}
Therefore, to show \eqref{key2suffcond}, we only need to show for sufficiently small $t_n$,
\beq \label{cond_tau+theta}
n^{-1}\max_{ij}|\tau_{ij}(\bz)|\leq \lambda_1\rho'_1(t_n), \quad n^{-1}\max_{ij}|\theta_{ij}(\bz)|\leq \lambda_2\rho'_2(t_n).
\eeq

Note that $\bz=\bX^T\bepsilon + \boldsymbol{\eta} + \boldsymbol{\eta}^m$, where $\boldsymbol{\eta}=\bX^T\bX(\bbeta^*-\bbeta^0)$ and $\boldsymbol{\eta}^m = \bX^T\bX(\bbeta^m-\bbeta^*)$. It is seen that $\Vert\boldsymbol{\eta}^m \Vert\leq \lambda_{\max}(\bX^T\bX)\Vert\bbeta-\bbeta^*\Vert\leq \lambda_{\max}(\bX^T\bX)t_n$. So $\tau_{ij}(\bz)=\tau_{ij}(\bX^T\bepsilon+\boldsymbol{\eta})+rem$, where the remainder term is uniformly bounded by $g(t_n)$ with $g(0)=0$. Similar situations are observed for $\theta_{ij}(\bz)$. As a result, to show \eqref{cond_tau+theta}, it suffices to show
\beq  \label{cond_theta2}
n^{-1}\max_{ij}|\theta_{ij}(\bX^T\bepsilon+\boldsymbol{\eta})| < \lambda_2\rho'_2(0+),
\eeq
and
\beq \label{cond_tau2}
n^{-1}\max_{ij}|\tau_{ij}(\bX^T\bepsilon+\boldsymbol{\eta})| < \lambda_1\rho'_1(0+).
\eeq

First, consider \eqref{cond_theta2}. Let $E'_{31}$ be the event
\[
n^{-1}\max_{i,j\in A_k}|\theta_{ij}(\bX^T\bepsilon)|\leq \sqrt{c_3^{-1}\log(2p)/(n|A_k|)}, \quad \text{for all }k.
\]
Note that $\theta_{ij}(\bX^T\bepsilon)=\tfrac{1}{|A_k|}(\bx_i-\bx_j)^T\bepsilon$, where $\Vert \bx_i-\bx_j\Vert=\sqrt{2n}$. Moreover, the number of such pairs is bounded by $|A_k|^2/2\leq p^2/2$. Applying Condition 3.3 and the union bound, we see that  $P((E'_{31})^c)\leq p^{-1}$.
Moreover, $|\theta_{i,j}(\boldsymbol{\eta})|\leq \tfrac{1}{|A_k|}\max_{i'}|\eta_{i'}-\bar{\eta}_{kh}|$, where $\bar{\eta}_{kh}$ is the average of $\{\eta_i: i\in V_{kh}\}$. Note that $\max_{i\in V_{kh}}|\eta_i-\bar{\eta}_{kh}|\leq n\nu_k\Vert\bbeta^*-\bbeta^0\Vert$ and $\Vert\bbeta^*-\bbeta^0 \Vert\leq \Vert \bbeta - \bbeta^0\Vert$ because $\bbeta^*$ is the orthogonal projection of $\bbeta$ onto $\cM_A$. Noticing that $\bbeta\in \cB$, we obtain
\[
n^{-1}\max_{i,j}|\theta_{ij}(\boldsymbol{\eta})| \leq C\nu_k |A_k|^{-1} \sqrt{K\log(n)/n}.
\]
Combing the above results to the choice of $\lambda_2$ yields $n^{-1}\max_{i,j}|\theta_{ij}(\bz)|\ll \lambda_2$, and \eqref{cond_theta2} follows.

Next, consider \eqref{cond_tau2}.
In \eqref{tau(ij)}, the first term can be written as $\tfrac{1}{|V_{kh}||V_{k(h+1)}|}w_{kh}(\bz)$, where $w_{kh}(\bz)$ has a similar form to that of $w_j(\bz)$ in \eqref{reorder}. Let $E'_{32}$ be the event that
\beq  \label{w_term3}
n^{-1} \max_{k,h} |w_{kh}(\bX^T\bepsilon+\boldsymbol{\eta})| \leq C\left( \sqrt{\frac{\sigma_k|A_k|\log(p)}{n}} + \nu_k |A_k|\sqrt{\frac{K\log(n)}{n}}\right).
\eeq
It is easy to see that we can follow the steps of proving \eqref{w_term1} and \eqref{w_term2} to show $P((E'_{23})^c)< 2p^{-1}$.
Write the second term in \eqref{tau(ij)} as $\tfrac{1}{|V_{kh}||V_{k(h+1)}|}\tilde{w}_{jj'}(\bz)= \tfrac{1}{|V_{kh}||V_{k(h+1)}|}[\tilde{w}_{jj'}(\bX^T\bepsilon)+ \tilde{w}_{jj'}(\boldsymbol{\eta})]$. First, let $E'_{33}$ be the event that
\beq  \label{tw_term1}
n^{-1} \max_{j,j'}|\tilde{w}_{jj'}(\bX^T\bepsilon)|\leq C\sqrt{\sigma_k|A_k|\log(p)/n}.
\eeq
We observe that $n^{-1}\tilde{w}_{jj'}(\bX^T\bepsilon)=-\ba_{jj'}^T\bepsilon$, where
\[
\ba_{jj'} = n^{-1}\bigg[ \frac{|A^1_{kh}|}{|A_k|}\bX_{A^2_{kh}}(\boldsymbol{1}_{A^2_{kh}}+\be_{j'}) - \frac{|A^2_{kh}|}{|A_k|}\bX_{A^1_{kh}}(\boldsymbol{1}_{A^2_{kh}}+\be_j) \bigg].
\]
So $\Vert \ba_{jj'}\Vert^2\leq 2n^{-1}\sigma_k[\tfrac{2L_1^2}{|A_k|^2}(L_2+1)+\tfrac{2L_2^2}{|A_2|^2}(L_1+1)]\leq n^{-1}\sigma_k(|A_k|+4)$, where $L_1=|A^1_{kh}|$ and $L_2=|A^2_{kh}|$.
Similar to \eqref{prob_E3}, we can show $P((E'_{33})^c)\leq 2p^{-1}$.
Second, note that $|\tilde{w}_{jj'}(\boldsymbol{\eta})|\leq (\tfrac{L_1(L_2+1)}{|A_k|}+\tfrac{L_2(L_1+1)}{|A_k|}) \max_{i\in V_{kh}}|\eta_i-\bar{\eta}_{kh}|\leq C|A_k|\max_{i\in V_{kh}}|\eta_i-\bar{\eta}_{kh}|$, where $\max_{i\in V_{kh}}|\eta_i-\bar{\eta}_{kh}|\leq n\nu_k \Vert \bbeta^*-\bbeta^0\Vert\leq C\nu_k\sqrt{nK\log(n)}$. As a result,
\beq  \label{tw_term2}
n^{-1}\max_{j,j'}|\tilde{w}_{jj'}(\boldsymbol{\eta})|\leq C \nu_k|A_k|\sqrt{K\log(n)/n}.
\eeq
Let $E_3'=E'_{31}\cap E'_{32}\cap E'_{33}$, where $P((E_3')^c)\leq 5p^{-1}$. Combining \eqref{w_term3}-\eqref{tw_term2} gives
\begin{eqnarray*}
n^{-1}\max_{j,j'}|\tau_{jj'}(\bX^T\bepsilon+\boldsymbol{\eta})|
\leq C \max_{k,h}\left\lbrace \frac{1}{|V_{kh}|^2}\left( \sqrt{\frac{\sigma_k|A_k|\log(p)}{n}} + \nu_k |A_k|\sqrt{\frac{K\log(n)}{n}}\right) \right\rbrace,
\end{eqnarray*}
over the event $E_1\cap E_2\cap E_3$. By choice of $\lambda_1$, the right hand side is much smaller than $\lambda_1$. This proves \eqref{cond_tau2}.
\qed

\subsection{Proof of Theorem \ref{thm:normality}}

Write $\bmu^0=T(\bbeta^0)$ and $\hbmu=T(\hbbeta)=\hbbeta_A$. Let $Q_n^A(\bmu)=L_n^A(\bmu)+P_n^A(\bmu)$ be as in the proof of Theorem \ref{thm:rate}. Denote
\[
\cB_n^0=\big\{\bmu\in\mathbb{R}^K: \Vert\bD(\bmu-\bmu^0)\Vert < \sqrt{K\log(n)/n} \big\}
\]
be a neighbourhood of $\bmu^0$. We have seen: (i) $\hbmu \in \cB_n^0$ with probability tending to $1$; (ii) $P_n^A(\bmu)=0$ for $\bmu\in\cB_n^0$; (iii) $\hbmu$ is a strictly local minimum of $Q_n^A$ in $\cB^0_n$. Combining the above, we find that
\[
\frac{\partial}{\partial \bmu}L_n^A(\hbmu) = -\tfrac{1}{n}\bX_A^T(\by-\bX_A\hbmu)= \bzero.
\]
It follows from $\by=\bX_A\bmu^0+\bepsilon$ that
\[
\hbmu - \bmu^0 = (\bX_A^T\bX_A)^{-1}(\bX_A^T\bepsilon).
\]
Therefore, to show the claim, it suffices to show
$ \bB_n(\bX_A^T\bX_A)^{-1/2} \bX_A^T\bepsilon\overset{d}{\to} N(\boldsymbol{0}, \bH)$, i.e.,
for any $\ba\in\mathbb{R}^q$,
\beq  \label{normdist}
\ba^T \bB_n(\bX_A^T\bX_A)^{-1/2} \bX_A^T\bepsilon \overset{d}{\to} N(0, \ba^T\bH\ba).
\eeq

Let $\bv= \bX_A (\bX_A^T\bX_A)^{-1/2}\bB_n^T \ba$,  and write the left hand side of \eqref{normdist} as $\bv^T\bepsilon=\sum_{i=1}^n v_i\varepsilon_i$. The $v_i\varepsilon_i$'s are independently distributed with $E[v_i\varepsilon_i]=0$ and $E[|v_i\varepsilon_i|^2]=v_i^2$. Let $s_n^2=\sum_{i=1}^n E[|v_i\varepsilon_i|^2]$. By Lindeberg's central limit theorem, if for any $\epsilon>0$,
\beq  \label{Lindeberg}
\lim_{n\to\infty} s_n^{-2} E\big[ |v_i\varepsilon_i|^2 1\{ |v_i\varepsilon_i|>\epsilon s_n \}\big] =0,
\eeq
then $s_n^{-1}\sum_{i=1}^n v_i\varepsilon_i\overset{d}{\to} N(0,1)$. Since $s_n^2 = \ba^T\bB_n\bB_n^T\ba\to \ba^T\bH\ba$, \eqref{normdist} follows immediately from the Slutsky's lemma.

It remains to show \eqref{Lindeberg}. Using the formula $E[X 1\{ X>\epsilon\}]= \epsilon P(X>\epsilon)+ \int_{\epsilon}^\infty P( X>u)du$ for $X=|v_i\varepsilon_i|^2$, we have
\[
E \big[ |v_i\varepsilon_i|^2 1\{ |v_i\varepsilon_i|>\epsilon s_n \} \big]
= \epsilon^2s_n^2 P(|v_i\varepsilon_i| > \epsilon s_n) + \int_{\epsilon s_n}^{\infty} P(|v_i\varepsilon_i|>\sqrt{u})du
\]
From Condition 3.3, $P(|v_i\varepsilon_i| > \epsilon s_n)\leq 2\exp(- c_3\epsilon^2 s_n^2/|v_i|^2)$ and $\int_{\epsilon s_n}^{\infty} P(|v_i\varepsilon_i|>\sqrt{u})du\leq 2\int_{\epsilon s_n}^{\infty}\exp(- c_3u/|v_i|^2)du = 2|v_i|^2/c_3\exp(-c_3\epsilon s_n/|v_i|^2)$. Note that $\exp(-x)\leq x^{-k}$ for any $x>0$ and positive integer $k$. It follows that
\[
\frac{1}{s_n^2}\sum_{i=1}^n E \big[ |v_i\varepsilon_i|^2 1\{ |v_i\varepsilon_i|>\epsilon s_n \} \big] \leq \frac{1}{s_n^2}\sum_{i=1}^n \Big( 2\epsilon s_n \frac{|v_i|^4}{c_3^2\epsilon^4 s_n^4} + \frac{2|v_i|^2}{c_3}\frac{|v_i|^2}{c_3\epsilon s_n} \Big) \leq C\max_{i}|v_i|^2,
\]
where in the last inequality we have used the facts that $s_n=\sum_{i=1}^n |v_i|^2$ and $s_n^{-1}= O(1)$.
Note that $\Vert \bv \Vert_{\infty}\leq \Vert \bX_A(\bX_A^T\bX_A)^{-1}\Vert_\infty \Vert \bB_n^T \Vert_{2,\infty} \Vert \ba\Vert = o(1)$.
\qed

\subsection{Proof of Corollary \ref{coro:compare}}

It is easy to see that the asymptotic variance of $\ba_n^T(\hbbeta^{ols}-\bbeta^0)$ is $\ba_n^T(\bX^T\bX)^{-1}\ba_n=v_{1n}$. To compute the asymptotic variance of $\ba_n^T(\hbbeta-\bbeta^0)$, note that
\[
\ba_n^T(\hbbeta-\bbeta^0)= \ba_n^T\bM_n\bD(\bX_A^T\bX_A)^{-1/2}(\bX_A^T\bX_A)^{1/2}(\hbbeta_A-\bbeta^0_A),
\]
where $\bD=\text{diag}(|A_1|^{1/2},\cdots, |A_K|^{1/2})$. Applying Theorem \ref{thm:normality}, the asymptotic variance is
\[
\ba_n^T\bM_n\bD(\bX_A^T\bX_A)^{-1}\bD\bM_n^T\ba_n.
\]
Since $\bX_A=\bX\bM_n\bD$, the above quantity is equal to $\ba_n^T\bM_n(\bM_n^T\bX^T\bX\bM_n)^{-1}\bM_n^T\ba_n=v_{2n}$.

Next, we show $v_{1n}>v_{2n}$. There exists an orthogonal matrix $\bQ$ such that $\bM_n$ is equal to the first $K$ columns of $\bQ$. Write $\bb=\bQ^T\ba_n$ and $\bG=\bQ^T\bX^T\bX\bQ$.
Direct calculations yield $v_{1n}=\bb^T\bG^{-1}\bb$ and $v_{2n}=\bb_1^T\bG_{11}^{-1}\bb_1$, where $\bb_1$ is the subvector of $\bv$ formed by its first $K$ elements and $\bG_{11}$ is the upper left $K\times K$ block of $\bG$. From basic algebra, $v_{1n}\geq v_{2n}$.
\qed

\subsection{Proof of Theorem \ref{thm:cards+sparse}}

It suffices to show that $\hbbeta^{oracle}$ is a strictly local minimum of $Q_n^{sparse}$ with probability at least $1-\epsilon_0-n^{-1}K- 2p^{-1}$. First, there exists an event $E_1$ such that $P(E_1^c)<\epsilon_0$ and $\cB$ preserves the order of $\bbeta^0$ over the event $E_1$. Second, for a sufficiently large constant $C$, define $\cB$ as the set of all $\bbeta$ such that $\Vert \bbeta-\bbeta^0\Vert\leq C\sqrt{K\log(n)/n}$. By recalling the proof of Theorem \ref{thm:fuse}, we see that there exists an event $E_2$ such that $P(E_2^c)\leq n^{-1}K$ and $\hbbeta^{oracle}\in\cB$ over the event $E_2$. Third, for any $\bbeta\in \cB$, let $\bbeta_S$ be the vector such that $\beta_{S,j}=\beta_j1\{j\in S\}$, where $S$ is the support of $\bbeta^0$; and let $\bbeta_S^*$ be the orthogonal projection of $\bbeta_S$ onto $\cM_A^*$. We aim to show there exists an event $E_3$ such that $P(E_3^c)\leq 2p^{-1}$ and over the event $E_1\cap E_2\cap E_3$:
\beq  \label{sparsekey1}
Q_n^{sparse}(\bbeta^*_S)\geq Q_n^{sparse}(\bbeta^{oracle}), \qquad \text{for any }\bbeta\in\cB,
\eeq
and the inequality is strict whenever $\bbeta^*_S\neq \hbbeta^{oracle}$; for a positive sequence $\{t_n\}$,
\beq \label{sparsekey2}
Q_n^{sparse}(\bbeta_S)\geq Q_n^{sparse}(\bbeta_S^*), \qquad \text{for any }\bbeta\in\cB \text{ and } \Vert \bbeta_S-\hbbeta^{oracle}\Vert\leq t_n,
\eeq
and the inequality is strict whenever $\bbeta_S\neq \bbeta^*_S$; for a positive sequence $\{t'_n\}$,
\beq \label{sparsekey3}
Q_n^{sparse}(\bbeta)\geq Q_n^{sparse}(\bbeta_S), \qquad \text{for any }\bbeta\in\cB \text{ and }\Vert \bbeta-\hbbeta^{oracle}\Vert\leq t'_n,
\eeq
and the inequality is strict whenever $\bbeta\neq \bbeta_S$.

Suppose \eqref{sparsekey1}-\eqref{sparsekey3} hold. Consider the neighborhood of $\hbbeta^{oracle}$ defined as $\cB_n=\{\bbeta\in\cB: \Vert \bbeta-\hbbeta^{oracle}\Vert \leq \min\{t_n, t'_n\}\}$. It is easy to see that $\Vert \bbeta-\hbbeta^{oracle}\Vert\leq t'_n$ and $\Vert \bbeta_S-\hbbeta^{oracle}\Vert\leq \Vert \bbeta-\hbbeta^{oracle}\Vert\leq t_n$ for any $\bbeta\in\cB_n$. As a result, $Q_n^{sparse}(\bbeta)\geq Q_n^{sparse}(\hbbeta^{oracle})$ for $\bbeta\in\cB_n$, and the inequality is strict except that $\bbeta=\bbeta_S=\bbeta_S^*=\hbbeta^{oracle}$. It follows that $\hbbeta^{oracle}$ is a strictly local minimum of $Q_n^{sparse}$.

Now, we show \eqref{sparsekey1}-\eqref{sparsekey3}. The proofs of \eqref{sparsekey1} and \eqref{sparsekey2} are exactly the same as those of \eqref{key1} and \eqref{key2}, by noting that $Q_n^{sparse}(\bbeta)=Q_n(\bbeta)$ for any $\bbeta$ whose support is contained in $S$. To show \eqref{sparsekey3}, write
\[
Q_n^{sparse}(\bbeta)- Q_n^{sparse}(\bbeta_S) = -\frac{1}{n}(\by-\bX\bbeta^m)^T\bX(\bbeta-\bbeta_S) + \lambda\sum_{j\notin S}\bar{\rho}(\beta^m_j)\beta_j,
\]
where $\bbeta^m$ lies in the line between $\bbeta$ and $\bbeta_S$. First, note that $\sgn(\beta^m_j)=\sgn(\beta_j)$ for $j\notin S$. Second, $\Vert \bbeta^m - \bbeta_S\Vert\leq \Vert \bbeta-\bbeta_S\Vert\leq t'_n$. Hence, for $j\notin S$, $|\beta_j^m|\leq t'_n$. By the concavity of $\rho$, $\rho'(|\beta_j^m|)\geq \rho'(t'_n)$. Third, write $z= \bX^T(\by - \bX\bbeta^m) = \bX^T\bepsilon + \boldsymbol{\eta}+ \boldsymbol{\eta}^m$, where $\boldsymbol{\eta}= \bX^T\bX(\bbeta^0 - \bbeta_S)$ and $\boldsymbol{\eta}^m = \bX^T\bX(\bbeta_S-\bbeta^m)$. Combining the above,
\[
Q_n^{sparse}(\bbeta)- Q_n^{sparse}(\bbeta_S)
\geq \sum_{j\notin S} [\lambda \rho'(t_n') - n^{-1}z_j]|\beta_j|
\geq \sum_{j\notin S} \big[\lambda \rho'(0+) - \Vert\tfrac{1}{n}\bX^T\bepsilon\Vert_\infty - g_n('_n)  \big)\big]|\beta_j|,
\]
where $g_n(t'_n)= \lambda[\lambda\rho'(0+)- \lambda\rho'(2t_n)] + n^{-1}\eta_j^m$ satisfying $g_n(0)=0$. First, from Condition 3.1, $\Vert\tfrac{1}{n}\bX^T\bepsilon\Vert_\infty\leq \sqrt{c_3^{-1}\log(2p)/n}$, except for a probability of $2p^{-1}$. Since $\lambda_n\gg \sqrt{\log(p)/n}$, when $n$ is sufficiently large, $C\sqrt{\log(p)/n}\leq \lambda_n/2\rho'(0+)=\lambda_n/2$. Second, since $g_n(0)=0$, we can always choose $t'_n$ sufficiently small so that $g_n(t)\leq \lambda_n/4$ for any $0\leq t\leq t'_n$. Combining the above gives
\[
Q_n^{sparse}(\bbeta)- Q_n^{sparse}(\bbeta_S)\geq \sum_{j\notin S} \frac{\lambda_n}{4}|\beta_j|.
\]
Then \eqref{sparsekey3} follows immediately. The proof is now complete.
\qed

\bibliography{homo}
\bibliographystyle{plainnat}

\end{document}